\documentclass[3p]{elsarticle}

\usepackage[colorlinks=true,linkcolor=blue,urlcolor=blue,citecolor=blue]{hyperref}

\usepackage[utf8x]{inputenc}
\usepackage{color}
\usepackage{bbm} 

\usepackage{amsfonts,amsmath,amssymb,stmaryrd}

\usepackage{graphicx}
\usepackage{subfigure}  
\usepackage{bbm} 
\usepackage{hyperref}
\usepackage{epsfig}
\usepackage{mathrsfs}
\usepackage{verbatim}
\usepackage{centernot}
\usepackage{ulem}
\usepackage{nicefrac}

\usepackage{tabu}

\renewcommand{\l}{\left(}
\renewcommand{\r}{\right)}

\newcommand{\bra}[1]{\langle#1|}
\newcommand{\ket}[1]{|#1\rangle}

\renewcommand{\ij}{{\langle \vec{i}, \vec{j} \rangle}}

\renewcommand{\H}{\hat{\mathcal{H}}}

\renewcommand{\c}{\hat{c}}

\newcommand{\s}{\hat{s}}
\newcommand{\cd}{\hat{c}^\dagger}

\newcommand{\f}{\hat{f}}
\newcommand{\fd}{\hat{f}^\dagger}

\newcommand{\sd}{\hat{s}^\dagger}
\newcommand{\bd}{\hat{b}^\dagger}

\renewcommand{\b}{\hat{b}}

\newcommand{\n}{\hat{n}}

\newcommand{\PG}{\hat{\mathcal{P}}_\mathrm{G}}
\renewcommand{\S}{\hat{S}}

\newcommand{\hc}{\text{h.c.}}

\newcommand{\U}{\hat{U}}

\newcommand{\Ud}{\hat{U}^\dagger}

\newcommand{\Jperp}{J_\perp}
\newcommand{\Jpar}{J_\parallel}
\newcommand{\tpar}{t_\parallel}

\usepackage{array}

\usepackage{cancel,ifthen}
\newcommand{\cmnt}[2][NoInPuT]{\ifthenelse{\equal{#1}{NoInPuT}}{}{{\color{red}\sout{#1}}} {\color{blue} #2}}

\usepackage{bm}	
\renewcommand{\vec}[1]{\bm{#1}}

\begin{document}
\normalem	

\title{Exploration of doped quantum magnets with ultracold atoms}

\author[TUM,MCQST,ITAMP,Harvard]{Annabelle Bohrdt}
\author[LMU,MCQST]{Lukas Homeier}
\author[LMU,MCQST]{Christian Reinmoser}
\author[Harvard,ETH]{Eugene Demler}
\author[LMU,MCQST]{Fabian Grusdt}

\address[TUM]{Department of Physics and Institute for Advanced Study and Technical University of Munich, 85748 Garching, Germany}
\address[MCQST]{Munich Center for Quantum Science and Technology (MCQST), Schellingstr. 4, 80799 M\"unchen, Germany}
\address[ITAMP]{ITAMP, Harvard-Smithsonian Center for Astrophysics, Cambridge, MA 02138, USA}
\address[Harvard]{Department of Physics, Harvard University, Cambridge, MA 02138, USA}
\address[ETH]{Institute for Theoretical Physics, ETH Zurich, 8093 Zurich, Switzerland}
\address[LMU]{Department of Physics and Arnold Sommerfeld Center for Theoretical Physics (ASC), Ludwig-Maximilians-Universit\"at M\"unchen, Theresienstr. 37, 80333 M\"unchen, Germany}

\date{\today}

\begin{abstract}
In the last decade, quantum simulators, and in particular cold atoms in optical lattices, have emerged as a valuable tool to study strongly correlated quantum matter. These experiments are now reaching regimes that are numerically difficult or impossible to access. In particular they have started to fulfill a promise which has contributed significantly to defining and shaping the field of cold atom quantum simulations, namely the exploration of doped and frustrated quantum magnets and the search for the origins of high-temperature superconductivity in the fermionic Hubbard model. Despite many future challenges lying ahead, such as the need to further lower the experimentally accessible temperatures, remarkable studies have already emerged. Among them, spin-charge separation in one-dimensional systems has been demonstrated, extended-range antiferromagnetism in two-dimensional systems has been observed, connections to modern day large-scale numerical simulations were made, and unprecedented comparisons with microscopic trial wavefunctions have been carried out at finite doping. In many regards, the field has acquired new realms, putting old ideas to a new test and producing new insights and inspiration for the next generation of physicists. 

In the first part of this paper, we review the results achieved in cold atom realizations of the Fermi-Hubbard model in recent years. We put special emphasis on the new probes available in quantum gas microscopes, such as higher-order correlation functions, full counting statistics, the ability to study far-from-equilibrium dynamics, machine learning and pattern recognition of instantaneous snapshots of the many-body wavefunction, and access to non-local correlators. Our review is written from a theoretical perspective, but aims to provide basic understanding of the experimental procedures. We cover one-dimensional systems, where the phenomenon of spin-charge separation is ubiquitous, and two-dimensional systems where we distinguish between situations with and without doping. Throughout, we focus on the strong coupling regime where the Hubbard interactions $U$ dominate and connections to $t-J$ models can be justified.

In the second part of this paper, with the stage set and the current state of the field in mind, we propose a new direction for cold atoms to explore: namely mixed-dimensional bilayer systems, where the charge motion is restricted to individual layers which remain coupled through spin-exchange. These systems can be directly realized experimentally and we argue that they have a rich phase diagram, potentially including a strongly correlated BEC-to-BCS cross-over and regimes with different superconducting order parameters, as well as complex parton phases and possibly even analogs of tetraquark states. In particular, we propose a novel, strong pairing mechanism in these systems, which puts the formation of hole pairs at experimentally accessible, elevated temperatures within reach. Ultimately we propose to explore how the physics of the mixed-dimensional bilayer system can be connected to the rich phenomenology of the single-layer Hubbard model.
\end{abstract}

\maketitle

\newpage 

\tableofcontents

\section{Introduction}
\label{SecIntro}
As early as 1959, Philip W. Anderson suggested a model which describes electrons hopping on a lattice with a strong on-site interaction \cite{Anderson1959}. Four years later, John Hubbard formally introduced the nowadays well-known Fermi-Hubbard model \cite{Hubbard1963} in order to describe electrons in solids. 
While the model was originally mainly used to study magnetism and was not considered to be relevant to superconductivity, Scalapino et al. \cite{Scalapino1986} showed in 1986 that close to a spin-density-wave instability, one can find d-wave pairing in the three dimensional Fermi-Hubbard model. 
The discovery of high-temperature superconductivity in the cuprate materials earlier the same year \cite{Bednorz1986} led to an increasing interest in strongly correlated electronic systems. Arguably the most important model in this context is the Fermi-Hubbard model, which is the conceptually simplest model describing strongly correlated electrons, and has been the subject of intensive studies over the past decades. Today, the two-dimensional Fermi-Hubbard model is widely believed to capture many of the essential aspects of the cuprate phase diagram. 

Among Anderson's most influential, but also most controversial, ideas was the proposal that resonating valence bond (RVB) states may describe the physics of the doped Hubbard model,  and by extension high-temperature superconductivity \cite{Anderson1987}.  Although this view has been questioned by many and cannot be viewed as generally accepted, the RVB paradigm has undoubtedly had a profound influence on the way we think about the doped Hubbard model and strongly correlated quantum matter in general. 
Much like at the time when Anderson actively worked on the high-Tc problem, understanding the rich phenomenology of high-temperature superconductors still remains one of the biggest outstanding puzzles today.

\begin{figure}[t!]
\centering
\epsfig{file=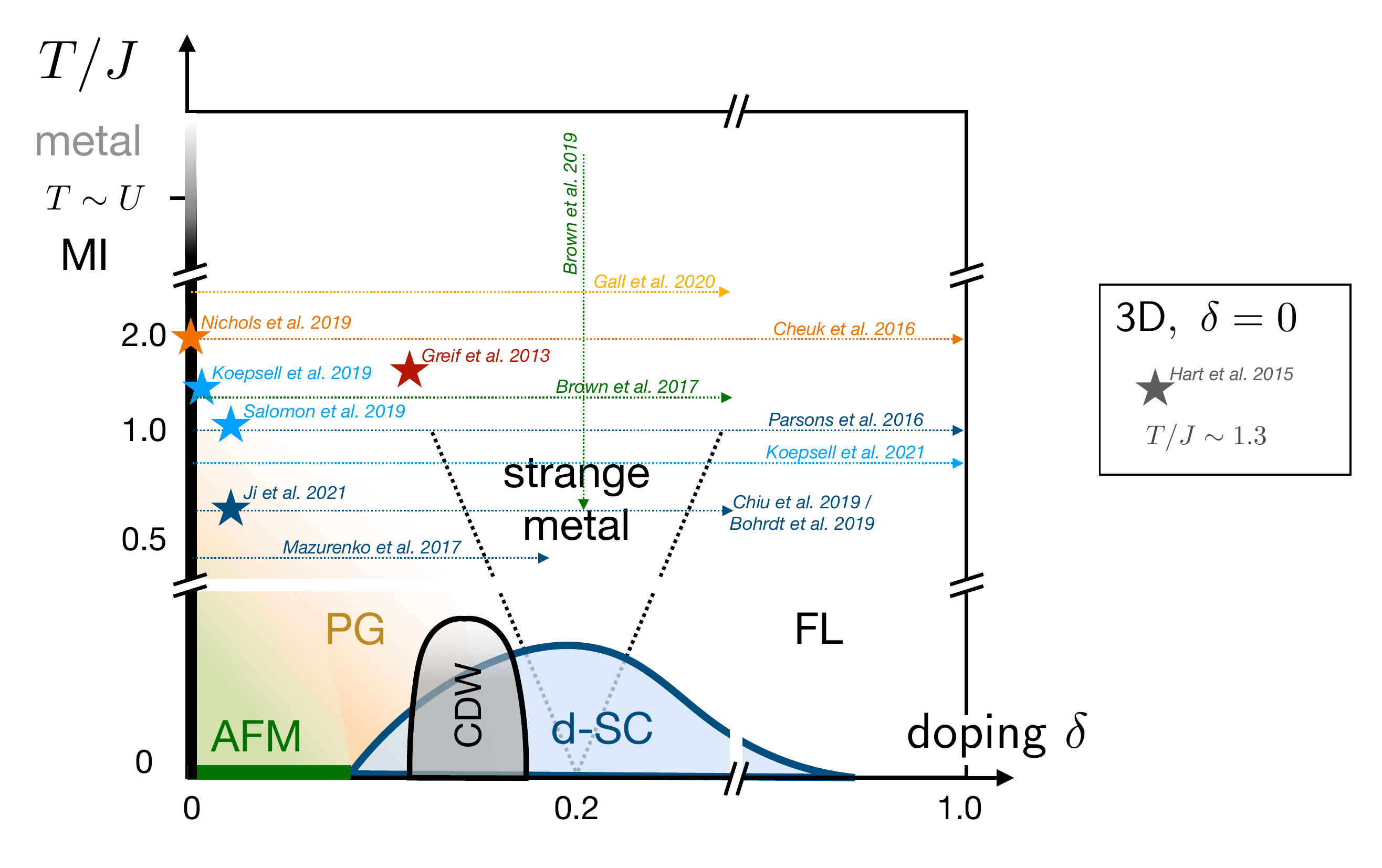, width=0.75\textwidth}
\caption{\textbf{Exploration of the doped  Hubbard model with ultracold atoms.} The phase diagram shows the distinct regimes believed to exist in the strictly two-dimensional Fermi Hubbard model: Mott insulator (MI), antiferromagnet (AFM), pseudogap (PG), charge density wave (CDW), $d$-wave superconductor (d-SC) and Fermi liquid (FL). Note that some phases may require specific coupling strengths to be stabilized \cite{Qin2019}, and additional phases may exist. Ultracold atom experiments have started to explore the clean two-dimensional Hubbard model in the indicated parameter regions. Among others, the publications marked by arrows/ symbols will be reviewed in this article, including from the Harvard group (dark blue - Mazurenko et al. 2017 \cite{Mazurenko2017}, Chiu et al. 2019 \cite{Chiu2019Science} and Bohrdt et al. 2019 \cite{Bohrdt2019_ML}, Ji et al. 2021 \cite{Ji2020}, Parsons et al. 2016 \cite{Parsons2016}), the Munich group (light blue - Koepsell et al. 2021 \cite{Koepsell2020_FL}, Salomon et al. 2019 \cite{Salomon2019}, Koepsell et al. 2019 \cite{Koepsell2019}), the Princeton group (green - Brown et al. 2019 \cite{Brown2019a}, Brown et al. 2017 \cite{Brown2017}), the Z\"urich (ETH) group (red - Greif et al. 2013 \cite{Greif2013}), the MIT group (orange - Nichols et al. 2019 \cite{Nichols2018}, Cheuk et al. 2016 \cite{Cheuk2016}) and the Bonn group (yellow - Gall et al. 2020 \cite{Gall2020}). Pioneering work in three dimensions by the Houston (Rice) group is also mentioned (gray -- Hart et al. 2015 \cite{Hart2015}).}
\label{fig1Overview}
\end{figure}

Recently, an entirely new kind of experiments \cite{Greiner2002a,Greiner2003,Bloch2008,Bakr2009,Sherson2010,Bernien2017} has started to explore this decades old problem from a fresh perspective: Quantum simulators are now emulating the Fermi-Hubbard model, and closely related cousins thereof, in a clean setting with fully tunable model parameters \cite{Esslinger2010,Hart2015,Cheuk2015,Parsons2015,Omran2015,Cocchi2016,Salfi2016,Dehollain2020,Arute2020}. 
Arguably the most promising approach to quantum simulating the Fermi-Hubbard model currently are cold atoms in optical lattices, which provide an accurate experimental realization of this paradigmatic model.
In these systems the ratio of interaction to tunneling can be controlled using Feshbach resonances and the strength of the optical lattice.   
With the advent of quantum gas microscopy, single atom- combined with single site resolved imaging of these systems has become possible, paving the way for an entirely new perspective on the study of strongly correlated systems, and the Fermi-Hubbard model in particular, see Fig.~\ref{fig1Overview}. For example, the projective measurements performed with a quantum gas microscope directly enable the study of higher order correlations between spin and charge \cite{Hilker2017,Koepsell2020_FL,Bohrdt2020_gauss}, including the search for specific patterns motivated by theoretical descriptions \cite{Chiu2019Science}. This is in contrast with traditional condensed matter physics, which can be understood from the perspective of two-point correlation functions and momentum space probes, as they are directly accessible  in experiments such as neutron scattering and angle resolved photoemission spectroscopy (ARPES). 

For some problems, however, a real space description can be more illuminating from a theory perspective. One such example is the motion of a single hole in an antiferromagnet, which lies at the heart of understanding the doped Fermi-Hubbard model. Already in some of the earliest theoretical work on this problem, a real space approach was taken \cite{Bulaevskii1968, Sachdev1989}. With the advent of quantum gas microscopes, which naturally probe the system in real space and time, this perspective has become directly accessible in experiments. Theoretical predictions for the real space correlations of a single hole with the spin background can now be confirmed experimentally \cite{Grusdt2019,Koepsell2019,Chiu2019Science}. With this perspective in mind, it becomes clear that quantum simulators, and in particular quantum gas microscopes, can go beyond reproducing the experimental results available in solid state electronic systems, but rather provide a complementary experimental platform with its own powerful toolbox. The one-hole problem mentioned above is particularly interesting in this regard, because like polaronic problems in general \cite{Franchini2021} it allows broad theoretical and experimental insights.

A typical approach to understand the strongly correlated electron systems studied in solid state experiments consists of formulating simplified models and solving them as accurately as possible. 
If the theoretical solution of such a simplified model does not agree with the experimental results, there are two possibilities: either the solution is incorrect or incomplete, or the simplified model does not capture the relevant physics of the electron system. In contrast using quantum simulators, and in particular cold atoms in optical lattices, we exactly know the model that we are analyzing. We can therefore put theoretical solutions to a detailed test by comparing directly to the Fermi-Hubbard model. Almost certainly, this clean model will not capture all the features observed in experiments on the cuprate materials. A detailed analysis and understanding of the Fermi-Hubbard model as well as comparison to results from solid state experiments will be crucial to determining which features are captured, and which are not, and moreover, what is missing: whether it is phonons, details of the band structure, or something else.

\begin{figure}[b!]
\centering
\epsfig{file=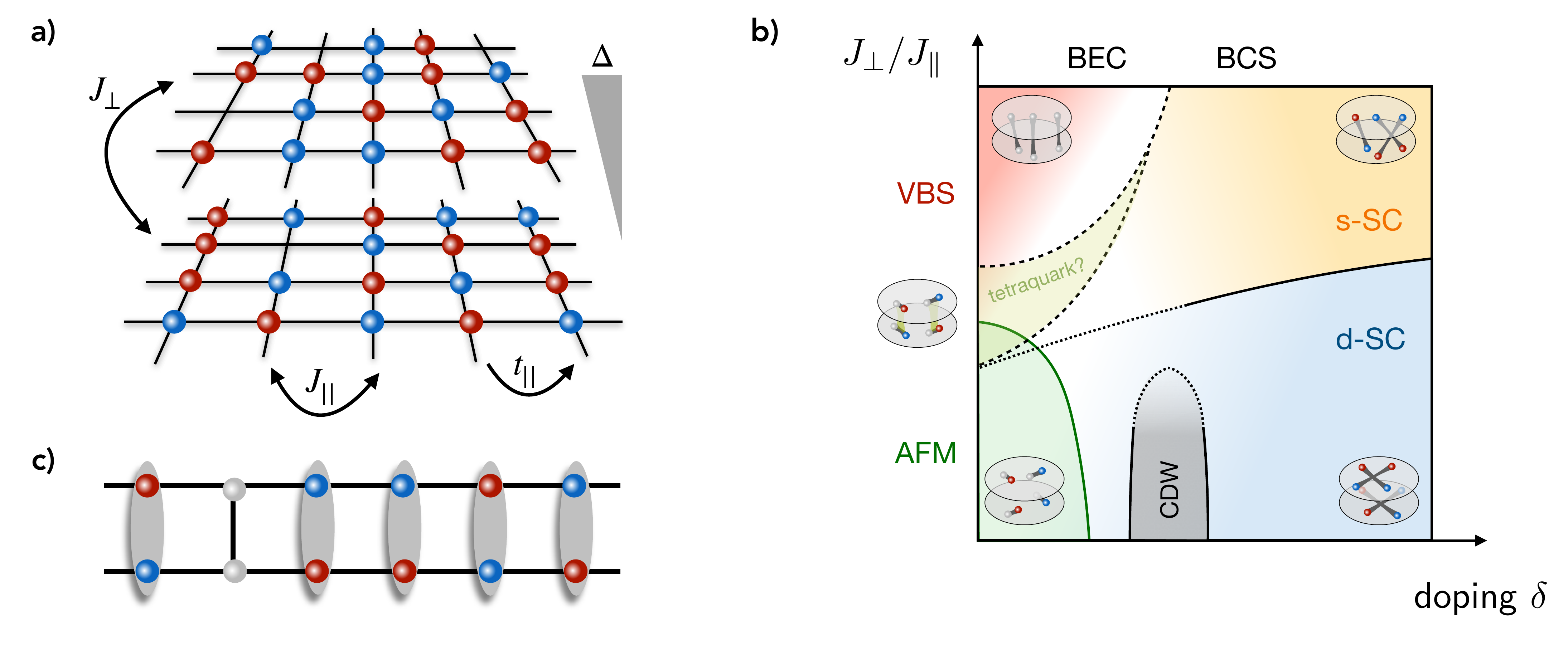, width=0.95\textwidth}
\caption{\textbf{New direction: Mixed-dimensional bilayer system.} a) Sketch of the mixed-dimensional bilayer model: two Fermi-Hubbard layers are coupled vertically. A potential gradient $\Delta$ inhibits tunneling between the layers, but still allows for a tunable superexchange coupling $J_\perp$. b) The competition of $J_\perp$ with the intra-layer tunneling $t_\parallel$ and superexchange $J_\parallel$ is expected to lead to a rich phase diagram: We predict that it includes a BEC regime of tightly bound hole pairs at low doping (red area) when the spins form a valence-bond solid (VBS) of vertical rung singlets. When $t_\parallel \gg J_\perp$ a cross-over to a BCS regime at higher doping takes place. We further speculate that the formation of a long-range AFM within the layers, upon tuning $J_\perp / J_\parallel$ \cite{Gall2021}, may be accompanied with the appearance of exotic tetra-quark states involving two chargons and two spinons. The physics at intermediate values of $J_\perp/J_\parallel$ and doping $\delta$ remains unclear. c) As a key feature of the mixed-dimensional bilayer model, it contains a powerful pairing mechanism for holes in the VBS regime: due to singlet formation between the layers, two holes are tightly bound in the limit of $J_\perp \gg J_\parallel$, while a competing kinetic energy along the rungs is absent.}
\label{fig1Bilayer}
\end{figure}

Apart from new probes, cold atom experiments also provide new possibilities in terms of system parameters and geometries. For example, tunable brick-wall, honeycomb \cite{SoltanPanahi2011,Tarruell2012} and Kagome \cite{Jo2012} lattices have been implemented, and recently the triangular lattice has been realized for fermions in a quantum gas microscope \cite{Yang2021}. Frustrated magnetism in a triangular lattice has earlier been explored using lattices of tubes filled with bosons \cite{Struck2011}. 
Another particularly interesting scenario is provided by the realization of mixed-dimensional settings, where the dynamics of one sector of the Hilbert space is restricted to a lower dimension. For example, for an embedding in two dimensions, the spins interact along both directions, whereas the charge can only move along one dimension. In the context of a single hole in an antiferromagnet, this model was already discussed in Ref.~\cite{Bulaevskii1968} as a well-suited test-case for the parton theory proposed there. Recently, it has been shown that parton gases and stripes can be realized in this setting \cite{Grusdt2020}. In a setup with cold atoms in optical lattices, such a mixed dimensional system can be realized by applying a potential gradient along one direction, which inhibits tunneling of the particles but still allows for tunable spin exchange interactions along the gradient \cite{Trotzky2008,Chen2011c,Dimitrova2019,Sun2020}.

Recently, the bilayer Fermi-Hubbard model has been realized in several quantum gas microscopy experiments \cite{Koepsell2020,Hartke2020,Gall2021} using a superlattice in the vertical direction. With this setup, a mixed-dimensional bilayer system, where the tunneling between the layers is suppressed by a strong gradient, can be directly realized through a potential offset $\Delta$ between the layers, see Fig.~\ref{fig1Bilayer} a). The model parameters, such as the inter- ($J_\perp$) and intra-layer ($J_\parallel$) spin exchange and hopping amplitude $t_\parallel$ within the planes can be tuned over a wide range in such an experiment. We speculate that this model may potentially yield a rich phase diagram with exotic physics, including a BEC-to-BCS crossover from low to intermediate doping, or possibly even a tetraquark phase at low dopings, see Fig.~\ref{fig1Bilayer} b). At high dopings we perform a more rigorous mean-field analysis in terms of a BCS description, which reveals a tunable transition between different pairing symmetries, namely inter-layer $s$-wave and intra-layer $d_{x^2-y^2}$ that could be explored in the future.

Most significantly, here we propose to study the mixed-dimensional bilayer setup as a new route to achieve hole pairing at the currently experimentally achievable temperatures, somewhat below but on the order of the super-exchange energy $J$. The mechanism we suggest relies on breaking up strong singlets between the two layers, which form in the undoped ground state when $J_\perp \gg J_\parallel$, see Fig.~\ref{fig1Bilayer} c). Since tunneling between the layers is strongly suppressed by the gradient, removing both spins from the singlet costs the same amount of energy as removing just one spin, assuming $t_\parallel \ll J_\perp$. This gives rise to a strong binding energy $E_{\rm bdg} = - J_\perp$, significantly larger than in a bilayer \emph{without} the gradient where the kinetic energy $-t_\perp$ leads to a much closer energetic competition of one- and two-hole states. We can give a physical picture for the pairing mechanism in mixed dimensions: When two fermions form an inter-layer singlet along a rung, they can regain some kinetic energy from $t_\perp$ by virtually tunneling to the doubly occupied state; the gain is $t_\perp^2/(U \pm \Delta)$ where $\pm$ refers to which one of the layers is doubly occupied. On the other hand, unpaired fermions on a rung give an energy shift $\pm t_\perp^2 / \Delta$, and contributions from opposite layers cancel to yield zero net energy gain.

For fixed $J_\perp \gg J_\parallel$ we predict strong hole pairing at low doping even in the limit when $t_\parallel \gg J_\perp$ \cite{Bohrdt2021pairingPrep}, although in this case the size of the bound state extends beyond one rung. Hence, for low doping, a condensate of tightly bound bosonic pairs -- in the BEC regime -- is expected to form. For the same parameters but at large doping, we still expect a mean-field BCS description to apply. This leads to our conjecture that the system should host a BEC-to-BCS crossover as a function of doping. This scenario is particularly appealing, since the mixed-dimensional bilayer model features a fully tunable parameter $J_\parallel / J_\perp$ which allows to connect continuously to the limit of weakly coupled layers of $t_\parallel-J_\parallel$ systems, relevant to cuprate compounds and high-Tc superconductivity. In the latter the low-doping regime features a pseudogap, reminiscent of the well-understood pseudogap from BEC-to-BCS crossovers, but its microscopic origin and relation to the underlying pairing mechanism in cuprates remains debated. Thus our proposal constitutes a new experimental route to explore strong hole pairing in a system with purely repulsive microscopic interactions, which moreover is readily accessible by current experiments.

The remainder of this paper is organized as follows. We start by giving a brief introduction into cold atom experiments, and in particular quantum gas microscopy in section \ref{secColdAtoms}. We then proceed by reviewing the progress made in cold atom simulations of the Fermi-Hubbard model in the past few years, building upon earlier excellent reviews, such as \cite{Esslinger2010,Bloch2012,Tarruell2018}. We apologize in advance for the many excellent works we could not include in our article. Here, we start with one-dimensional systems in section \ref{sec1D}, reviewing experimental results on the Fermi-Hubbard model in one dimension in- and out-of equilibrium.
We then proceed in section \ref{sec2DHalffilling} to two dimensions at half filling and continue with the discussion of a single dopant in- and out-of equilibrium.
Section \ref{sec2DDoping} covers the experiments performed on the Fermi-Hubbard model at finite doping, ranging from conventional probes such as two-point correlation functions and transport experiments to higher-order correlations and machine learning techniques.
Our focus here is on new probes in terms of observables, such as higher-order correlations, as well as exploring dynamics beyond the linear response regime. 
In this context, we provide new results on the application of machine learning techniques to Fock space snapshots of doped one-dimensional systems as a particularly illustrative example.
Finally, we introduce the mixed-dimensional bilayer system in section \ref{secPairing}, which can be directly realized in current experimental setups, and analyze its phase diagram in more detail. We summarize the state of the field and conclude with an outlook in section \ref{secSummary}.

\section{Cold atoms and quantum gas microscopy}\label{secColdAtoms}

In the past two decades, cold atoms have emerged as one of the most promising platforms for analog quantum simulation. With the advent of quantum gas microscopy of bosonic \cite{Bakr2009,Sherson2010} as well as fermionic \cite{Parsons2015,Omran2015,Cheuk2015,Haller2015,Edge2015} atoms in optical lattices, a completely new toolbox of manipulation and analysis methods has become available. Here, we give a very brief overview of the experimental capabilities from a theorist's perspective, in particular in the context of the Fermi-Hubbard model, before discussing specific physical results obtained in these systems in the subsequent sections of this paper. We refer to the many excellent reviews, such as \cite{Esslinger2010,Tarruell2018,Gross2017,Gross2020}, for more details on technical implementations.

The experimental procedure typically consists of cooling and trapping a cloud of ultracold atoms and subsequently loading them into an optical lattice created by retroreflected lasers. The spatial arrangement of the lasers can be chosen to realize one-, two-, or three-dimensional square lattices as well as more complicated lattice geometries \cite{Sbroscia2020,Yang2021}. Here, we only consider the one- and two-dimensional square lattice. The Gaussian envelope of the optical beams creates a harmonic trapping potential. In different experimental setups, this is either compensated for by the imaging system of the microscope to realize a flat potential \cite{Mazurenko2017}, or used to obtain a range of dopings in different spatial regions of the system in a single experimental run by means of a local density approximation \cite{Cheuk2016}.

The Fermi-Hubbard model is described by the following Hamiltonian,
\begin{equation}
    \hat{\mathcal{H}} = - t \sum_{\ij} \sum_\sigma \l \cd_{\vec{i},\sigma} \c_{\vec{j},\sigma} + \hc \r + U \sum_{\vec{j}} \hat{n}_{\vec{j},\uparrow} \hat{n}_{\vec{j},\downarrow}, \qquad \text{with}~~ \hat{n}_{\vec{j},\sigma} = \cd_{\vec{j},\sigma} \c_{\vec{j},\sigma}
\end{equation}
where $t$ and $U$ denote nearest neighbor tunneling and the on-site interaction strength, respectively. The two-component fermions $\c_{\vec{j},\sigma}$ have a site index $\vec{j}$ and spin index $\sigma=\uparrow, \downarrow$. 
In order to realize this model experimentally, two hyperfine states of an atom -- typically light atoms such as Lithium or Potassium -- realize spin up and spin down fermions. While it is possible to realize $SU(N)$ symmetric systems by addressing more internal states \cite{Honerkamp2004,Ozawa2018}, we here focus on the spin-$1/2$ case. Using two hyperfine states of bosonic atoms, it is also possible to realize the spin-$1/2$ Bose-Hubbard model  \cite{Duan2003,Trotzky2008,Trotzky2010,Brown2015,Jepsen2020}. 

Depending on the specifics of the loading sequence, spin-balanced as well as spin-imbalanced systems can be studied \cite{Brown2017,Nichols2018}.
The model parameters are tunable with a large degree of control, for example the interaction strength $U$ can be changed by tuning a magnetic field across a Feshbach resonance \cite{Bloch2008}. This also allows to control the resulting super-exchange interactions 
\begin{equation}
    J = 4 t^2 / U
\end{equation}
between the spins, which play an important role in the regime mostly considered in this paper where $U \gg t$.
Here, we only consider on-site interactions between fermions with different spin, however, long-range interactions can also be realized using for example dipolar atoms \cite{Lahaye2009,Lu2012,Aikawa2013}, molecules \cite{Ospelkaus2006,Moses2015}
or Rydberg dressing \cite{GuardadoSanchez2020Ry}. 
The chemical potential or doping is effectively controlled through the number of atoms loaded into the optical lattice during the initial stage of the experiment. In order to determine the chemical potential corresponding to half-filling, i.e. one atom per site, one can thus tune the chemical potential $\mu$ and for each value of $\mu$, count the number of singly occupied sites. At half-filling, this number is maximized. 

In a quantum gas microscope, atoms are imaged with single-site and single-atom resolution with the help of fluorescence imaging. So far, in most experiments, one obtains an image of the singly occupied sites, since doubly occupied sites undergo light assisted collisions and therefore appear empty. In order to gain information on the spin sector, two different procedures have been realized: one can either remove one of the two hyperfine states using a short pulse of resonant light, resulting in images where only atoms in the other hyperfine state are visible \cite{Parsons2016}. The other option is to split each site adiabatically in two while applying a magnetic field gradient, thus performing a local Stern-Gerlach experiment \cite{Boll2016,Koepsell2020}. The latter technique enables \emph{full resolution}, meaning that both spin states are imaged simultaneously.

The imaging system of the microscope itself can also be used to manipulate the potential landscape for the atoms with the help of a digital micromirror device (DMD), which consists of an array of adjustable mirrors that are illuminated with light. DMDs can be used to cancel the harmonic potential from the trap as well as to alter the potential on individual sites \cite{Zupancic2016} or create box potentials. As another important example, it is possible to initialize a state with a pinned hole \cite{Ji2020} or a density modulation \cite{Brown2019a}. 

Apart from the preparation of such initial equilibrium states, one may apply sudden perturbations including the removal of a single atom using a near-resonant laser beam \cite{Vijayan2020}, or transferring an atom to a non-interacting final hyperfine state to perform what would be the analogue of (angle-resolved) photoemission spectroscopy \cite{Chin2004,Stewart2008,Brown2019}. Similarly, lattice modulations can be used as another way to perform spectroscopy \cite{Torma2000,Jordens2008,Greif2011}, including in quantum gas microscope settings \cite{Bohrdt2018}. Local probes \cite{Kollath2007}, the analogues of scanning tunneling microscopy (STM), and multi-photon spectroscopy \cite{Bohrdt2021arXiv} have also been proposed. 

The site resolved imaging of atoms opens up completely new ways to analyze doped quantum magnets: using the instantaneous snapshots taken with a quantum gas microscope, density correlation functions up to arbitrary order \cite{Schweigler2017,Koepsell2020_FL} can be studied. These snapshots enable new possibilities to compare to theories, for example by using experimental snapshots in one regime to make theoretical predictions in another regime \cite{Chiu2019Science}, or by applying machine learning techniques \cite{Bohrdt2019_ML}.

The first experimental results on the Fermi-Hubbard model consisted in the observation of the reduction of double occupancy and compressibility \cite{Jordens2008,Schneider2008,Duarte2015} and short-range antiferromagnetic correlations \cite{Greif2013,Hart2015}. In 2015, several groups obtained their first site resolved images of fermionic atoms with a quantum gas microscope \cite{Cheuk2015,Parsons2015,Omran2015,Edge2015,Haller2015}, thus paving the way for the direct observation of two-dimensional (2D) fermionic Mott insulators \cite{Greif2016,Cheuk2016b} and long-range antiferromagnetic correlations \cite{Mazurenko2017} and the exploration of the phase diagram of the doped Fermi-Hubbard model, which is the topic of the review part of this paper -- see Fig.~\ref{fig1Overview} for an overview. More recently, the crossover into the Mott regime \cite{Hofrichter2016} and short-range anti-ferromagnetic correlations \cite{Ozawa2018} have also been observed in  $SU(N)$ symmetric incarnations of the Fermi-Hubbard model. 

In order to determine the doping without full resolution, the singles density is compared to exact numerical simulations. 
Similarly, the temperature is typically determined by comparison to numerical results, for example quantum Monte Carlo calculations. Depending on the temperature range, charge observables, such as the singles density or charge correlations, or spin correlations are used. In parameter regimes where numerical simulations become challenging, such as low temperatures and finite doping, the temperature is determined for half-filling and assumed to be constant as the system is doped. This procedure is justified if the doped region is in thermal equilibrium with a half-filled region in the system. Recently, a model-free thermometer, based on the fluctation-dissipation theorem, has been demonstrated \cite{Hartke2020}.

So far, the lowest temperatures reached in cold atom realizations of the Fermi-Hubbard model are about half of the superexchange energy scale. The antiferromagnetic correlations at half-filling in this case extended across the observation region of roughly ten by ten sites \cite{Mazurenko2017}. 
In condensed matter systems, the focus is typically on the long wavelength, low energy features, since they are expected to contain information about universal aspects, such as broken symmetries. 
Cold atom platforms are better suited for analyzing the short-range correlations, which are crucial for computing expectation values of the Hamiltonian. Moreover, a comparison to numerical results is usually only possible in terms of short-range correlations, also due to the finite system sizes. While this aspect makes it difficult to distinguish states with no long-range order, but slowly decaying spatial correlations, it should be sufficient to address the most important questions for the Hubbard model regarding the nature of the dominant instability.
Furthermore we expect system sizes to increase in the coming years and new generations of experiments, as there are no intrinsic limitations. In Ref.~\cite{GuardadoSanchez2020} for example, a system of $30\times 36$ sites was studied. Cold atom realizations of the Fermi-Hubbard model without microscopes already reach system sizes of several thousand sites \cite{Gall2021}.

\section{Quantum magnets in 1D: spin-charge separation}\label{sec1D}

Using optical lattices, one-dimensional systems can be realized in a well-controlled setting by ramping up the lattice depth in the two perpendicular spatial directions. In this section we discuss experiments on one-dimensional Fermi Hubbard chains which demonstrated spin-charge separation over a wide range of energies in and out-of-equilibrium, by employing observables directly accessible in quantum gas microscopes. 

Spin-charge separation is a remarkable effect of strong interactions in one spatial dimension, where charge and spin degrees of freedom effectively decouple. Although the underlying constituents are fermions, for which  spin and charge quantum numbers come together, the elementary excitations are solitons, which carry exclusively one of them, either spin (spinons) or charge (holons). This phenomenon is well understood theoretically in the low-energy limit, where it is described by Luttinger liquid theory \cite{Giamarchi2003}. However, it has been predicted to remain robust and valid far from equilibrium even at high excitation energies \cite{Kollath2005,Kollath2006}, and in the spin-incoherent Luttinger liquid regime \cite{Fiete2007} where spin degrees of freedom form an incoherent bath.

In condensed matter systems, spin-charge separation has been probed indirectly at low temperatures and energies with spectroscopic \cite{Kim1996,Segovia1999,Kim2006} or transport measurements \cite{Auslaender2005,Jompol2009}. Using quantum gas microscopes, more direct real space probes through non-local correlation functions are possible \cite{Endres2011} over a wider range of energies and temperatures, as we discuss below.

\subsection{Equilibrium}

\begin{figure}
\centering
  \includegraphics[width=0.99\linewidth]{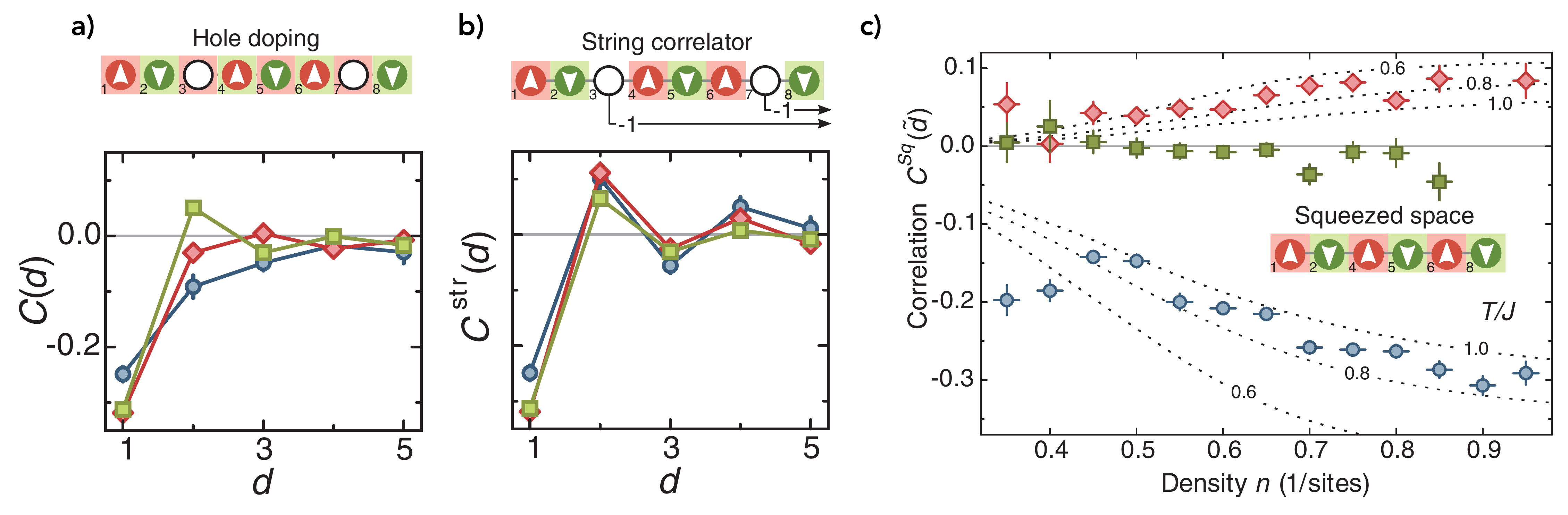}
\caption{\textbf{Spin-charge separation.} a) Spin correlations as a function of distance $d$ for densities $\left<n\right> = 0.4$ (blue), $\left<n\right> = 0.7$ (red), and $\left<n\right> = 1$ (green). b) String correlator (defined in the main text) as a function of distance. c) Squeezed space spin correlations as a function of density for distances $\tilde{d}=1$ (blue), $\tilde{d}=2$ (red), and $\tilde{d}=3$ (green). Dotted lines correspond to the spin correlations $C(d=1)$ and $C(d=2)$ in the Heisenberg model with coupling constant $J_{\text{eff}}(n)$, obtained by exact diagonalization. Figures extracted from \cite{Hilker2017}. }
\label{fig:chains1D}
\end{figure}

\subsubsection{String-order parameters and squeezed space}

In a finite temperature Fermi-Hubbard chain at zero doping, the antiferromagnetic spin correlations decay exponentially with distance. For temperatures below the super-exchange energy as realized in the experiments, correlations extending over several lattice spacings can be measured. Upon doping, these correlations quickly loose their antiferromagnetic character and decay to zero at currently achievable temperatures \cite{Hilker2017}, see Fig.~\ref{fig:chains1D} a). As summarized next, ultracold atom experiments have demonstrated that this strong suppression of spin-correlations is caused primarily by the motion of the doped holes. 

As a consequence of spin-charge separation, the antiferromagnetic correlations are however not lost, but rather hidden \cite{Ogata1990,Kruis2004}. In \cite{Hilker2017}, this hidden order was revealed by taking the hole positions into account. First it was shown that across each hole the spin-spin correlations remain negative, i.e. $C_{\rm SH}(2) = 4 \langle \hat{S}^z_i \hat{n}^h_{i+1} \hat{S}^z_{i+2} \rangle_{\bullet \circ \bullet} < 0$ where $\hat{n}^h_i$ is the hole density operator. This reflects the fact that each hole can be associated with an antiferromagnetic parity domain wall.

To include the effect of multiple holes, one can construct a string-order correlator to take the hole positions -- and thus the antiferromagnetic parity domain walls associated therewith -- into account as follows \cite{Kruis2004}:
\begin{equation}
C^{str}(d) = 4\left<\hat{S}_i^z \left( \prod_{j=1}^{d-1}(-1)^{1-\hat{n}_{i+j}} \right)\hat{S}_{i+d}^z 
\right>_{\bullet_i \bullet_{i+d}},
\end{equation}
where the full circles indicate that the correlator is evaluated only on data where sites $i$ and $i+d$ are occupied by a single atom. This string order correlator adds a minus sign, equivalent to a domain wall, for each hole between the two sites under consideration. Comparing the conventional two-point spin correlation function and the string order correlator for different densities of $\left<n\right> = 0.4, 0.7, 1.0$ shows how accounting for the hole positions almost exactly restores the antiferromagnetic order in the spin sector, see Fig.~\ref{fig:chains1D}.

Another way to reveal the hidden antiferromagnetic order in a one-dimensional system is to work in \emph{squeezed space}: in each snapshot taken with the quantum gas microscope, all holes are removed from the chain, and the remainder of the system is squeezed together, such that in the resulting image, each site is occupied by either a spin up or a spin down atom. The spins in squeezed space can then be described as a Heisenberg spin chain with an effective exchange interaction $J_{\rm{eff}}(n^h)$. In \cite{Hilker2017}, the resulting spin correlations as a function of doping show excellent agreement with theoretical results on the undoped Heisenberg model, see Fig.~\ref{fig:chains1D} c).

When the system exhibits spin-charge separation, the spin and charge sectors approximately factorize and can be considered separately when taking into account their geometric entanglement in squeezed space. Increasing the number of holes in the system effectively decreases the coupling $J_{\rm{eff}}(n^h)$ between the spins in squeezed space since they only exhibit exchange interactions if they are located on neighboring lattice sites. The density of the holes as well as the correlations between holes have to be taken into account in order to derive an expression for $J_{\rm{eff}}(n^h)$ \cite{Hilker2017}.

\subsubsection{Incommensurate magnetism}

\begin{figure}
\centering
  \includegraphics[width=0.99\linewidth]{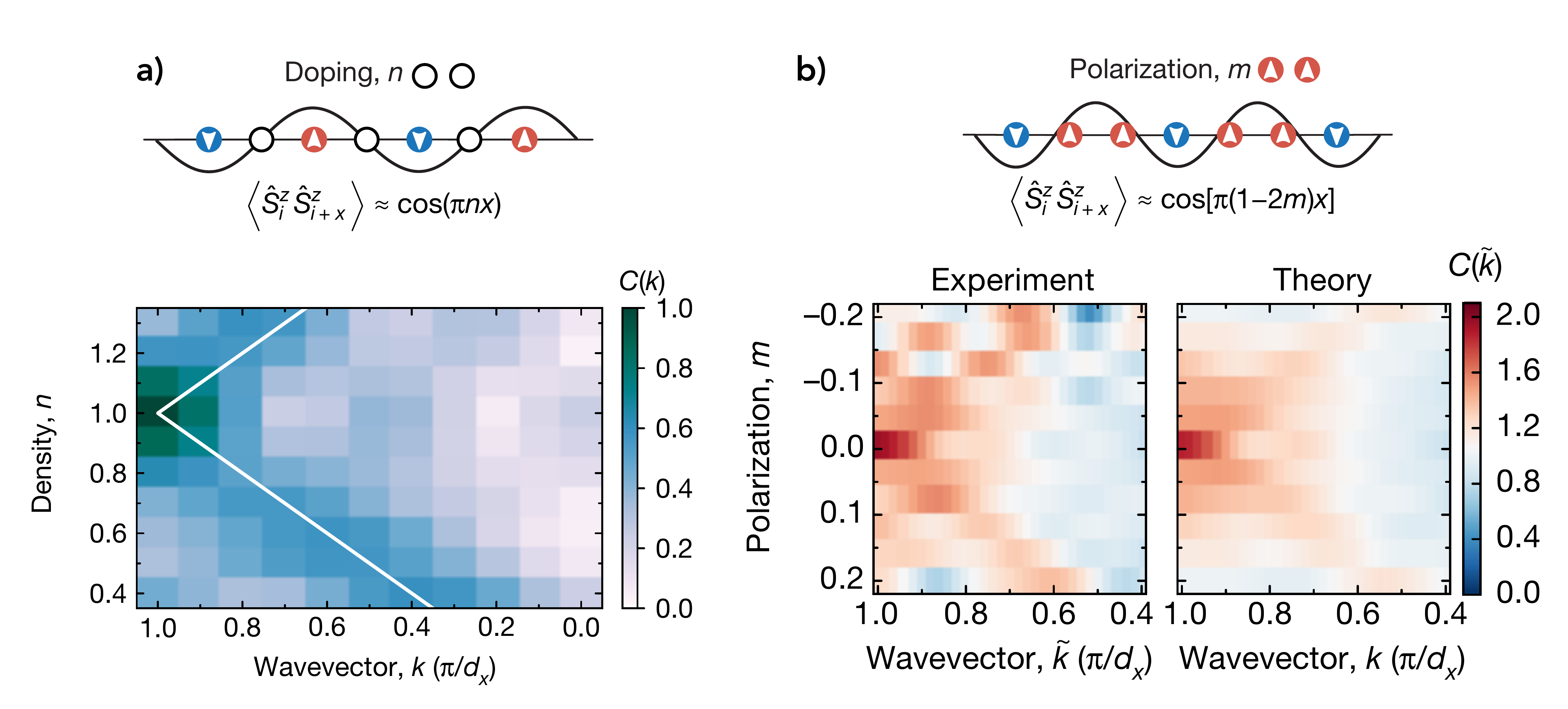}
\caption{\textbf{Incommensurate magnetism in 1D.} a) The normalized Fourier transform of the spin correlations $C(k)$ as a function of wavevector and density. A linear increase of the position of the peak with density is observed. The white line corresponds to the Luttinger liquid theory prediction $k_{\rm{SDW}} = \pi n$. b) Normalized Fourier transform $C(\tilde{k})$ of the squeezed space spin correlations $C(\tilde{x})$ as a function of wavevector $\tilde{k}$ and polarization $m$. Theory results are obtained with exact diagonalization calculations of the Heisenberg chain at $T/J=0.7$. Figures extracted from \cite{Salomon2019}.}
\label{fig:incommensurate1D}
\end{figure}

Another perspective on the influence of doping on the spin correlations is presented in \cite{Salomon2018}: at half-filling, the spin correlations form at a commensurate wavevector of $\pi$, meaning that $\langle\hat{S}_i^z\hat{S}_{i+x}^z\rangle \propto \cos(\pi x)$.  Luttinger liquid theory predicts a spin density wave with wavevector $k_\text{SDW} = 2k_F = \pi n$, where $n$ is the total fermion density and $k_F$ the Fermi momentum. Therefore, upon doping, the delocalization of dopants leads to incommensurate spin correlations with a wavevector of $\pi n$, such that $\langle\hat{S}_i^z\hat{S}_{i+x}^z\rangle \propto \cos(\pi n x)$. This behavior can be probed by considering the normalized Fourier transform of the spin correlations, which reveal a linear increase of the spin density wavevector $k_\text{SDW}$ with density, Fig.~\ref{fig:incommensurate1D}a).

Similarly, a finite spin polarization (magnetization) leads to incommensurate spin correlations due to the excess spin, with a wavevector of $\pi(1-2m)$. Here $m=S^z/N_s$ is the polarization, $S^z$ is the total spin along $z$ and $N_s$ is the number of singly occupied sites. In the experiment \cite{Salomon2018}, this incommensurate magnetism was analyzed in squeezed space, as discussed above, in order to separate the effect of spin imbalance from the effect of non-zero doping.
Considering the Fourier transform of the squeezed space spin correlations as a function of the polarization, the linear dependence of the wavevector was revealed, Fig.~\ref{fig:incommensurate1D}b). The excess spins -- similar as the doped holes -- thus act as delocalized domain walls for the antiferromagnetic order.

\subsection{Dynamical spin-charge fractionalization}\label{sec1Ddyn}

\begin{figure}
\centering
  \includegraphics[width=0.99\linewidth]{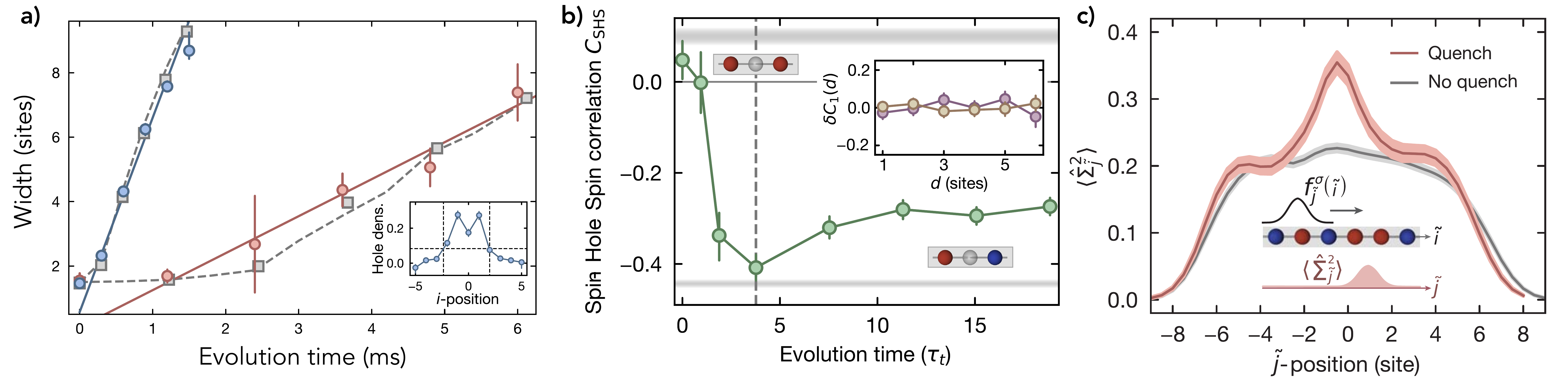}
\caption{\textbf{Dynamical spin-charge fractionalization.}  Dynamics after the creation of a hole in a Fermi-Hubbard chain at half-filling: a) spinon and chargon dynamics are tracked through nearest neighbor spin correlations in squeezed space and hole position, respectively. From the width of the distribution, a velocity is extracted, which scales linearly with $J$ and $t$, respectively. b) Spin-hole-spin correlations $C_{\text{SHS}}$ averaged over the entire chain as a function of time after the quench. The correlator starts with a positive value consistent with the next-nearest neighbour spin correlations $C(2)$ in the absence of the quench (top grey shaded region) and turns negative, approaching the nearest neighbour spin correlations $C(1)$ without the quench (bottom grey shaded region). Inset: Normalized deviation from the mean nearest neighbour correlations
$\delta C_1(d)$ as a function of the distance $d$ from the hole.
c) The local spin fluctuations are probed through an envelope function, defined in the main text. The comparison to the case without a quench (gray) indicates the excess spin excitation. Figures taken from \cite{Vijayan2020}.}
\label{fig:dynamics1D}
\end{figure}

So far, we discussed probes for spin-charge separation in equilibrium, where the phenomenon is well-described by Luttinger liquid theory. Cold atom experiments are very well isolated from the environment and are therefore perfectly suited to study coherent quantum dynamics. In \cite{Vijayan2020}, the dynamics after the high-fidelity removal of an atom from a single site in a one-dimensional Fermi-Hubbard system was investigated, revealing dynamical spin-charge separation beyond the low-energy regime. This experiment is summarized below, and followed earlier theoretical proposals by Kollath et al. \cite{Kollath2005,Kollath2006}.

Removing an atom from the system corresponds to the simultaneous creation of a spin-$1/2$ and a charge-$1$ excitation, see Fig.~\ref{fig:dynamics1D}. Both can be directly probed with a quantum gas microscope. The charge excitation is straightforwardly tracked by measuring the spatially resolved hole density distribution and was found to exhibit a light-cone-like ballistic propagation of the wave front. The coherent evolution of the hole leads to an evolving interference pattern and is in excellent agreement with a prediction by a single-particle quantum walk at zero temperature. 

In order to probe the spin excitation, the nearest-neighbor spin correlation was measured in squeezed space. Considering the nearest-neighbor spin correlation has the advantage that it is sizeable even at the comparably high temperatures of $T/J \approx 0.75$ in this experiment. The removal of a single atom leads to a local reduction of antiferromagnetic order due to the creation of a spinon, which corresponds intuitively to a domain wall of two aligned spins. The region with reduced antiferromagnetic correlations was found in \cite{Vijayan2020} to spread through the system in time with a light-cone-like propagation of the wavefront. Comparing the dynamics of the squeezed space spin correlations to exact numerical results for an undoped Heisenberg spin chain yields excellent agreement, showing that it is a very good approximation to describe the spin and charge sectors independently from one another. 

By monitoring the spatial width of the squeezed space spin correlation and the hole distribution, velocities of the spin and charge excitations were extracted, Fig.~\ref{fig:dynamics1D} a). The maximum expected group velocities allowed by the respective dispersions of the charge and spin quasiparticles scales linearly with the hopping strength $t$ and the exchange coupling $J$, respectively.  
Making use of the tunability of the cold atom setup, the hopping strength was varied, leading to relative interaction strengths between $U/t=8$ and $U/t=20$. The extracted velocities were found to scale linearly with $t$ and $J$, and follow the theoretically expected predictions $v_h=2 t /\hbar$ and $v_s=\frac{\pi}{2} J / \hbar$ respectively. 

The dynamical deconfinement between spin and charge excitations was further quantified by the spin correlation across the propagating hole as a function of time in terms of the spin-hole-spin correlator
\begin{equation}
C_{\rm{SHS}}  = 4 \left< \hat{S}_i^z \hat{n}_{i+1}^h \hat{S}_{i+2}^z\right>.
\end{equation}
This correlator is a first indicator as to whether the spin excitation is spatially separated from the charge excitation. Experimentally, a fast decay from the value of the next nearest neighbor correlation $C(2)$ in an undoped chain, to the value of the nearest neighbor correlation $C(1)$ was observed as the hole propagates away from the initially created spinon, Fig.~\ref{fig:dynamics1D} b). The spin excitation, which corresponds to reduced antiferromagnetic correlations, is thus not in close proximity to the charge excitation. 
Taking this analysis one step further, the normalized deviation of the mean nearest neighbor correlation was studied as a function of the distance $d$ to the hole,
\begin{equation}
\delta C_1(d) = \left< \frac{\hat{S}_i^z\hat{S}_{i+1}^z}
{\left< \hat{S}_i^z\hat{S}_{i+1}^z  \right>} - 1\right>_{\bullet_i \bullet_{i+1} \circ_{i+1+d \vee i-d}},
\end{equation}
where the empty (full) circle indicates that the corresponding site must be empty (occupied) in the underlying dataset. The absence of binding between spin and charge excitations beyond the immediate vicinity of the hole is shown as $\delta C_1(d)$ shows no dependence on the distance to the hole, see Fig.~\ref{fig:dynamics1D} b).

Finally, the position of the excess spin excitation was located by quantifying the local spin fluctuations in squeezed space. 
To this end, the operator
\begin{equation}
\hat{\Sigma}_j^2 = \left( \sum_{\tilde{i}} \hat{S}_{\tilde{i}}^z f_{\tilde{j}}^\sigma(\tilde{i})\right)^2 \qquad \text{with a smooth window function} \qquad f_{\tilde{j}}^\sigma(\tilde{i}) = \exp \left(-\frac{(\tilde{i}-\tilde{j})^2}{2\sigma^2}\right)
\end{equation}
of characteristic size $\sigma$ is defined. At zero temperature, the operator $\hat{\Sigma}_j^2$ captures local fractional quantum numbers \cite{Kivelson1982}. 
The thermal fluctuations in the system lead to a significant background signal for this operator. Moreover, due to the finite temperature initial state, the local quench in \cite{Vijayan2020} creates a spinon excitation only with $50\%$ probability. The experimentally measured increase in $\hat{\Sigma}_j^2$ due to the quench is in accordance with these considerations, Fig.~\ref{fig:dynamics1D} c).

Taken together, the experimental results in \cite{Vijayan2020} demonstrated the fractionalization of a localized fermionic excitation into spinons and holons, in a far-from equilibrium setting.

\subsection{Other dynamical probes}\label{sec1Ddyn2}

Other dynamical probes in one dimension include the dynamics in Heisenberg spin chains, such as the observation of free and bound magnon excitations \cite{Fukuhara2013,Fukuhara2013a}.
Recently, XXZ spin chains, described by the Hamiltonian
\begin{equation}
    \hat{\mathcal{H}} = \sum_{\left<ij\right>} \left[ J_{xy} (\hat{S}_i^x \hat{S}_j^x + \hat{S}_i^y\hat{S}_j^y) + J_z \hat{S}_i^z \hat{S}_j^z \right],
\end{equation}
have been realized using bosonic Lithium atoms, where the on-site interactions between atoms in the same hyperfine state can be tuned to a different value than the interactions between atoms in different hyperfine states \cite{Jepsen2020} in order to realize an anisotropy in the spin couplings. By tuning the coupling constant $J_z/J_{xy}$ from $0$ to $1$, a crossover from ballistic to diffusive behavior has been observed \cite{Jepsen2020} in the dynamics following a global quench after a spin spiral was prepared. For anisotropies $J_z/J_{xy} > 1$, even subdiffusive transport behavior has been found. 
In the two-component Bose-Hubbard model, an effective magnetic field can be realized through the different scattering lengths for the two hyperfine states. In the relaxation dynamics of an initially prepared spin spiral, the presence of mobile holes leads to a fluctuating effective magnetic field, providing an additional dephasing mechanism \cite{Jepsen2021}.

Recently, experiments with a strong potential gradient have been also performed in the one-dimensional Fermi-Hubbard model, where the relaxation of a charge density wave is significantly slowed down. In certain regimes of gradient and interaction strength, non-ergodic behavior due to Hilbert space fragmentation was observed \cite{Scherg2020}.  

Another line of research is concerned with the transport through a one-dimensional wire between two large atom reservoirs at different chemical potentials \cite{Krinner2017,Lebrat2018}. In particular, the transport properties in the presence of a weak periodic potential within the 1D wire and attractive interaction between the fermions have been probed. The experimental results strongly indicate the existence of a Luther-Emery liquid in the one-dimensional wire.

\subsection{Open questions in one dimension}\label{sec1DopenQ}
Several future directions of doped one-dimensional spin chains remain unexplored. A particularly important class of experiments are spectroscopic probes, using e.g. momentum resolved probes such as ARPES or the dynamical spin structure factor, or local probes such as STM. For example, it has been proposed that ARPES measurements in an undoped spin chain can be utilized to directly reveal the emergent spinon Fermi surface \cite{Bohrdt2018}. An extension of such probes to dimensional cross-overs, where weak tunneling along a second direction is added \cite{Salomon2019}, would be very interesting to explore how spin-charge separation breaks down and how the spinon fermi surface would be affected \cite{Bohrdt2020_ARPES}. 

Other interesting future directions involve dynamical probes far-from equilibrium, strong frustration or extensions from $SU(2)$ to $SU(N)$ invariant \cite{Scazza2014,Ozawa2018} doped spin chains. Furthermore, topological order can be explored; a first step in this direction has recently been taken by the realization of the symmetry protected Haldane phase in Fermi-Hubbard ladders \cite{Sompet2021}.

\section{Quantum magnets in 2D: half-filling and single dopant}\label{sec2DHalffilling}

\subsection{Half filling: long-range AFM}\label{secAFM}
\begin{figure}
\centering
  \includegraphics[width=0.95\linewidth]{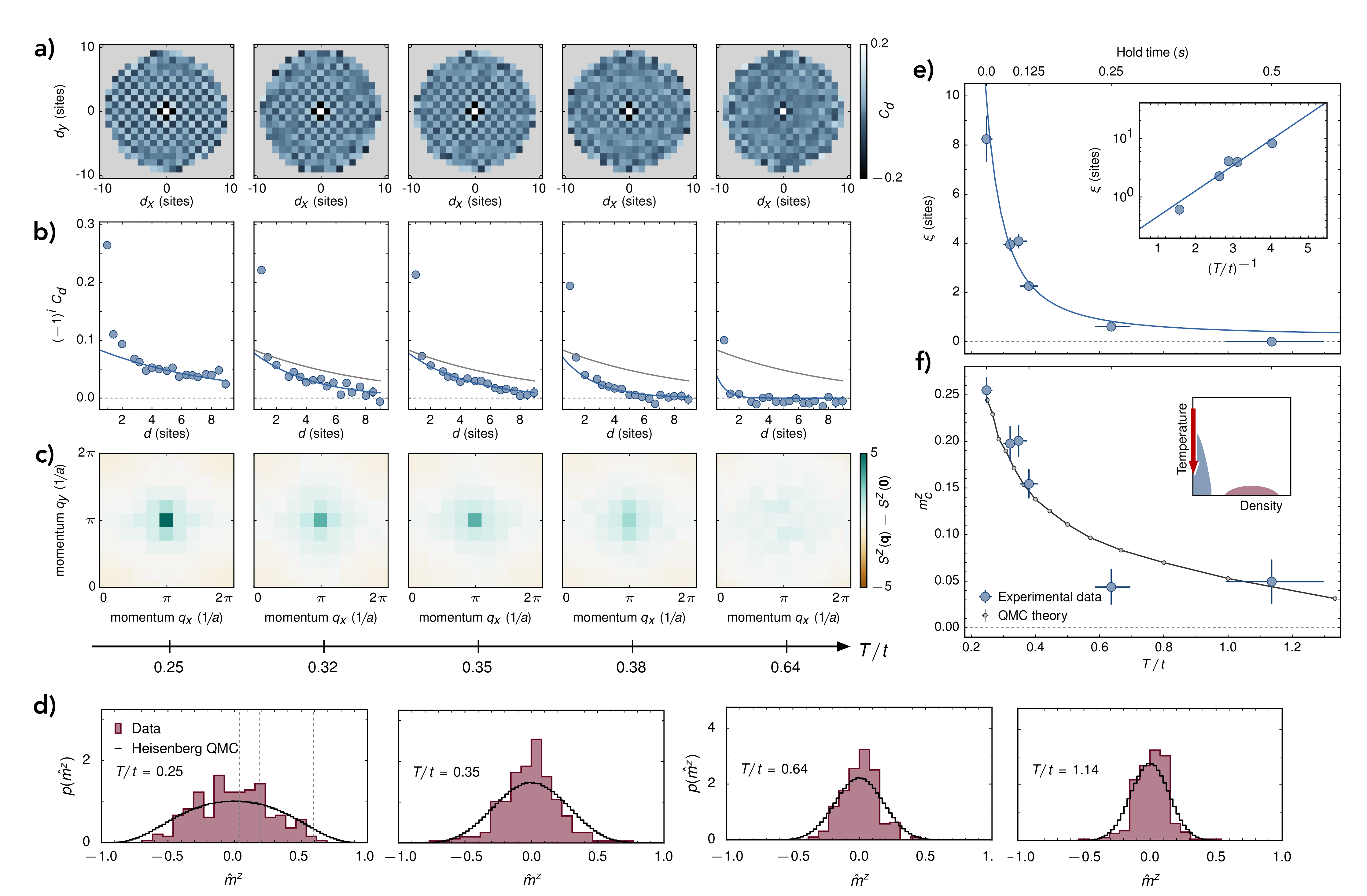}
\caption{ \textbf{Long-range antiferromagnetic correlations in a quantum gas microscope.} a) Spin correlations $C(\mathbf{d})$ for different distances $\mathbf{d} = (d_x,d_y)$. b) Sign corrected spin correlations as a function of distance. c) Static structure factor $S^z(\vec{q}) - S^z(\vec{0})$ for temperatures from $T/t = 0.25$ to $T/t=0.64$. d) Full counting statistics of the staggered magnetization for temperatures between $T/t=0.25$ and $T/t=1.14$. e) The correlation length and f) the average value of the staggered magnetizaton in $z$-direction as a function of temperature. Figure taken from  \cite{Mazurenko2017}.}
\label{fig:longrangeAFM}
\end{figure}

The most challenging step to probe the phase diagram of the two-dimensional Fermi-Hubbard model with cold atoms consists in reaching sufficiently low temperatures to explore the most interesting and highly debated regimes. The first measurements of spin correlation functions in quantum gas microscopes were performed at temperatures around $T/t \approx 0.5$ \cite{Parsons2016} to $T/t \approx 1$ \cite{Cheuk2016}. At an interaction strength of $U/t \approx 8$ this is well in the Mott-insulating regime and corresponds to temperatures $T/J \approx 1$ to $T/J \approx 2$ in units of the antiferromagnetic superexchange $J$. 
At finite temperatures, the spin correlations in two dimensions decay exponentially over a correlation length 
\begin{equation}
\xi(T) = C_\xi \exp(2\pi\rho_S/T)
\label{eqxiTAFM}
\end{equation}
with $\rho_S$ the spin stiffness \cite{Manousakis1991}.
A temperature on the order of the superexchange is sufficient to observe antiferromagnetic correlations up to distances of three sites. Tuning the temperature between $T/t=0.45$ and $T/t=1.53$ led experimentally to correlation lengths between $0.5$ and $0.25$ sites. 

In cold atom experiments, the interaction strength $U/t$ can be easily varied using a Feshbach resonance by tuning a magnetic field. The entropy per particle is fixed, such that the temperature varies slightly for different interaction strengths. The overall chemical potential is fixed, but the central region on which experiments usually focus is coupled to its surrounding, where the trapping potential leads to smaller densities. Hence by tuning the interaction strength the number of atoms in the center of the trap can change to some degree. Upon tuning the interaction strength between $U/t=4$ and $U/t=24$, the experiments \cite{Cheuk2016,Parsons2016} find a maximum absolute value of the nearest neighbor spin correlation at $U/t\approx 8$, in good agreement with theoretical results \cite{Cheuk2016,Parsons2016}. One can understand this as follows: for large values of $U/t$, the antiferromagnetic correlations are suppressed because the superexchange $J$ becomes small compared to the temperature. For $U/t<8$ on the other hand, charge fluctuations increasingly play a role and tend to destroy magnetic correlations. 
Another tuning knob in cold atom experiments is the dimensionality. In \cite{Greif2015}, the nearest neighbor spin correlations were probed for a different number of nearest neighbors $Z$, going from $Z=1$ for dimers, to 1D systems ($Z=2$), ladders ($Z=3$), 2D systems ($Z=4$) and cubic three-dimensional systems ($Z=6$). Upon increasing the coupling between one-dimensional chains, the nearest neighbor spin correlations along the chains were found to decrease, while simultaneously the spin correlations between the chains increase. 

The earliest cold atom experiments did not have full resolution, meaning that doubly occupied sites appeared as empty in the snapshots taken with a quantum gas microscope. Experimentally, the charge fluctuations are thus mainly probed in terms of anti-moment correlations, defined as
\begin{equation}
C_h(d) = \frac{1}{N}\sum_{\mathbf{i}} \left(
 \left<
  \left(  1-\hat{n}_{s,\mathbf{i}} \right)
   \left(1-\hat{n}_{s,\mathbf{i+d}}\right) 
   \right> - 
\left< \left(1-\hat{n}_{s,\mathbf{i}} \right)\right>
 \left< \left(1-\hat{n}_{s,\mathbf{i+d}}\right) \right> \right),
\end{equation}
where $\hat{n}_{s,\mathbf{i}}$ is the single occupation density at site $\mathbf{i}$ and $N$ is the total number of sites.
The charge fluctuations observed experimentally are mainly virtual doublon-hole pairs, leading to comparably strong positive anti-moment correlations for nearest neighbors \cite{Cheuk2016,Chiu2019Science}. Doublon-hole pairs beyond nearest neighbors become increasingly unlikely.  In a later experiment, the doublon-hole correlations were resolved with full resolution, revealing a peak at filling one and distance one \cite{Hartke2020}. As a function of $(t/U)^2$, the number of doublon-hole pairs exhibits a linear dependence \cite{Hartke2020}, as expected from second order perturbation theory.

While the Mermin-Wagner-Hohenberg theorem implies that there is no long-range spin-order in the thermodynamic limit at finite temperature in two dimensions, in a finite size system, the correlation length can become comparable or larger than the system size. In this case, the $z$-component of the staggered magnetization, defined for a spin-$S$ system as
\begin{equation}
m^z = \sqrt{\left< \left(\frac{1}{N} \sum_{\mathbf{i}} (-1)^{\mathbf{i}} \frac{1}{S} \hat{S}_{\mathbf{i}}^z \right)^2 \right>}
\end{equation}
reaches values of order unity.
In 2017, sufficiently cold temperatures of $T/t=0.25$ at an interaction strength of $U/t=7.2$ were achieved in a system of uniform density, yielding the first observation of antiferromagnetic correlations in ultracold atoms across the entire system of approximately ten sites in diameter \cite{Mazurenko2017}, see Fig.~\ref{fig:longrangeAFM}. Here, a digital micromirror device (DMD) in the image plane of the microscope was used as a spatial light modulator in order to cancel out the underlying harmonic potential. The filling in the region of interest of the system is thus highly uniform and at the same time tunable to realize arbitrary doping values.

The measurements performed with a quantum gas microscope yield not only the average value of the staggered magnetization (or any other observable under consideration), but also the full counting statistics $p(m^z)$, i.e. a histogram of how often a specific outcome is obtained in individual measurements. In these experiments, full distribution functions are very natural quantities to consider, as each snapshot corresponds to a single value of the staggered magnetization, and one has to average over many snapshots to obtain the expectation value of $m^z$ in the first place. 
At high temperatures, without antiferromagnetic correlations, the full counting statistics of the staggered magnetization is peaked at zero with a width given by $1/\sqrt{N}$. As the temperature is decreased and antiferromagnetic correlations extend across the entire system, the distribution attains a substantially larger width, see Fig.~\ref{fig:longrangeAFM} d). 

In solid-state systems, one typically detects antiferromagnetic long-range order with neutron scattering or magnetic X-ray scattering experiments, which yield the spin structure factor as an observable. In order to bridge the gap to these solid state experiments, one can also Fourier transform the real space spin correlations considered so far to obtain the static spin structure factor,
\begin{equation}
    S^z(\vec{q}) = \frac{1}{L^2} \sum_{\vec{i},\vec{j}} \frac{1}{S^2} \langle \hat{S}^z_{\vec{i}} \hat{S}^z_{\vec{j}} \rangle \exp \left[ i \vec{q} \cdot \l \vec{i} - \vec{j} \r \right],
\end{equation}
where $L$ is the linear system size. The spin correlations, or equivalently the spin structure factor, show clear antiferromagnetic correlations at the lowest temperatures achieved in the cold atom setup \cite{Mazurenko2017}. As the temperature is increased, the correlation length decreases and simultaneously the peak at momentum $\vec{q} = (\pi,\pi)$ in the spin structure factor vanishes, Fig.~\ref{fig:longrangeAFM} c).

Even without single-site resolution, the static structure factor at arbitrary wave vectors can be probed. One option is to use spin-sensitive Bragg scattering, which was used to reveal short-range antiferromagnetic correlations in a three-dimensional Hubbard model \cite{Hart2015}. Another option is to use coherent manipulation of the spins in a Ramsey-type scheme: after ramping up the lattice to inhibit tunneling, a $\pi/2$ radiofrequency (RF) pulse is applied to the spins. Then, the spins evolve under a magnetic field gradient for a time $t_{sp}$, which determines the wave vector that is probed. During this time evolution, the spins precess in the $xy$-plane. Finally, a second $\pi/2$ RF pulse is applied in order to map the transverse magnetization onto the magnetization in $z$-direction, which can then be directly probed. This technique was used in \cite{Wurz2018} to probe the static structure factor as a function of wave vector and chemical potential. At half-filling, a clear peak appears at momentum $\mathbf{q} = (\pi,\pi)$ at the experimental temperature of $T/t =  0.57(3)$ and interaction strength $U/t=8$. As the chemical potential is decreased away from $U/2$, the staggered ($\mathbf{q} = (\pi,\pi)$) as well as the uniform ($\mathbf{q} = \mathbf{0}$) static structure factor decay.

\subsubsection{Spin-imbalanced system: canted AFM}
\label{secCantedAFM}
\begin{figure}
\centering
  \includegraphics[width=0.85\linewidth]{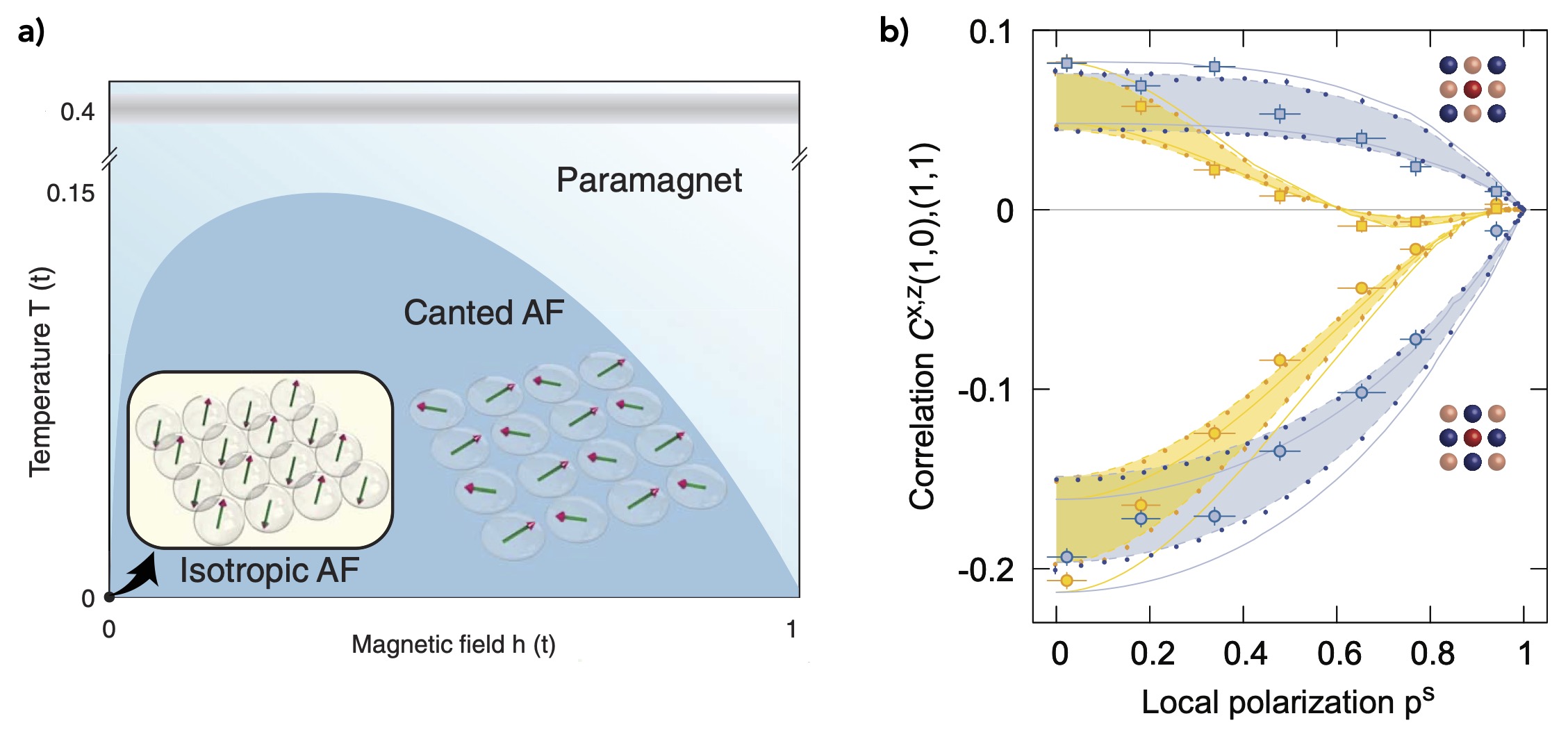}
\caption{ \textbf{Spin-imbalanced system.} a) Phase diagram of the Heisenberg model as a function of temperature $T$ and effective magnetic field $h$. The effective magnetic field is experimentally realized through a spin-imbalance. The shaded gray band indicated the experimentally realized temperature in \cite{Brown2017}. b) Nearest (circles)  and diagonal next-nearest neighbor (squares) spin correlations $C^x$ (blue) and $C^z$ (yellow) as a function of the local polarization. Figure taken from  \cite{Brown2017}.}
\label{fig:cantedAFM}
\end{figure}

Going beyond the standard spin-balanced equilibrium case, the spin physics in the half-filled Fermi-Hubbard model was further probed by studying its properties in the spin-imbalanced case \cite{Brown2017}. In a cold atom setup, the numbers of spin up and spin down fermions are separately conserved. While loading the optical lattice, one can prepare a spin-imbalanced system, for example by using a microwave pulse to resonantly drive the transition between the two relevant hyperfine states or by applying a magnetic gradient during evaporation, leading to preferential loss of one of the two spin states \cite{Brown2017}. By the latter technique, the global spin imbalance
\begin{equation}
P = (N_\uparrow − N_\downarrow)/(N_\uparrow + N_\downarrow)
\end{equation}
was varied contiuously from zero to $\approx 0.9$ in Ref.~\cite{Brown2017}.

The finite magnetization in the system is akin to a non-zero magnetic field, breaking down the global $SU(2)$ to a global $U(1)$ spin symmetry. The Heisenberg model in the presence of a non-zero field hence exhibits a finite temperature Berezinskii-Kosterlitz-Thouless (BKT) transition to a canted antiferromagnet \cite{Koetsier2010}. The canted antiferromagnet accomodates the finite magnetization in the direction of the field, but at the same time maximizes the superexchange interaction energy through long-range antiferromagnetic correlations of the spin components perpendicular to the magnetization.
An inhomogeneous density of atoms introduces additional complications into this scenario. Spin polarization can be distributed unequally between the Mott insulating and compressible regions of the ensemble. The latter are expected to have a larger spin susceptibility and can therefore accommodate a larger part of the spin imbalance in the system \cite{Wunsch2010}.

At the temperatures in the experiment \cite{Brown2017}, $T/t \approx 0.4$, the antiferromagnetic correlations extend across a few lattice sites. These antiferromagnetic correlations become canted in the presence of a spin-imbalance: the spin correlations are stronger in the direction orthogonal to the magnetization, see Fig.~\ref{fig:cantedAFM}. The rotational anisotropy of the spin correlations increases with polarization. Note that the spin correlations in $x$-direction, orthogonal to the magnetization, can be measured by applying a global RF pulse to rotate all spins before imaging \cite{Brown2017}. 

At lower temperatures than experimentally realized, the BKT transition takes place at a critical temperature $T_{\rm BKT}(P)$. It reaches a maximum $T_{\rm BKT}(P_{\rm opt}) \approx 0.15 t$ for some optimal polarization $0 < P_{\rm opt} < 1$, and vanishes at $T_{\rm BKT}(P=0)=T_{\rm BKT}(P=1)=0$. Below the transition temperature, the spin correlations perpendicular to the field become quasi-long ranged with a power-law decay, while the correlations in the direction of the field remain short-ranged and decay exponentially with distance. Correspondingly, the anisotropy in the spin correlations $\langle \hat{S}^x_{\vec{d}} \hat{S}^x_{\vec{0}} \rangle / \langle \hat{S}^z_{\vec{d}} \hat{S}^xz_{\vec{0}} \rangle$ discussed above was also found to increase with the distance $|\vec{d}|$ between the spins \cite{Brown2017}. 

A spin-imbalance in the repulsive Fermi-Hubbard model corresponds to a finite doping in the attractive model. The $S^zS^z$ spin correlations in the repulsive model can be mapped to the doublon-doublon correlations in the attractive version. The measurement of the doublon-doublon correlations for an interaction of $U/t=-5.7$ at a temperature of $T/t=0.45$ in \cite{Mitra2017} thus shows the same qualitative features as the $S^zS^z$ spin correlations shown in Fig.~\ref{fig:cantedAFM}b).

\subsubsection{Spin transport}
\begin{figure}
\centering
  \includegraphics[width=0.99\linewidth]{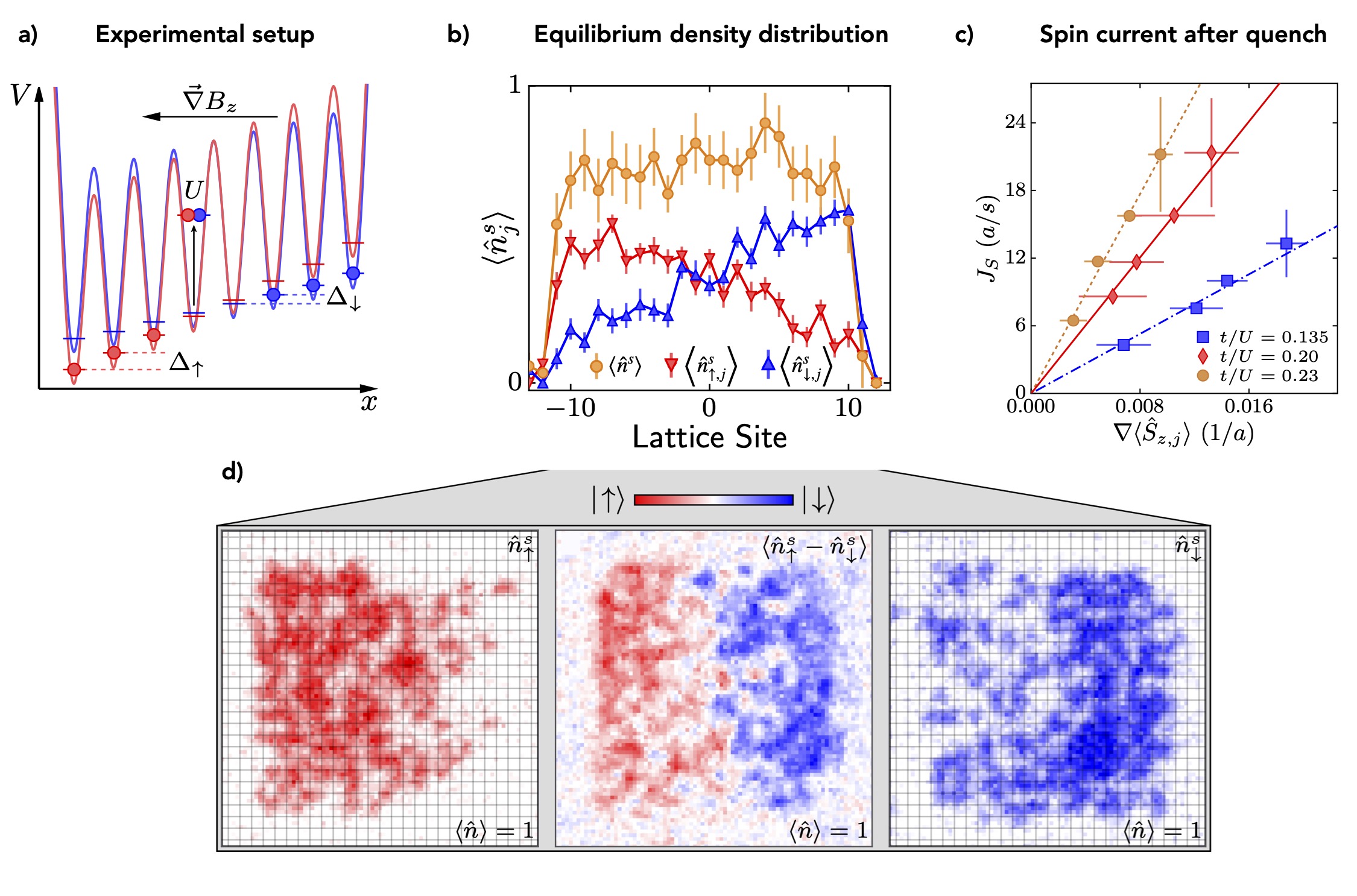}
\caption{\textbf{Spin transport.} a) Experimental setup: a spin-dependent potential gradient is applied during the experiment. b) The equilibrium density distribution of single occupied sites as well as spin up and spin down fermions under a magnetic field gradient is used to extract the spin susceptibility $\chi_S$. Shown here is the singlets density, as well as the spin-resolved singles-density, averaged over four independent realizations at $U/t=38.5$. c) The magnetic field gradient is suddenly removed. The resulting spin current exhibits a linear dependence on the gradient of the magnetization. d) Left, Right: single raw images of $\left<\hat{n}_\uparrow^s\right>$, $\left<\hat{n}_\downarrow^s\right>$ in equilibrium in the presence of a spin-dependent potential at half-filling for $U/t=14.9$. Middle: Difference between up and down density distributions as shown in left and right averaged over six independent realizations.
Figures extracted from \cite{Nichols2018}.}
\label{fig:spin_transport}
\end{figure}

At half filling and for temperatures $T\ll U$, charge transport is strongly suppressed, rendering this an interesting regime to study spin transport. In \cite{Nichols2018}, the spin susceptibility as well as spin transport coefficients were probed with the help of magnetic field gradients, Fig.~\ref{fig:spin_transport} a). Here, Potassium atoms are used, which in comparison to Lithium atoms have the advantage that magnetic field gradients can be applied during the experiment itself \footnote{For Lithium atoms, the large magnetic fields required to tune close to the repulsive Feshbach resonance bring the experiment into the Paschen-Back regime where the two pseudo-spin states used in the quantum simulation have virtually the same magnetic moment. Hence they are no longer susceptible to a magnetic field gradient.}.
In the first experiment \cite{Nichols2018}, the equilibrium spin distribution under a magnetic field gradient was measured, Fig.~\ref{fig:spin_transport} b), yielding the spin susceptibility 
\begin{equation}
\chi_S = \frac{\Delta \left<\hat{S}^z\right>}{\Delta \mu}
\end{equation}
as the ratio of the change in the local magnetization, $\Delta \langle \hat{S}^z \rangle$, induced by a change in the chemical potential between the two spin species at the same position,  $\Delta \mu = \mu_\uparrow - \mu_\downarrow$.

When this magnetic field gradient is suddenly turned off, the system relaxes to a new equilibrium state with local spin balance, $\langle \hat{S}^z \rangle \approx 0$. During this relaxation dynamics, one can define a spin current
\begin{equation}
J_S(\tau) = -\frac{a}{2} \frac{d}{dt} \mathcal{I}(t)|_\tau
\end{equation}
with the imbalance $\mathcal{I}(t)$ defined as difference in spin between the left and right half of the system; $a$ denotes the lattice constant. The experiment \cite{Nichols2018} showed a linear dependence of the spin current $J_S(\tau)$ on the spin density gradient $\nabla_x \langle \hat{S}^z_{x} \rangle$, Fig.~\ref{fig:spin_transport} c), showing that the spin dynamics is diffusive:
\begin{equation}
J_S = D_S  \l \nabla_x \langle \hat{S}^z_{x} \rangle \r,
\end{equation}
with a spin diffusion constant $D_S$. This result holds for a range of values of $U/t = 4$ to $17$. For strong interactions $U/t \geq 8$, the extracted spin diffusion constant depends linearly on $t^2/U$, and thus on the superexchange $J=4t^2/U$. For the Heisenberg model, the spin diffusion constant at temperatures $T \gg J$ is $\hbar D_S = 4\sqrt{\pi/20}a^2t^2/U = 1.585... a^2t^2/U$ \cite{Bennett1965,Sokol1993,Bonca1995}, which is significantly lower than the experimentally observed value of $\hbar D_S = 6.2(5)a^2t^2/U$. In \cite{Nichols2018}, this is attributed to doublon-hole fluctuations in the Fermi-Hubbard model. For weaker interactions, $U/t \leq 8$, the diffusivity increases faster than linearly with $t^2/U$. In the same line of reasoning as above, this is attributed to the increasing number of doublon-hole fluctuations in this regime. It remains an open question, however, why the experimental results do not approach the Heisenberg limit for $U \rightarrow \infty$, where doublon-hole fluctuations are strongly suppressed. 

Finally, the experiment \cite{Nichols2018} starts without a tilt and a magnetic field gradient is slowly turned on in order to study the spin conductivity $\sigma_S$. The magnetic field gradient here consists of a spin- and position- dependent potential $(-1)^\sigma \Delta_\sigma(x)$, where $\Delta_\sigma(x) = \Delta_\sigma \cdot x$. 
The initial spin current in this setup is directly proportional to the applied spin-dependent force, obtained as the derivative of the spin-dependent potential, with the spin conductivity as constant of proportionality, i.e.
\begin{equation}
J_S(\tau=0) = \sigma_S \frac{\Delta_\downarrow \langle \hat{n}_\downarrow \rangle - \Delta_\uparrow \langle \hat{n}_\uparrow \rangle}{a}.
\end{equation}
The value for $\sigma_S$ measured with this experiment agrees well with the value obtained through the Einstein relation as $\sigma_S = D_s\chi$, with $D_S$ and $\chi$ measured independently. 

The experimentally measured spin conductivities for various values of $U/t$ are below the Mott-Ioffe-Regel limit. This behavior is expected in the regime considered here, where quasiparticles are ill-defined and therefore Drude-Boltzmann theory does not apply.

A different approach to separate spin and charge transport was taken in \cite{Brown2015}, where a potential offset on one sublattice was applied in the two-dimensional system in order to suppress charge motion, thus making superexchange the dominant effect. Superexchange can on the other hand be suppressed through a spin-dependent potential on one sublattice. Therefore, the effects of charge motion and superexchange on the magnetization dynamics can be probed separately. 
In \cite{Brown2015}, a spin-$1/2$ system was simulated using two hyperfine states of bosonic atoms. In the limit of large on-site interactions $U/t$, the model can be mapped to a bosonic $t-J$ model, as discussed in section \ref{sec1Ddyn2}. 
In this experiment, a N\'eel state was prepared as initial state. Its decay under the time evolution with the two-component Bose-Hubbard Hamiltonian was then observed without single-site resolution by mapping the states in one sublattice to two separate hyperfine states, and then performing a Stern-Gerlach scheme on both sublattices separately. The observed relaxation rate is governed by two separate rates, which scale with the superexchange and tunneling, respectively.

\subsection{Magnetic polaron}\label{secMagneticPolaron}
So far,  we discussed one-dimensional Fermi-Hubbard chains at finite doping as well as the half-filling limit of the two-dimensional Fermi-Hubbard model.  Both of these cases are fairly well understood theoretically.  We now venture into far more disputed territory by going away from half-filling,  beginning with a single dopant.  While there is a vast amount of theoretical literature on the topic of a single dopant in a quantum antiferromagnet, the so-called magnetic or spin polaron, and several of its properties,  such as its dispersion relation,  are well established,  quantum gas microscopes provide an entirely new perspective,  for example by enabling the direct microscopic imaging of mobile dopants and their dressing cloud \cite{Koepsell2019}. The real space imaging in quantum gas microscopes renders it very natural to think about the problem of a single dopant in a quantum antiferromagnet in real space, which moreover provides an intuitive understanding through the perspective of the geometrical string picture that we present below.

\subsubsection{Theoretical background}
\begin{figure}
\centering
  \includegraphics[width=0.95\linewidth]{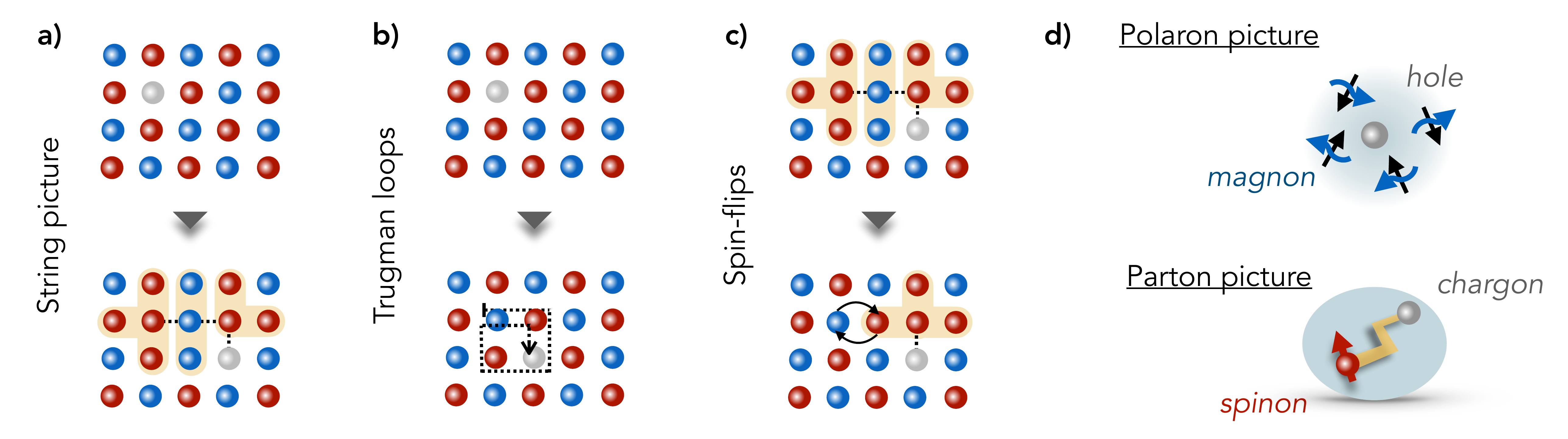}
\caption{ \textbf{Theory of mobile dopants in an antiferromagnet.} a) The motion of a doped hole creates a string of displaced, or flipped, spins. For most configurations, the energy of the string grows approximately linearly with its length. b) Self-trapping of the dopant can be avoided by performing Trugman loops. While these allow a hole to move freely through the antiferromagnet in principle, the mechanism is very slow since Trugman loops are costly in terms of the localization energy they require. c) Spin-exchange processes $\propto J$ allow to re-trace the string and provide a much more efficient means for the dopant to propagate in the antiferromagnet. d) Two theoretical pictures of magnetic polarons are currently used. They emphasize different constituents of the quasiparticle and should not be viewed as mutually exclusive, but rather complementary approaches to the problem with their own strengths and weaknesses.}
\label{fig:theoryMagPol}
\end{figure}

One of the earliest treatments of a mobile dopant in an antiferromagnet was by Bulaevskii et al. \cite{Bulaevskii1968} in 1968, before the discovery of high-temperature superconductivity in cuprates. The authors considered the case of a perfect N\'eel state in two dimensions and argued that a mobile hole should self-localize because its motion leaves behind a string of displaced, or flipped, spins costing an amount of energy proportional to the length of this string, see Fig.~\ref{fig:theoryMagPol} a). The spectral function in such a model was then calculated by Brinkman and Rice \cite{Brinkman1970}. Later on, Trugman clarified \cite{Trugman1988} that the loop configurations now named after him allow to remove the string and lead to a very slow \cite{Poilblanc1992,Chernyshev1999,Grusdt2018tJz} but free motion of the dopant, see Fig.~\ref{fig:theoryMagPol} b). These works \cite{Bulaevskii1968,Brinkman1970,Trugman1988} established the basis for the \emph{string picture} of magnetic polarons.

In the $SU(2)$ invariant $t-J$ or Hubbard model quantum fluctuations of the spin background need to be included. They provide a much more efficient way to remove the string created by the hole than Trugman loops, and introduce motion of the entire quasiparticle, see Fig.~\ref{fig:theoryMagPol} c). Shortly after the discovery of high-temperature superconductivity, quantitative descriptions of these quantum fluctuations were put forward, which have laid the foundation for the magnetic polaron picture. This work goes back to the seminal papers by Schmitt-Rink, Varma and Ruckenstein \cite{SchmittRink1988}, by Kane, Lee and Read \cite{Kane1989} and related work by Sachdev \cite{Sachdev1989} which describe how the motion of the dopant couples to spin-wave excitations (magnons) of the surrounding antiferromagnet. 

The philosophy of the \emph{polaron picture} is that the dopant acts as a mobile impurity which gets dressed by a cloud of magnetic excitations. In the weak coupling regime when $t < J$, where the mapping between the Hubbard and $t-J$ models is no longer valid, the dressing is weak and the theory can be relatively easily solved. However, at strong couplings when $t \gtrsim 2 J$, significant efforts were required to successfully solve the problem \cite{Martinez1991,Liu1992}. In this case, the quasiparticle properties are strongly renormalized, as has been confirmed by controlled numerical simulations such as large-scale exact diagonalization (ED) \cite{Elser1990,Dagotto1990,Leung1995,Beran1996,Wang2015a}, quantum Monte Carlo \cite{Brunner2000,Mishchenko2001} and more recently tensor network studies, e.g. \cite{White2001,Mezzacapo2011,Bohrdt2020_ARPES}. Several other variational and numerical descriptions of magnetic polarons have been put forward that are worth mentioning, including: the semi-classical description \cite{Shraiman1988}, the small polaron theory \cite{Auerbach1991} and the variational Monte Carlo theory by Bonisegni and Manousakis \cite{Boninsegni1992}.

Another perspective, complementary to the polaron picture, is the \emph{parton theory} of magnetic polarons. It starts from the strong coupling regime, $t > J$, and assumes that mobile dopants can be understood as bound states of two emergent partons, a spinon and a chargon. The latter are confined at low doping, rendering the magnetic polaron the analogue of a meson in particle physics. The parton picture was first proposed on phenomenological grounds by B\'eran et al. \cite{Beran1996} where it was already pointed out that the strong renormalization of doped holes can be understood by simple arguments from a parton ansatz. Several subsequent works, still on a phenomenological level, discussed the connection to observations in the low-doping regime \cite{Laughlin1997,Baskaran2007}, including a paper by P. Anderson \cite{Anderson2007}. Unlike the polaron theory, the parton picture can be directly connected to the one-dimensional case where spinons and chargons are deconfined, whereas magnons no longer constitute well-defined excitations. 

Recently new indications for the parton picture have been reported by advanced matrix product state (MPS) simulations \cite{Grusdt2019,Bohrdt2020_ARPES,Bohrdt2021arXiv}. Moreover, the string picture has been extended to include quantum fluctuations of the spin background \cite{Manousakis2007} by introducing so-called geometric strings \cite{Grusdt2018tJz,Grusdt2019}. This allowed to describe the binding of spinons and chargons on a microscopic level, starting from a Born-Oppenheimer wavefunction and using different levels of approximation \cite{Grusdt2019,Bohrdt2020_ARPES}: First level theories only include chargon fluctuations while a frozen spin background is considered \cite{Grusdt2018tJz}. In a second level description spinon dynamics are included \cite{Grusdt2019}, and  third-level theories include backactions on the entire spin background. 

It should be emphasized that the parton and polaron pictures, see Fig.~\ref{fig:theoryMagPol} d), of magnetic polarons are not strictly exclusive. Rather, they emphasize different aspects of the problem and approach it from different limits. While the polaron theory is simplest at weak coupling, $t<J$, the parton picture is a genuine strong coupling theory which works best for $t> J$. One of the most striking predictions by the parton picture to date is the existence of long-lived discrete rotational excitations. These were first proposed for holes in a classical N\'eel state \cite{Grusdt2018tJz}, but recently direct numerical evidence has been reported in large-scale MPS simulations of the $t-J$ model on four-leg cylinders and experimental probes for cold atoms and solids have been proposed \cite{Bohrdt2021arXiv}. Such rotational excitations are closely related to vibrational excitations of magnetic polarons. The latter have previously been  discussed in the context of the polaron picture \cite{Dagotto1990,Liu1992,Leung1995,Mishchenko2001,Manousakis2007,Wrzosek2020}, where their properties such as excitation energies can only be obtained from heavy numerics however.

Finally, for very strong coupling $t \ggg J$ the physics of a mobile dopant changes drastically. In this regime the spins around the impurity polarize and the dopant is self-trapped inside a ferromagnetic bubble. The precise transition point was determined accurately for the two-dimensional $t-J$ model by White and Affleck to be located at $J/t = 0.02$ to $0.03$ \cite{White2001}. This is a precursor of the famous Nagaoka effect \cite{Nagaoka1966}, which states that the addition of one single mobile hole is sufficient to drive a genuine phase transition of the entire system from an antiferromagnet to a ferromagnet in the asymptotic limit $U/t \to \infty$.

In the solid state context, the main emphasis has been on the following properties of magnetic polarons accessible by ARPES, see e.g. \cite{Wells1995,Ronning2005,Graf2007}: Their dispersion relation, location of the band minimum, and their quasiparticle weight. As we will discuss next, ultracold atoms provide access to complementary quantities characterizing magnetic polarons, which have also recently been analyzed theoretically: these include e.g. the microscopic structure and size of their dressing cloud \cite{Grusdt2019,Blomquist2019} and higher-order spin-charge correlations \cite{Bohrdt2020_gauss}.

\subsubsection{Experiments in equilibrium}
\begin{figure}
\centering
  \includegraphics[width=0.99\linewidth]{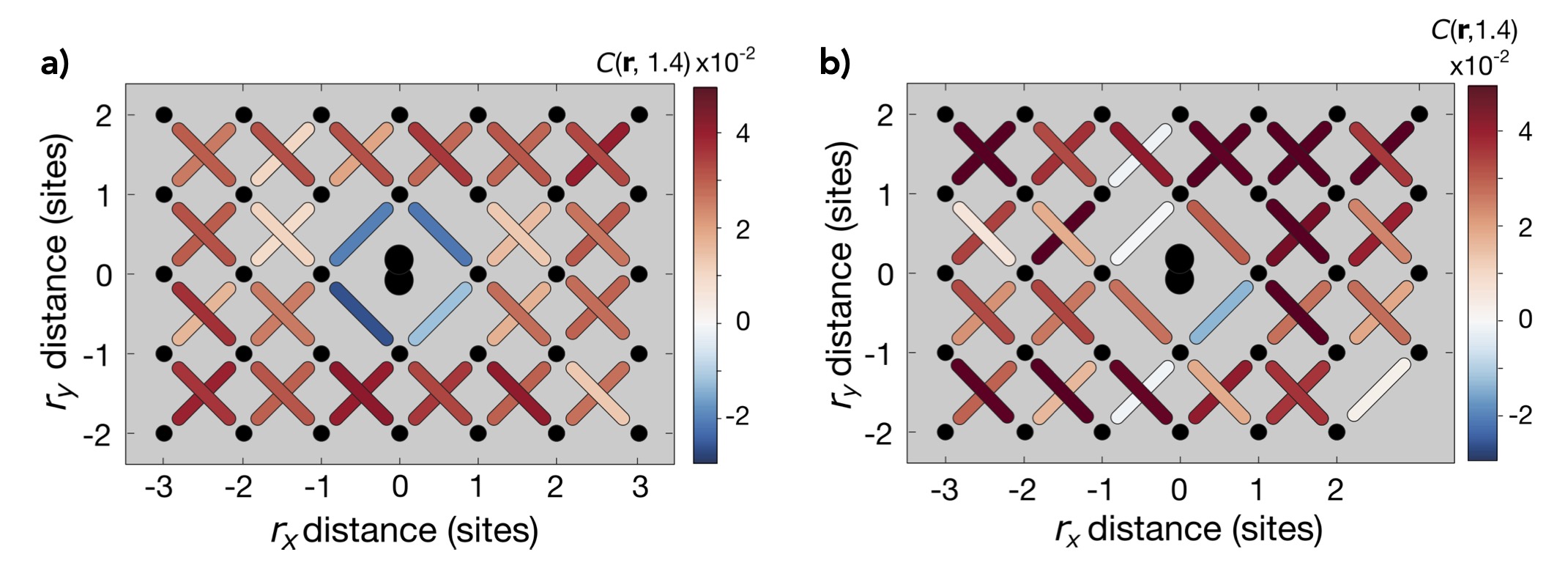}
\caption{ \textbf{Imaging a magnetic polaron and its dressing cloud.} Diagonal next nearest neighbor spin correlations measured around a) a mobile and b) a pinned doublon. Around the mobile dopant, the spin correlations are strongly sign-reversed (blue), a direct signature of the polaronic dressing cloud. Figures taken from \cite{Koepsell2019}.}
\label{fig:magneticPolaronDyn}
\end{figure}

In \cite{Salomon2018}, the behavior of the spin correlations around a mobile doublon\footnote{Note that, owing to the exact particle-hole symmetry of the Fermi Hubbard model with nearest-neighbor hopping only, doublon and hole doping are equivalent and can be used interchangeably.} was studied as the hopping $t_y/t_x$ was tuned,  thus going from decoupled chains to a two-dimensional system. While in the one-dimensional limit strong AFM correlations across the doublon are found, a significant suppression of the correlations was observed when the system becomes two-dimensional. This was the first direct indication for the formation of a magnetic polaron.

The magnetic environment of a single dopant in the two-dimensional system was studied in detail in \cite{Koepsell2019}.  Based on snapshots from a quantum gas microscope,  one can directly probe the dressing cloud of the magnetic polaron in terms of the spin correlations relative to the dopant.  In \cite{Koepsell2019} the chemical potential was set such that on average $1.95(1)$ delocalized doublons are present in a central region of $5 \times 3$ sites.  At a temperature of $T/J \approx 1.4$,  there is no long-range antiferromagnetic order in this experiment.  However,  the nearest and next nearest neighbor spin correlation functions are still significant in the undoped system.  The three-point doublon-spin correlator
\begin{equation}
C(\vec{r}_0;\vec{r}_1,\vec{r}_2) = 4\left<S_{\vec{r}_1}^z S_{\vec{r}_2}^z \right>_{\circ_{\vec{r}_0} \bullet_{\vec{r}_1} \bullet_{\vec{r}_2}}
\end{equation}
with $\vec{r}_{1,2}$ nearest or next nearest neighbors is therefore optimally suited to investigate the spin environment around the dopant. Here the open / filled circles indicated that the correlator is evaluated on datasets where the respective sites were empty / occupied by a fermion. 

In the immediate vicinity of the mobile dopant, the diagonal next nearest neighbor correlations were found to become antiferromagnetic \cite{Koepsell2019}.  Similarly,  the straight next nearest neighbor correlations across and next to the dopant were sign flipped.  These experimental results confirm the theoretical picture of a dopant dressed by a local spin distortion,  and gives microscopic insights into the details of this distortion.  By measuring the strength of the local spin correlations as a function of distance from the dopant,  the radial dependence of the polaronic dressing and thus the size of the magnetic polaron has been studied \cite{Koepsell2019}. As a result, a relatively small polaron radius between one and two lattice sites has been established. The experimental results were also found to be consistent with a geometric string description of the problem when the harmonic trapping potential was included. In particular, the geometric string picture explains the observed sign-reversal around the mobile dopant by a simple displacement of the neighboring spins which turns antiferromagnetic nearest neighbour correlations in the undoped system into negative diagonal correlations.

Using an optical tweezer, a doublon could also be confined to a single lattice site in Ref.~\cite{Koepsell2019}, rendering the dopant immobile.  The comparison between the spin environment around the pinned and unpinned dopants reveals the importance of its mobility: in the case of an immobile dopant,  the antiferromagnetic spin correlations surrounding the doublon are even enhanced.

Motivated by the capabilities of quantum gas microscopes,  new theoretical results on higher-order correlation functions  for the ground state of a single hole in the $t-J$ model have been obtained \cite{Grusdt2019,Blomquist2019,Bohrdt2020_gauss,Wang2021}. These studies include cases beyond the three point correlation functions studied here and investigate e.g. a five point correlator capturing the four spins surrounding a hole. Related higher-order correlation functions have also recently been measured experimentally \cite{Koepsell2020_FL}.

\subsubsection{Experiments out-of equilibrium}
\begin{figure}
\centering
  \includegraphics[width=0.99\linewidth]{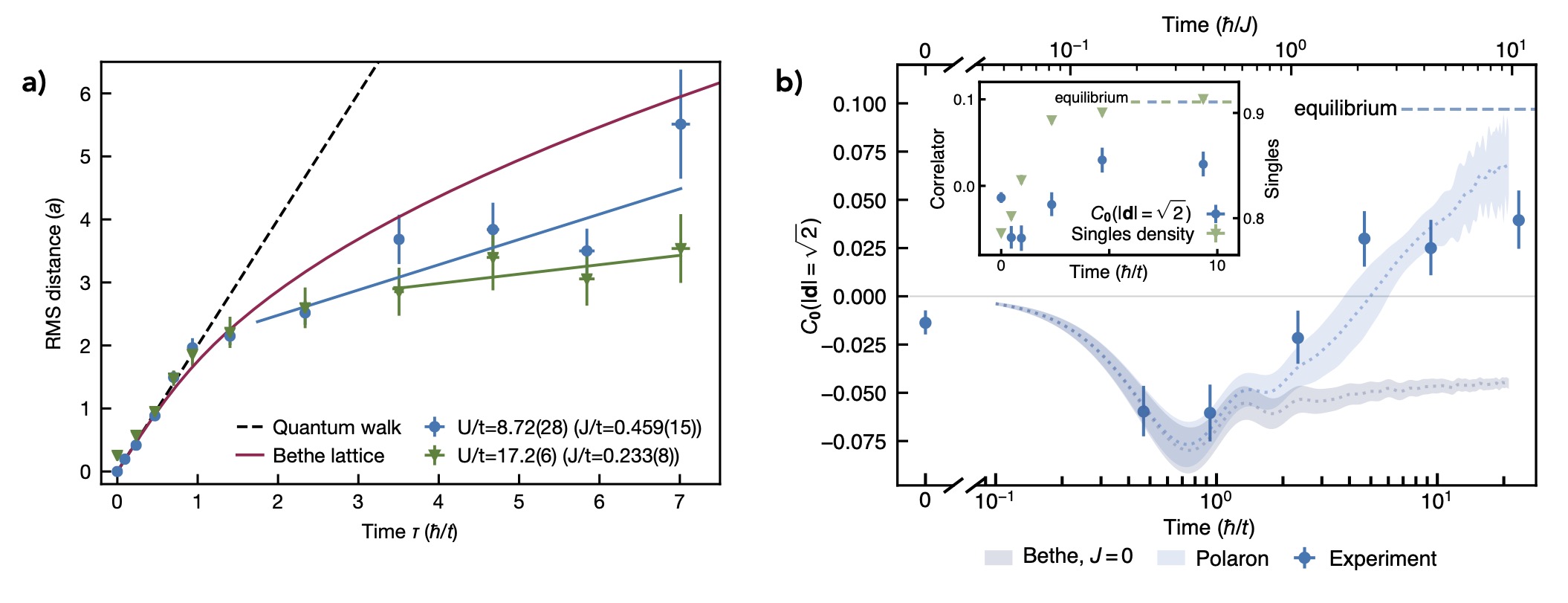}
\caption{\textbf{Formation and spreading of a magnetic polaron.} a) Root mean square distance of the hole from the origin, where it was initially pinned, as a function of time. Measurements were performed for two different interaction strengths, $U/t =8.7$ and $U/t = 17.2$. b) The diagonal next nearest neighbor spin correlations across the site where the hole was initially created show fast initial dynamics, followed by a slow recovery at long times. The observations are in good agreement with theoretical expectations from a polaron model (blue band). Figures taken from \cite{Ji2020}.}
\label{fig:magneticPolaronDyn}
\end{figure}

Similar to the experiment described in section \ref{sec1Ddyn},  a single hole can also be created or released in a two-dimensional system.  This was done in \cite{Ji2020}, where a second digital micromirror device (DMD) was used to prevent the occupation of a single site when the atoms are adiabatically loaded into the optical lattice,  resulting in an initial state with a pinned hole.  Note that unlike in a one-dimensional spin system, where  a pinned hole leads to an initial state of two completely disconnected chains,  all spins are coupled to each other through paths around the pinned hole in this case. 
In \cite{Ji2020},  the pinned hole was released by quickly shutting off the light illuminating the second DMD.  Snapshots with the quantum gas microscope were then taken after variable evolution times $\tau$.  In this particular experiment,  either a single spin state or the parity projected singles population could be observed.  

Due to the parity projected imaging,  doublon-hole pairs cannot be distinguished from the released hole.  The average background density is thus subtracted from the measured density to reveal the resulting hole motion.  In the measured density distribution,  coherent features of the propagation were observed,  for example oscillations in the return probability.  
In the initial dynamics of the hole,  the details of its motion depend on the presence or absence of quantum interference effects due to the (in-)distinguishable background spins.  A comparison to numerically simulated dynamics in an infinite temperature spin background \cite{Carlstrom2016PRL,Nagy2017PRB},  as well as in a ferromagnetic background,  shows that the hole motion is affected by the antiferromagnetic spin background in the experiment \cite{Ji2020}. 

At times shorter than the superexchange time scale $1/J$ associated with the spins,  the hole was found to spread out ballistically with a velocity $v = 2t/\hbar$ given by the hole hopping $t$.  Subsequently,  the spreading of the hole slows down.  In order to analyze this behavior further,  a Feshbach resonance was used to double the on-site interaction $U$ and thus halve the superexchange $J$ relative to $t$.  Comparison of the hole dynamics for the two different values of $t/J$ showed that the slower motion at longer times happens with a velocity controlled by $J$,  see Fig.~\ref{fig:magneticPolaronDyn} a).  The same result was obtained in tensor network simulations in (quasi-) two dimensional systems \cite{Bohrdt2020_NJP,Hubig2020} for the dynamics of a hole created locally in the undoped ground state, where the velocity was found to be directly proportional to $J$ at long times for a four-leg cylinder \cite{Bohrdt2020_NJP}.  The experimentally observed long-time behavior is consistent with a quasiparticle bandwidth proportional to the superexchange, as predicted theoretically for the the $t-J$ model, see e.g. \cite{Martinez1991}.  

The entire dynamics thus consists in the fast formation and subsequent slow spreading of a magnetic polaron.  This characteristic two stage dynamics \cite{Mierzejewski2011,Eckstein2014,Golez2014,Grusdt2018tJz,Bohrdt2020_NJP} has also been observed in the local spin correlations \cite{Ji2020}.  The nearest and next nearest neighbor spin correlations in the vicinity of the initial hole location similarly exhibit a fast evolution away from, and a subsequent slow return to equilibrium, see  Fig.~\ref{fig:magneticPolaronDyn} b).

The observed spin relaxation dynamics can be quantitatively described with an analytical model,  where the hole motion is mapped to a single particle moving on a Bethe lattice and the origin of this Bethe lattice corresponds to the position of the spinon created initially.  The spinon itself propagates ballistically with a velocity given by the superexchange energy \cite{Bohrdt2020_NJP}.  Importantly,  the snapshots taken with a quantum gas microscope allow for a direct comparison of this theoretical model to the experimental data \cite{Ji2020}: experimental snapshots after the quench are compared to snapshots before the quench, in which the hole motion is added by hand according to the theoretic prediction as follows: In the effective model, the spinon dynamics is accounted for by shifting the experimental pictures according to the predicted spinon motion,  while the hole motion is included by displacing the spins according to a distribution of hole trajectories predicted by the analytic Bethe-lattice model.

\section{Doped quantum magnets in 2D}\label{sec2DDoping}

Arguably one of the biggest promises of quantum simulation has been to enable insights into models that are numerically difficult or impossible to simulate.  Arguably most prominently,  this entails the simulation of the Fermi-Hubbard model at finite doping and low temperatures.  In this section,  we will discuss a variety of experiments which have addressed this regime with quantum gas microscopes in the past five years. 

Experimentally,  finite doping is realized by reducing the density of atoms in the system,  for example by adding a potential offset in the observation region \cite{Mazurenko2017}.
In order to determine the doping realized in an experiment without simultaneous spin- and charge resolution, one has to compare the single-particle density to exact numerical results. Note that the number of doublon-hole pairs -- and thus the number of sites detected as  in this case -- depends on the doping.

\subsection{Conventional probes}\label{ConventionalProbes}

In this subsection we start by discussing conventional probes of the Fermi-Hubbard model,  such as two-point correlation functions and measurements motivated by solid state experiments,  for example angle resolved photoemission spectroscopy (ARPES). 

\subsubsection{Spin-spin correlations}
\label{secTwoPointSpinCorrEquil}

\begin{figure}[t]
\centering
  \includegraphics[width=0.99\linewidth]{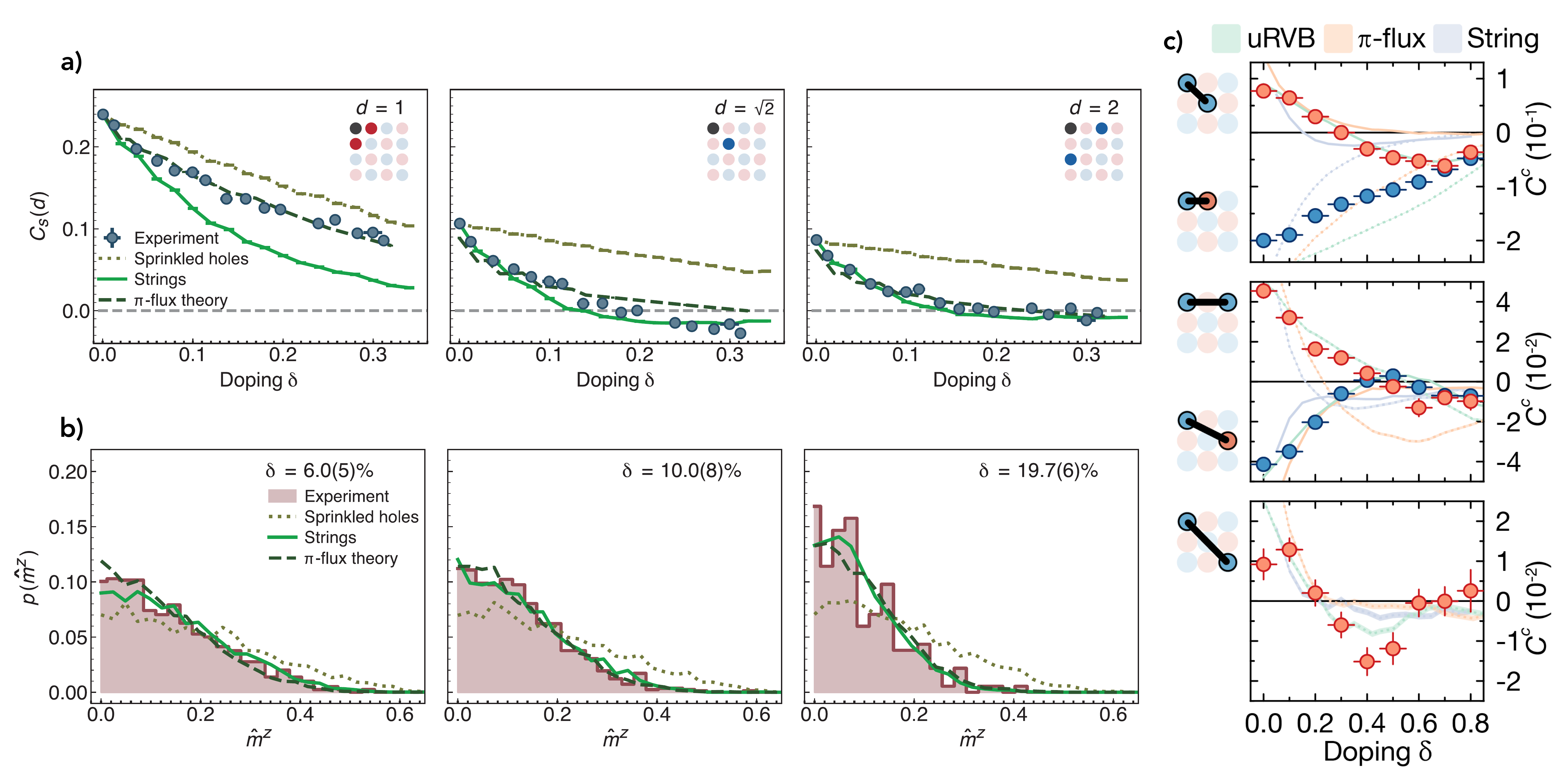}
\caption{ \textbf{Spin-dependent observables at finite doping.} a) Sign-corrected spin correlations as a function of doping and b) the full counting statistics of the staggered magnetization measured  at a temperature of $T/J = 0.65(4)$ in units of the super-exchange $J$. In b) the doping levels $\delta$ are indicated. These measurements were performed in the Harvard group \cite{Chiu2019Science}. c) Similar measurements (without sign-correction) performed in Munich \cite{Koepsell2020_FL} at a slightly higher temperature of $T/t=0.43(3)$ in units of the tunneling $t$. In part c) of the figure, the spin correlations are normalized by single site occupation. Part a),b) taken from \cite{Chiu2019Science}, c) from \cite{Koepsell2020_FL}. }
\label{fig:spincorrelations}
\end{figure}

In order to study how the antiferromagnetic order vanishes upon doping,  two-point spin-spin correlations are perhaps the most obvious quantity to look at.  In early experiments,  the temperatures were relatively high,  such that only short-range spin correlations were sizable.  One observation that was already made in those experiments was however that the next-nearest neighbor correlation changes its sign at roughly 20\% hole doping \cite{Cheuk2016,Parsons2016}. 
This sign change of the spin correlations at approximately 20\% doping was later observed at lower temperatures of $T/J = 0.65(4)$ \cite{Chiu2019Science} and $T/t = 0.43(3)$ \cite{Koepsell2020_FL} in correlations up to distances of $d=\sqrt{8}$ between the spins, Fig.~\ref{fig:spincorrelations}. Only the nearest-neighbor correlator remains antiferromagnetic and does not change sign. 

Qualitatively, this effect is captured by the geometric string theory \cite{Grusdt2019,Chiu2019Science}, where each dopant causes a string of displaced spins. This displacement of spins leads to a mixing of the different correlation functions, see also section \ref{secMagneticPolaron}, and therefore causes the correlations to change sign once a certain threshold of mixing is reached. The geometric string picture describes a gas of magnetic polarons and thus constitutes a low-doping theory. On the other hand, coming from the high-doping side the observed sign-change can be interpreted as a departure from a weakly interacting Fermi-liquid description of the original fermions \cite{Cheuk2016,Koepsell2020_FL}.

The behavior of the spin correlations as the doping is increased  also appears to be consistent with incommensurate magnetism,  as observed in cuprate materials \cite{Yamada1998}.  While finite size effects and finite temperatures obscure the observation of clearly resolved peaks in the static structure factor at momenta close but not equal to $\vec{q}_{\rm{AFM}} = (\pi,\pi)$ \cite{Mazurenko2017,Koepsell2020_FL},  a clear shift of the peaks from $(\pi,\pi)$ to $(\pi,0)$ was observed in \cite{Koepsell2020_FL},  where the doping is varied from $0$ to almost $100\%$.  

The fluctuation-dissipation relation allows one to extract the uniform spin susceptibility directly from the measurements of the static spin structure factor.  Starting from high dopings the susceptibility was found in \cite{Koepsell2020_FL} to first increase when the doping is lowered. But then at about $30\%$ doping, it was found to stop increasing \cite{Koepsell2020_FL}, see section \ref{secSusceptibility} for an extended discussion. This behavior is reminiscent of the pseudogap phenomenology. Note that in cold atom experiments, the \emph{dynamical} spin structure factor is accessible through Bragg spectroscopy. A corresponding experiment has been for example performed for a one-dimensional system of Lithium atoms in Ref.~\cite{Yang2018}, yielding excellent agreement with predictions from Tomonaga-Luttinger theory. 

Apart from the two-point spin correlation functions, the loss of antiferromagnetic ordering with doping can also be probed through the full counting statistics of the staggered magnetization, where the staggered magnetization operator is defined as
\begin{equation}
    \hat{m}^z = \frac{1}{N} \sum_{\mathbf{i}} (-1)^{|\mathbf{i}|} \hat{S}_{\mathbf{i}}^z
\end{equation}
for a system with $N$ lattice sites. 
As doping increases, the full counting statistics of the staggered magnetization was found in \cite{Chiu2019Science} to narrow, see Fig.~\ref{fig:spincorrelations} b). Note that the increasing number of empty sites is not sufficient to explain this narrowing, as the comparison to sprinkled, i.e. randomly placed holes according to the doping level, shows. Similarly, a narrowing of the distribution was observed at half filling with increasing temperature \cite{Mazurenko2017}.

\subsubsection{Charge-charge correlations}

Motivated by the idea that interactions between doped holes,  or even hole pairing,  could manifest itself in bunching of holes in real space,  several experiments have studied charge-charge correlations \cite{Cheuk2016,Chiu2019Science,Koepsell2020_FL}.  
In experiments without full resolution,  where doublons are imaged as empty sites,  care must be taken when charge-charge correlations are evaluated.  Typically in this case,  the correlation between empty sites,  the anti-moment correlation \cite{Cheuk2016}, is considered,
\begin{equation}
C_h(|\vec{d}|) = \bigl\langle(1-\hat{n}_{s,\vec{i}}) (1-\hat{n}_{s,\vec{i}+\vec{d}}) \bigr\rangle - \bigl\langle(1-\hat{n}_{s,\vec{i}})\bigr\rangle \bigl\langle (1-\hat{n}_{s,\vec{i}+\vec{d}})\bigr\rangle,
\label{eq:holehole}
\end{equation}
where $\hat{n}_{s,\vec{i}}$ is the single particle occupation on site $\vec{i}$ and an average over all sites $\vec{i}$ is typically performed.  
This correlator contains three different contributions: \emph{(i)} correlations between actual dopants,  \emph{(ii)} correlations between a doped hole and a hole or doublon belonging to a doublon hole pair,  and \emph{(iii)} correlations between holes and dopants both belonging to a doublon hole pair.  Effects of doublon-hole pairs are particularly strong at low doping,  where there are more or comparably many empty sites due to doublon hole pairs than due to doping.
The dominant contribution of doublon hole pairs results in strong bunching of empty sites on nearest neighboring sites,  see also section \ref{secAFM}.  This effect is also explicitely observed by studying correlations between doublons and holes using full resolution in \cite{Hartke2020}.  
At distances larger than one,  doublon hole pairs mainly contribute to the average number of empty sites,  which is subtracted in \eqref{eq:holehole}.  

For straight and diagonal next nearest neighbor anti-moment correlations,  anti-bunching has been observed at low temperatures of $T/J=0.65(4)$ \cite{Chiu2019Science}.  
In a different experiment with full resolution,  anti-bunching between holes was observed also on nearest neighboring sites \cite{Koepsell2020_FL}.  In said experiment,  no anti-correlations were found at larger distances.  However,  this might be due to the considerably higher temperature of $T/t=0.52(5)$.
Negative anti-moment correlations on nearest neighboring sites were also observed without full resolution at comparably high temperatures of $T/t = 1.22$ for dopings $\delta \geq 40\%$,  where the signal due to doublon-hole pairs is negligible \cite{Cheuk2016}. In this regime, the formation of a Pauli correlation hole expected from a free-fermion theory explains the observations.  

The experimental observation of anti-bunching between holes at the lowest currently achievable temperatures \cite{Chiu2019Science} may appear paradoxical at first, given that the repulsive Hubbard model is considered a promising candidate to explain the underlying pairing mechanism of high-Tc superconductors. However, if pairing of fermions is possible in this model, it takes place at significantly lower temperatures. At the relatively high temperatures realized in current experiments, it is thus not surprising that repulsive interactions dominate. As an example, consider fermion correlations on the BEC-side and in the vicinity of a Feshbach resonance in the continuum. At the lowest energies, a shallow bound state exists, however at elevated temperatures the associated repulsive scattering length leads to dominantly repulsive correlations.

The measurement of density-density correlations and density fluctuations has also been used as a model-free thermometer through the fluctuation-dissipation theorem,  which relates the compressibility $\kappa = d n/d\mu$ and density correlations to the temperature \cite{Hartke2020}. 
In \cite{Drewes2016}, the compressibility as well as the local density fluctuations were measured for interaction strengths $U/t$ between $1.6(2)$ and $12.0(7)$ and temperatures between $T/t \approx 0.5$ and $T/t \approx 4$. Using the fluctuation-dissipation theorem, the contribution of non-local density correlations was determined. It was found that close to half-filling, interactions lead to a strong reduction of non-local density correlations, as expected for a Mott insulator. As doping is increased to $50\%$ and beyond, the different interaction strengths yield similar results and are consistent with an ideal Fermi gas on a lattice.

\subsubsection{Susceptibility and compressibility}\label{secSusceptibility}
\begin{figure}[t]
\centering
  \includegraphics[width=0.99\linewidth]{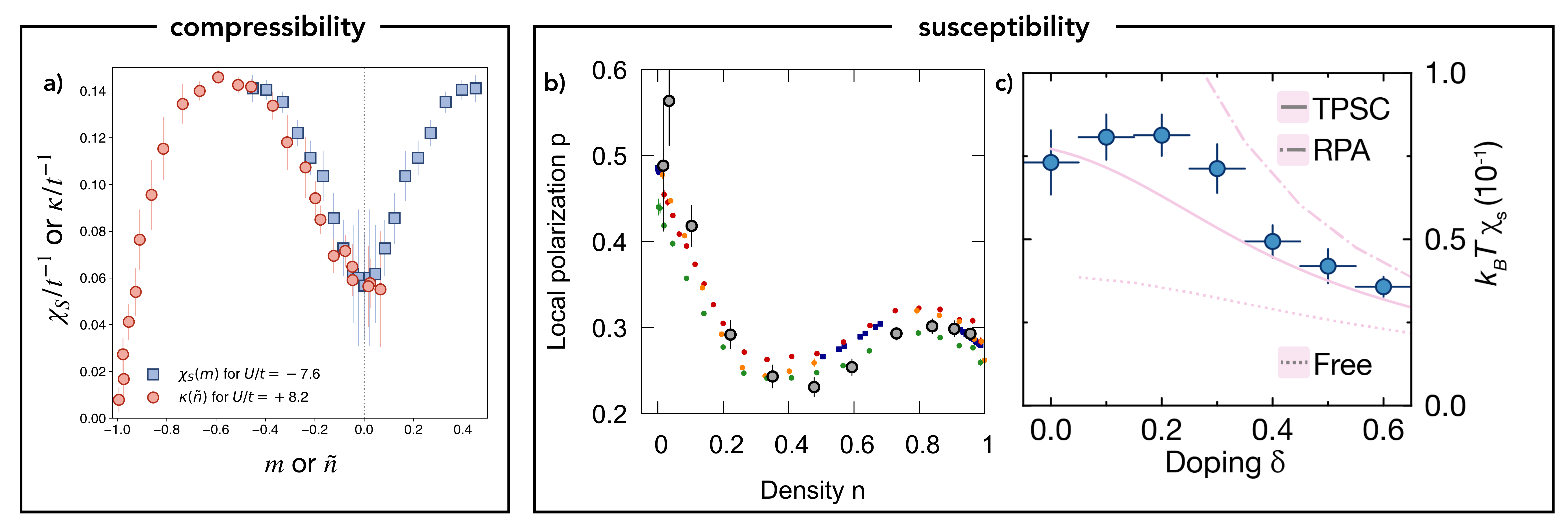}
\caption{\textbf{Compressibility and susceptibility.} a) Compressibility $\kappa / t^{-1}$ in units of $1/t$ as function of hole doping $\tilde{n} = - \delta$ in the repulsive Fermi-Hubbard model and spin susceptibility $\chi_S / t^{-1}$ as a function of magnetization $m$ in the attractive model; Figure taken from \cite{Gall2020}. b) Local polarization $p$ as a function of density $n=1-\delta$ for $U/t = 14.7(8)$ at an effective field of $h/t = 0.2(1)$, realized through a global spin imbalance; Figure taken from \cite{Brown2017}. The large gray dots are experimental measurements, colored data points are NLCE and DQMC results at different temperatures and finite magnetic field. c) Doping dependence of the uniform magnetic susceptibility, obtained from spin correlations through the fluctuation-dissipation theorem; Figure taken from \cite{Koepsell2020_FL}. Experimental data (blue symbols) is compared to a two-particle self consistent approach (TPSC), random-phase approximation (RPA), and free fermion calculations (solid, dashed and dotted pink curves).}
\label{fig:susceptibility}
\end{figure}

In \cite{Gall2020}, the spin susceptibility $\chi_S = \partial m/ \partial h$ of the \emph{attractive} Fermi-Hubbard model was measured using two hyperfine states of Potassium, where an effective Zeeman $h$ field was applied during the experiment. 
Due to the particle-hole symmetry, the spin susceptiblity in the attractive Fermi-Hubbard model directly corresponds to the compressibility $\kappa = \partial n/\partial \mu$ in the repulsive Fermi-Hubbard model. Both the susceptibility and the compressibility can be measured experimentally as numerical derivative with respect to the effective magnetic field and the chemical potential, respectively, and their correspondence has been directly observed, see Fig.~\ref{fig:susceptibility} a). The measurements of the compressibility in \cite{Gall2020} confirm the expected dip in the Mott insulating regime of the repulsive Hubbard model  \cite{Jordens2008,Schneider2008}. Similar results have been obtained for a range of interactions $U/t$ and temperatures $T/t \geq 0.6$ in the repulsive Fermi-Hubbard model \cite{Cocchi2016}. 

With similar techniques as in \cite{Gall2020}, the compressibility in the \emph{attractive} Fermi-Hubbard model has been measured in \cite{Chan2020} as a function of density. Note that doping in the attractive model translates to spin imbalance in the repulsive model. Through the fluctuation dissipation relation, the compressibility can be related to the pair correlation function. From the latter, a pair correlation length was extracted, which was found to increase with decreasing filling, temperature, and absolute value of the interaction strength $|U/t|$ \cite{Chan2020}.

An effective magnetic field can also be realized in the repulsive Hubbard model through a spin imbalance in a system in equilibrium. Due to the harmonic confinement in \cite{Brown2017}, the density of atoms increases towards the center of the trap, and therefore different doping regimes are realized simultaneously. For a fixed global spin imbalance, the local polarization $p = (n_\uparrow - n_\downarrow)/(n_\uparrow + n_\downarrow)$ can then be obtained as a function of the density $n$. In \cite{Brown2017}, the local polarization was obtained as a proxy for the spin susceptibility, $\chi_S \propto n p$ in a range of densities from $n=0$ to $n=1$  for an interaction strength of $U/t = 14.7(8)$, Fig.~\ref{fig:susceptibility} b).  

The non-monotonic behavior of the local polarization as a function of doping resembles the behavior of the magnetic susceptibility observed in the cuprates in the metallic phase with a weak peak around $20\%$ doping \cite{Torrance1989,Johnston1989,Takagi1989}. This is considered a hallmark signature of the pseudogap phase in cuprates.  

The uniform ($\mathbf{q} = \mathbf{0}$) spin susceptibility $\chi_S(\vec{0})$ was also extracted from equal-time spin correlation functions through the fluctuation-dissipation theorem in \cite{Koepsell2020_FL}, see Fig.~\ref{fig:susceptibility} c). It was found to exhibit similar behavior at low doping as reported in \cite{Brown2017}. In particular, it is consistent with the non-monotonic behavior described above with a weak maximum at approximately $20\%$ doping. The comparison to Fermi liquid behavior shown in Fig.~\ref{fig:susceptibility} c) emphasizes that the decrease of the spin susceptibility with decreasing doping at small doping is an anomalous behavior. 
In \cite{Drewes2017}, the uniform spin susceptibility was similarly extracted from measurements of the static structure factor at $\mathbf{q}=\mathbf{0}$ as a function of doping and for temperatures $T/t \geq 0.63$. For a constant temperature, the static structure factor increases with decreasing doping and reaches a plateau at around $20\%$ doping for the lowest temperatures considered. The onset of this plateau shifts to smaller doping values as the temperature is increased.

Other thermodynamic quantities can be measured as well. For example, in Ref. ~\cite{Cocchi2017} the dependence of the density on the chemical potential, related to the compressibility, was used to determine a pressure $P$. The temperature dependence of the pressure then yields the thermodynamic entropy per site $s$ through the relation 
\begin{equation}
    s = a^2 \left. \frac{\text{d}P}{\text{d}T} \right|_{\mu=\text{const.}},
\end{equation}
where $a$ is the lattice constant. At strong interactions and temperatures $T\lesssim t$, the entropy per site as a function of the chemical potential $\mu$ exhibits a non-monotonous behavior with a local minimum at half-filling ($\mu = U/2$) \cite{Cocchi2017}.

\subsubsection{Transport}
\begin{figure}[t]
\centering
  \includegraphics[width=0.99\linewidth]{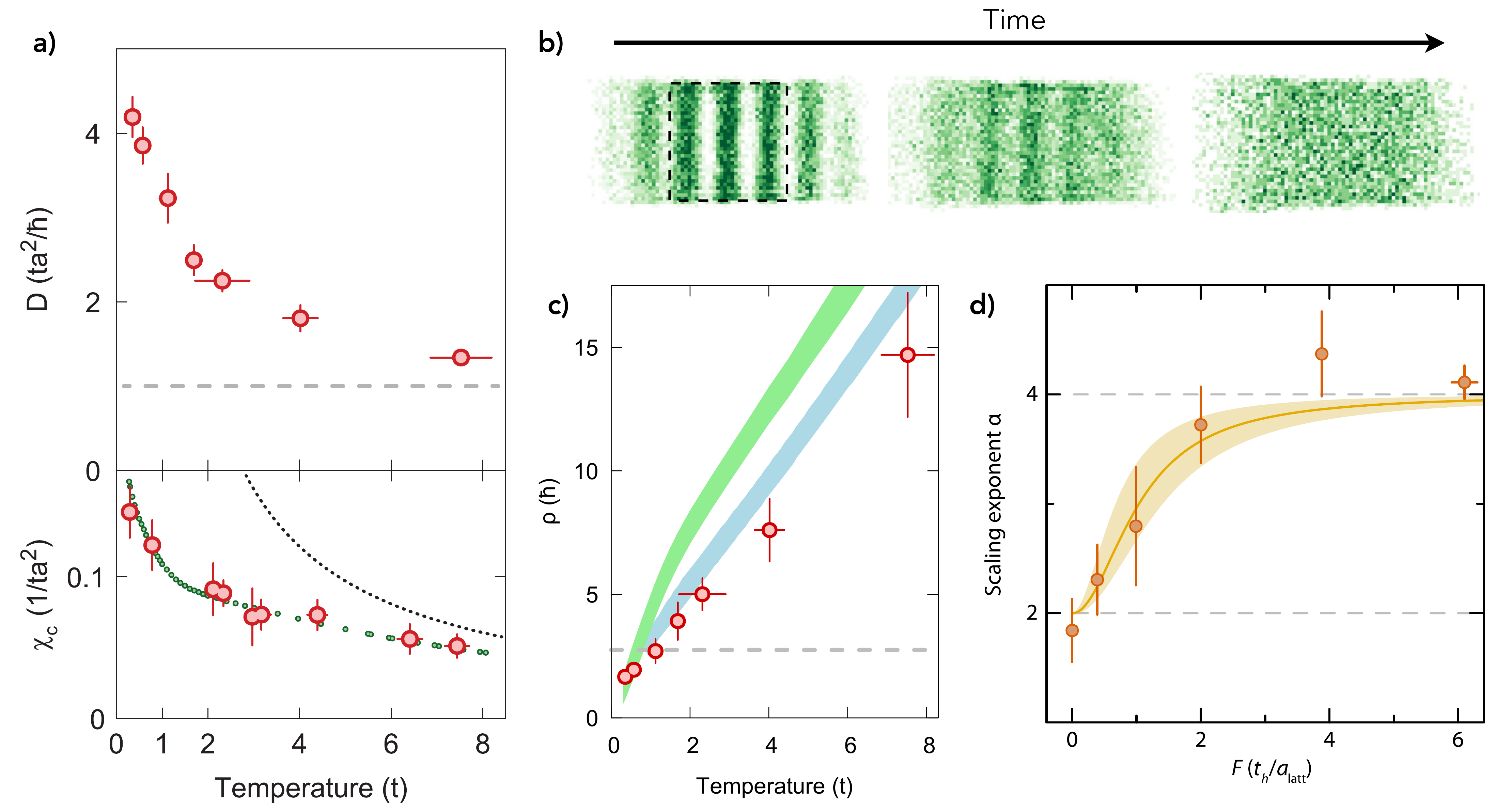}
\caption{ \textbf{Diffusion and subdiffusion.} Decay of an initially prepared charge density wave without a tilt \cite{Brown2019a}, panel a) and c), and with an additional tilt \cite{GuardadoSanchez2020}, panel d). b) The system is prepared in a charge density wave (CDW) state along one direction. The ensuing time dynamics is then probed. a) In a system without a tilt, the relaxation of the CDW leads to diffusive behavior, with a diffusion constant $D$. The dashed gray line (top panel) corresponds to the Mott-Ioffe-Regel limit, which provides a lower bound for the diffusion constant. Combined with the measurement of the charge compressibility $\chi_C$ (lower panel), the resistivity can be extracted. c) Resistivity as a function of temperature, experimental data (red) compared to a finite temperature Lanczos method (blue) and DMFT (green). The Mott-Ioffe-Regel limit $\sigma_{\text{MIR}} \hbar = \sqrt{n/2\pi}$ is shown as gray dashed line. 
d) In the presence of a strong tilted potential, the relaxation dynamics crosses over from diffusive to subdiffusive. In order to extract the scaling exponent $\alpha$, with decay time $\tau = \lambda^\alpha$, for a given tilt strength, the decay for CDWs of different wavelengths $\lambda$ is studied. The plot shows the scaling exponent $\alpha$ as a function of tilt strength, showing diffusive behavior with $\alpha=2$ for the non-tilt case. Figures extracted and adapted from \cite{Brown2019a} a), c), and \cite{GuardadoSanchez2020} b), d).}
\label{fig:diffusion}
\end{figure}

The strange metal phase constitutes one of the biggest mysteries of the cuprate phase diagram: at intermediate to high dopings,  above the superconducting dome,  the cuprate materials display curious transport properties, see e.g. \cite{Hussey2018}, which have led to the description as strange or bad metals.  In particular, the resistivity is $\propto T$ for low temperatures, \cite{Hussey2008} as opposed to $\propto T^2$ as expected for a Fermi liquid.

In \cite{Brown2019a}, the transport behavior was probed in the Fermi-Hubbard model by studying the relaxation of an initially prepared charge density wave at an average overall density of $\left<n\right> = 0.82$ and in a temperature range of $0.3 < T/t<1$, Fig.~\ref{fig:diffusion} a).  The decay of the charge density wave was modeled hydrodynamically, taking into account particle number conservation as well as a finite current relaxation rate,  in order to extract a charge diffusion constant $D$.  The diffusion constant can be related to the resistivity $\rho$ through the Nernst-Einstein equation,
\begin{equation}
\sigma = \chi_C D,
\label{eqNernstEinstein}
\end{equation}
with the compressibility $\chi_C=dn/d\mu|_T$ and the conductivity $\sigma = 1/\rho$. The compressibility was measured in a separate experiment, as described below. 

The diffusion constant was measured for initially superimposed sinusoidal patterns with different wavelengths, Fig.~\ref{fig:diffusion} a). The decay of the density wave pattern becomes consistent with diffusive transport expected from a hydrodynamic description at long wavelengths, where a density pattern with wave vector $k$ decays exponentially on a time scale given by $\tau = 1/Dk^2$, as can be seen in the limit of $F = 0$ (no applied force) in Fig.~\ref{fig:diffusion} c).
The temperature dependence of the diffusion constant is measured by controlled heating of the atomic cloud.  As the temperature is increased, the diffusion constant decreases rapidly, Fig.~\ref{fig:diffusion} b).  The Mott-Ioffe-Regel limit provides a lower bound on the diffusion constant, which is approached but not violated in \cite{Brown2019a} at high temperatures. 

Remarkably, the experiments in Ref.~\cite{Brown2019a} also observed damped oscillations of the decaying charge density wave patterns. This indicates the existence of a well-defined sound-like mode in the considered finite-doping regime.

In the harmonic trap in this experiment, the position can be converted to a chemical potential in the local density approximation.  The change of the density as a function of the position thus yields the charge compressibility.  The experimentally measured charge compressibility decreases with increasing temperature and approaches a high temperature limit, Fig.~\ref{fig:diffusion} b).  For low temperatures,  saturation is expected but not seen at the lowest experimentally accessible temperatures of $T/t=0.3$. 

From the Nernst-Einstein equation \eqref{eqNernstEinstein},  the resistivity $\rho$ was obtained through the experimentally measured diffusion constant and compressibility.  It was found that the resistivity increases with increasing temperature and does not saturate within the range of temperatures realized in \cite{Brown2019a}.  The resistivity violates the corresponding Mott-Ioffe-Regel bound for temperatures  higher than $T/t \approx 1.3$, which is consistent with observations in cuprate materials in the bad metal phase \cite{Gunnarsson2003}.

The same setup, but at a significantly weaker interaction strength of $U/t=3.9$,  was used in \cite{GuardadoSanchez2020} to study the thermalization behavior in the presence of a potential gradient,  leading to subdiffusive behavior for sufficiently strong tilts, Fig.~\ref{fig:diffusion} c). 
As the imprinted density profile decays,  local number density currents flow and the \emph{tilt energy} is reduced.  Since the total energy is conserved, \emph{non-tilt energy} has to flow correspondingly.  This flow can be described by a thermal diffusion constant $D_{th}$, which sets the rate for global relaxation of the system,  assuming that the system is locally thermalized with inverse temperature $\beta(x,t)$. 
This behavior is captured by a hydrodynamic description, where the potential gradient couples the conserved density and energy.  The local inverse temperature can be determined through measurements of the single occupancy density, and thus predictions from the hydrodynamic description for $\beta(x,t)$ and $n(x,t)$ are verified. 

\emph{Transport measurements in three dimensions.--}
In \cite{Xu2019}, a different approach was taken to measure the resistivity: in an initially spin polarized gas, a pair of Raman beams was used to transfer atoms to the other spin state ($\ket{\! \downarrow}$) while simultaneously introducing a momentum shift, which is determined by the difference in the wavevectors of the Raman beams. The relaxation of the average momentum of atoms in the spin down state then yields a transport lifetime $\tau$, from which the resistivity can be extracted. In the three-dimensional lattice studied in \cite{Xu2019}, the resulting resistivity scales quadratically with interaction strength $U/t$ and linearly with temperature $T/t$. 

The conductivity $\sigma_{\alpha \beta}(\omega)$ in a three-dimensional system has also been measured in the linear response regime \cite{Anderson2019} through Ohm's law, 
\begin{equation}
    \left< J_\alpha (\omega) \right> = \sum_\beta \sigma_{\alpha \beta} (\omega) F_\beta(\omega),
\end{equation}
where the force $F_\beta(\omega)$ stems from the periodic displacement of the laser beams forming the crossed dipole trap. The steady-state bulk current $\left< J_\alpha (\omega) \right>$ was determined as the time-derivative of the site-granulated centre-of-mass position operator. The latter was measured experimentally from images of the central four planes of the system realized in Ref.~\cite{Anderson2019}. 

Perhaps the most straightforward way to probe transport properties in a cold atom experiment is to observe the dynamics of the atoms after a global quench in the potential. 
In \cite{Schneider2012}, a harmonic confining potential was turned off. In the subsequent time evolution, it was observed that without interactions, atoms in a three-dimensional optical lattice spread ballistically. In the presence of interactions of either sign, the density distribution after some expansion time was found to be bimodal, with a ballistic and a diffusive contribution \cite{Schneider2012}. The ballistic contribution is due to the tails of the distribution, where the density is lower, and the atoms thus only scatter rarely. 
In \cite{Strohmaier2007}, the trap minimum was suddenly displaced and the center of mass dynamic of the atoms, again in a three-dimensional optical lattice, was monitored. Without interactions, the center of mass displacement exhibits weakly damped oscillations. Upon increasing the attractive interactions, a slow relaxational drift was observed instead. 

The techniques described above were applied in the three-dimensional Fermi-Hubbard model, but can in principle be also used to study the transport properties of the two-dimensional Fermi-Hubbard model with cold atoms. 

\subsubsection{ARPES}
\begin{figure}[t]
\centering
  \includegraphics[width=0.99\linewidth]{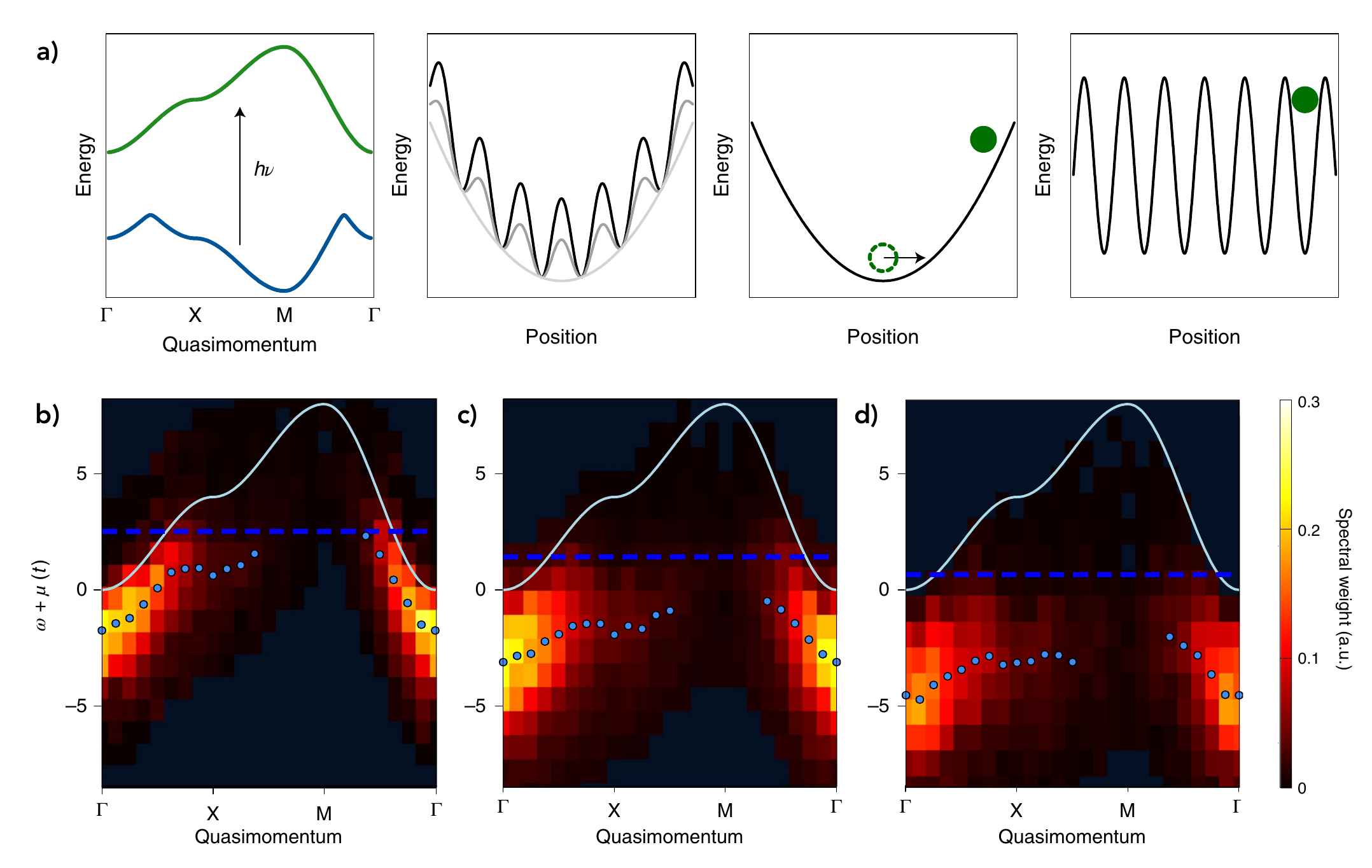}
\caption{ \textbf{ARPES in a quantum gas microscope.} 
a) Measurement scheme used in \cite{Brown2019} for extracting the spectral function in a quantum gas microscope: A radiofrequency photon with energy $h\nu$ leads to a transfer of an $\ket{\!\uparrow}$ atom to a third, non-interacting state $\ket{f}$. The quasimomentum is mapped to momentum through band mapping (second panel from left) by ramping down the lattice depth slowly. The momentum is then mapped to a real-space position through expansion for a quarter-period in a harmonic trap (third panel). An optical lattice is ramped up in order to freeze the position of the atoms, thus enabling imaging with the microscope. 
b)-d) Experimentally measured occupied spectral function as a function of frequency $\omega$ and quasimomentum along the high-symmetry lines of the Brillouin zone, with the chemical potential (dark blue dashed line) and the non-interacting dispersion relation (light blue solid line). For each momentum, a Lorentzian is fit to the energy distribution curve and the corresponding centre position is marked as blue circle in the plots.
Interaction strengths and temperatures are b) $U/t = -3.7(1)$ and $T/t=0.48(2)$, c) $U/t = −6.0(1)$, $T/t = 0.50(2)$, d) $U/t = −7.5(1)$, $T/t = 0.55(3)$. 
For the parameters in b), no pseudogap is present and the dispersion reaches the chemical potential. 
In c) and d), a pseudogap can be observed as the dispersion pulls back from the chemical potential. 
Figures extracted and adapted from \cite{Brown2019}.}
\label{fig:ARPES}
\end{figure}

Similar to angle resolved photoemission spectroscopy in solid state experiments,  the spectral function can be measured in a quantum gas microscope, see Fig.~\ref{fig:ARPES}a).  To this end, a particle is removed from the system either by photon-assisted tunneling to an empty probe system \cite{Kollath2007,Bohrdt2018}, or by applying a radiofrequency pulse that transfers one of the two interacting spin states to a third non-interacting (hyperfine) state \cite{Brown2019}.
In both cases, the quasimomentum of the final state has to be determined. This can be achieved using bandmapping or time-of-flight techniques that map momentum states to real space positions, which can then be detected with the quantum gas microscope.

In \cite{Brown2019},  the protocol described above,  using a third non-interacting hyperfine state, was applied in the attractive Fermi-Hubbard model. 
In this experiment, two hyperfine states of lithium realize the two spin states, as discussed before. A third hyperfine state can be rendered non-interacting by tuning the magnetic field controlling the interaction strength accordingly. This, however, is only possible for attractive interactions between the two hyperfine states used to represent the electronic spin. Therefore, the scheme employing a radiofrequency pulse to transfer one of the two interacting spins to a non-interacting state can only be applied in the attractive Fermi-Hubbard model. 
This model exhibits a BEC-to-BCS crossover as the interaction strength is tuned.  In \cite{Brown2019},  a pseudogap was observed in the occupied single particle spectral function for temperatures between $T/t=0.5$ and $T/t=5$. The term pseudogap here refers to a suppression of spectral weight around the chemical potential. In the experimental data in Fig.~\ref{fig:ARPES}, this effect can be seen by comparing the peak position (fit marked by blue circles) to the chemical potential (dashed line). For sufficiently strong interactions, Fig.~\ref{fig:ARPES}c) and d), the peak positions stay well below the chemical potential for all momenta. 

Note that the attractive Fermi-Hubbard model $\H^-$ can be mapped to its repulsive counterpart $\H^+$ by identifying empty and doubly occupied sites with spin up and spin down and vice versa. In particular, a particle-hole transformation on the spin-down fermions only,
\begin{equation}
    \hat{U}^\dagger \hat{c}_{\mathbf{i},\uparrow} \hat{U} = \hat{c}_{\mathbf{i},\uparrow}, \qquad \hat{U}^\dagger  \hat{c}_{\mathbf{i},\downarrow} \hat{U} =  \hat{c}_{\mathbf{i},\downarrow}^\dagger,
\end{equation}
transforms the attractive to the repulsive Fermi-Hubbard model, $\hat{U}^\dagger \H^+ \hat{U} = \H^-$ \cite{Auerbach1998}. Since the mapping can also be applied to the eigenstates and eigenenergies,
\begin{equation}
    \H^- \ket{\psi_n^-} = E_n\ket{\psi_n^-}, \qquad \H^+\ket{\psi_n^+} = \H^+ \hat{U} \ket{\psi_n^-}= E_n \hat{U} \ket{\psi_n^-} = E_n \ket{\psi_n^+},
\end{equation}
the spectral function in the attractive Fermi-Hubbard model directly corresponds to the spectral function in the repulsive Fermi-Hubbard model: In the time-domain one obtains
\begin{equation}
    A^-_{\mathbf{ij},\uparrow} (t) = 
    \frac{1}{Z} \sum_n e^{-(\beta-i t) E_n} \bra{\psi_n^-} \c_{\mathbf{j},\uparrow}^\dagger e^{-i\H^-t} \c_{\mathbf{i},\uparrow} \ket{\psi_n^-}
    = \frac{1}{Z} \sum_n e^{-(\beta-i t) E_n} \bra{\psi_n^-} \hat{U}^\dagger  \hat{c}_{\mathbf{j},\uparrow}^\dagger e^{-i\H^+ t}  \hat{c}_{\mathbf{i},\uparrow} \hat{U} \ket{\psi_n^-} =  A^+_{\mathbf{ij},\uparrow} (t).
\end{equation}
The measurements shown in Fig.~\ref{fig:ARPES} are averaged over the harmonic trap, and thus different densities in the attractive Fermi-Hubbard model, which corresponds to different magnetizations in the repulsive model. However, if one assumes that a large part of the signal stems from the half-filled center of the trap, because there the density is highest, the measured occupied spectral function approximately corresponds to the occupied spectral function of the half-filled, spin-balanced, repulsive Fermi-Hubbard model. As theoretically expected \cite{Brunner2000,Mishchenko2001,Bohrdt2020_ARPES}, the spectral weight vanishes at low energies at momentum $\mathbf{k} = (\pi,\pi)$. Measurements at finite doping in the repulsive model can be performed using the same scheme by preparing a spin-imbalanced system with attractive interactions. 

Once experiments reach colder temperatures,  ARPES in quantum gas microscopes can be used, for example, to probe the pseudogap phase in the repulsive Hubbard model in a well-controlled setting. In particular, this will allow to address the problem of Fermi-surface reconstruction in a clean Hubbard model.

\subsection{Higher-order correlations}\label{higherOrderCorrs}

In section \ref{ConventionalProbes}, we saw how different microscopic theories, such as a doped RVB state and geometric string theory, can lead to similarly good agreement with experimental data for two-point correlation functions. However, if one considers higher-order correlation functions, significant differences between theories and experiment can arise. Note that since the energy is only sensitive to two-point correlations, such higher-order correlations reveal differences between theoretical models which go beyond a mere comparison of their variational energies. Cold atoms, and quantum gas microscopes in particular, offer an unprecedented opportunity to investigate correlation functions up to almost arbitrary order.

\subsubsection{Spin-charge correlations}
\begin{figure}
\centering
  \includegraphics[width=0.99\linewidth]{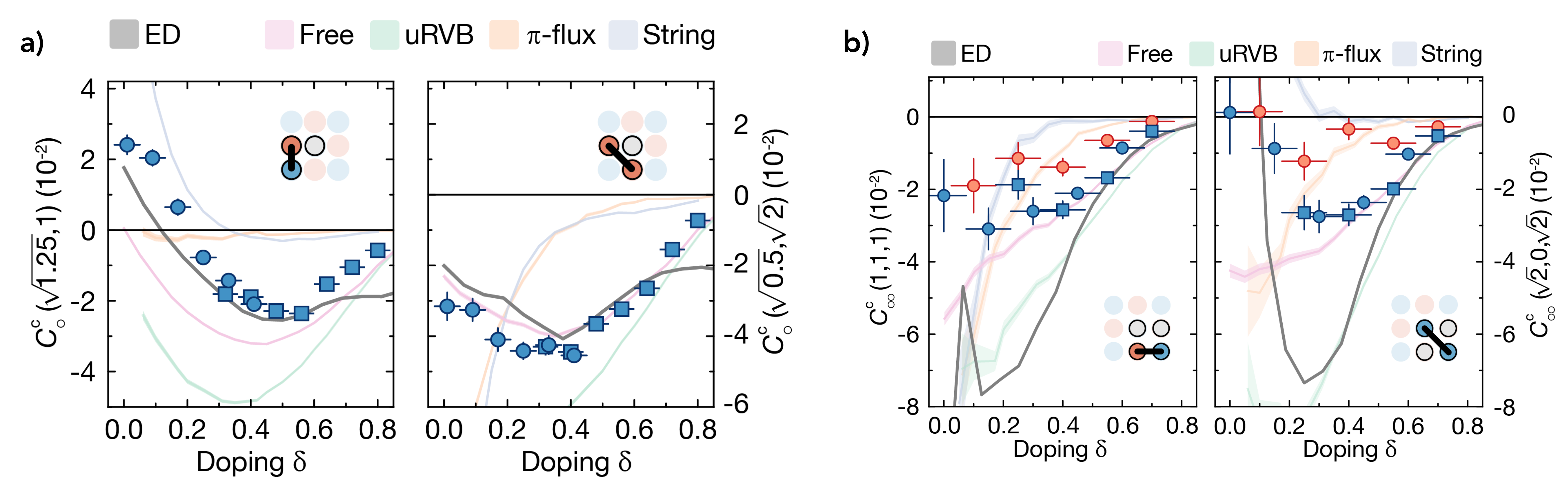}
\caption{ \textbf{Higher-order correlations.} a) Three-point spin-charge correlations: Connected nearest and next nearest neighbor spin correlations next to a hole as a function of doping. b) Four-point spin-charge correlations: Nearest neighbor connected spin correlations next to two nearest neighboring holes (left) and diagonal spin correlations across two holes (right) as a function of doping. Figures taken from \cite{Koepsell2020_FL}.}
\label{fig:higherOrderCorr}
\end{figure}

In \cite{Koepsell2020_FL}, third and fourth order correlations between spin and charge were studied across an extensive doping range from half-filling to $80\%$ hole doping at two values of $U/t$, namely $U/t=7.4(8)$ and $U/t = 8.9(5)$, and temperatures of $T/t=0.43(3)$ and higher. In order to reveal genuine higher order effects, the connected correlations were considered. 
By comparing experimental data to several different theories, the crossover from a polaronic to a Fermi liquid regime was observed in correlation functions of different order.

The magnetic polaron regime can be identified by considering the three-point correlations discussed in section \ref{secMagneticPolaron}, which probe the influence of the hole on the spin correlations in its vicinity. At low doping, the 
presence of the hole leads to a strong reduction of antiferromagnetic correlations in its surrounding, Fig.~\ref{fig:higherOrderCorr} a). As the doping is increased, this characteristic feature of the magnetic polaron disappears and changes sign, showing that the spin correlations around a given hole are no longer reversed relative to the background. This behavior is an indication of the crossover from a magnetic polaron- to a Fermi liquid regime with increasing doping. The doping level where this cross-over takes places agrees with its location in cuprate materials.

The experiment \cite{Koepsell2020_FL} constitutes a prime example for a case where only higher-order correlations allow to discriminate between different theoretical scenarios: As shown in Sec.~\ref{secTwoPointSpinCorrEquil}, the doped RVB and geometric string theories yield good agreement across all dopings and distances for two-point correlations. However, significant deviations from the experimental results are observed in the connected third-order correlations shown in Fig.~\ref{fig:higherOrderCorr} a). For example, the RVB states do not capture the sign-reversal of the nearest-neighbor spin correlations. 

In Ref.~\cite{White1997}, the ground state of two holes in the $t-J$ model was studied using DMRG, and negative binding energies are found. The authors suggested that \emph{the clearest ‘‘signature’’ of a bound pair of holes} is the emergence of a strong antiferromagnetic spin correlation close to the holes, in particular across a diagonal pair. While the density correlations measured in different cold atom experiments, see section \ref{ConventionalProbes}, do not exhibit any sign of pairing at the currently accessible temperatures, the four-point spin-charge correlation discussed in \cite{White1997} was measured in \cite{Koepsell2020_FL}, and indeed strong antiferromagnetic correlations are found on bonds directly neighboring two holes, see Fig.~\ref{fig:higherOrderCorr} b). Based on snapshots, the diagonal spin correlation can be directly evaluated for a pair of holes sitting on a diagonal bond, i.e. at distance $\mathbf{d} = (\pm 1,\pm 1)^T$, with a pair of spins on the crossing diagonal. This spin bond has the shortest possible distance to a configuration of two holes. As doping is increased, the corresponding connected four-point spin-charge correlator evolves from uncorrelated at half-filling, to antiferromagnetic with a peak at a doping of $\delta = 30\%$ \cite{Koepsell2020_FL}. Beyond its peak, the correlator is quantitatively captured by Fermi-liquid correlations.
For this correlation function, considering the connected part means that -- as all lower order correlations are subtracted -- one directly detects correlations which are linked to the presence of the holes as a pair, since the effect of simply adding two independent single holes is removed. Whether this experimental observation may be a precursor to true binding of holes at lower temperatures remains an open question.

\subsubsection{String patterns}
\begin{figure}
\centering
  \includegraphics[width=0.95\linewidth]{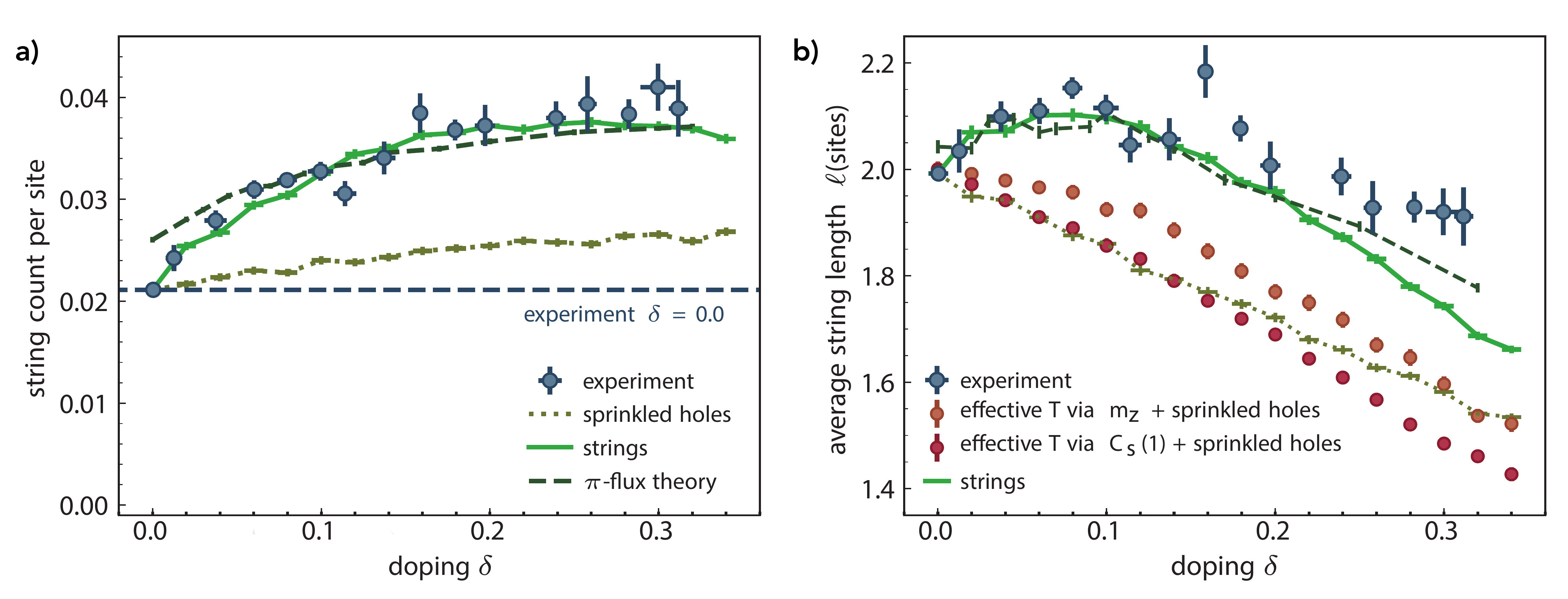}
\caption{\textbf{String patterns.} a) String pattern count and b) average string pattern length as a function of doping, extracted from experimental data (blue symbols) and string theory (green), $\pi$-flux theory (dark green, dashed), and sprinkled holes (brown, dashed-dotted). 
a) The background signal of the experiment at half-filling is marked by a gray dashed line. For small dopings, the string pattern count increases approximately linear. At around 15\%, the count starts to saturate. b) In the same doping regime, the average string pattern length is approximately constant. For dopings $\delta \geq 20\%$, the average length starts to decrease. The experimental and theoretical results for the average length are also compared to experimental data at a finite temperature, where the temperature is chosen at each doping level to best fit the average value of the staggered magnetization (orange) or the nearest neighbor spin correlator (red) with the corresponding number of holes added by hand.
Figures taken from \cite{Chiu2019Science,Christie_extra_Plot}.}
\label{fig:stringPatterns}
\end{figure}

Motivated by the geometric string theory briefly discussed in section \ref{secMagneticPolaron}, string patterns were directly probed in snapshots taken with a quantum gas microscope in Ref.~\cite{Chiu2019Science}. 
To that end, a pattern recognition algorithm for strings was designed, which identifies deviations from a perfect checkerboard pattern acting as a reference state. In the experimental setup of Ref.~\cite{Chiu2019Science}, doublons, holes and one of the two spin states are imaged as empty sites, such that a string pattern is identified as any non-branching deviation from the checkerboard pattern with an empty site -- consistent with having a hole -- on either of its two ends. 

Comparing the string patterns extracted from experimental snapshots at finite doping to the baseline at half-filling, shows a substantial increase of the number of detected string patterns with doping up to $\delta \approx 15\%$, see Fig.~\ref{fig:stringPatterns}a). This observation is expected from a picture of independent extended string objects that do not interact with each other, where for each dopant an additional string appears, as shown by the comparison of the experimental data to theoretical predictions (strings) in Fig.~\ref{fig:stringPatterns}a).
For larger doping values, the string pattern count saturates. This can be understood from the fact that the AFM background is at this point significantly scrambled, such that individual string patterns cannot be resolved anymore in the pattern recognition algorithm. 

The average length of string patterns was found to be constant up to the same doping value of $\delta \approx 15\%$, and then decreases with further increasing doping \cite{Chiu2019Science}, Fig.~\ref{fig:stringPatterns}b). The picture of non-interacting strings, predicting a constant string length, is not valid anymore once the strings start to overlap substantially, such that a deviation from constant string length and linear increasing string pattern count with increasing doping is expected. Notably the doping of $15-20\%$, at which string patterns start to show a deviation from the picture of independent string objects, coincides with the doping regime of $\delta \approx 20\%$ at which the magnetic polaron regime observed through spin-charge correlations in Ref.~\cite{Koepsell2020_FL} ends. 

The geometric string theory gives a prescription to relate finite doping to the parent AFM. Based on experimental data at half-filling, it is therefore possible to generate a new set of 'geometric string theory snapshots', where the strings are included by introducing and moving holes and displacing spins along their path by hand. Besides geometric string theory, the experimental results are compared to a doped RVB $\pi$-flux theory, and \emph{sprinkled holes}, where a number of holes corresponding to the doping under consideration are placed into experimental images at half-filling by hand but not moved around as in the case of geometric string theory. This data thus provides an estimate as to how the string pattern count changes simply due to an increasing number of empty sites in the system. The comparison to experimental data shows that sprinkled holes do neither account for the increase in the number of string patterns found, nor for the constant average string length as a function of doping. The two other theories, however, yield excellent agreement. Notably, the geometric string theory does not have any free fitting parameter.  

In Fig.~\ref{fig:stringPatterns} b) the average string pattern length is also compared to results from data at higher temperatures which was chosen to have the \emph{same} nearest-neighbor correlations $C_S(1)$ or mean staggered magnetization $|m_Z|$ as the actual experimental results. These curves serve for comparison and demonstrate that observed patterns contain additional information which is not captured by more traditional observables quantifying antiferromagnetic correlations in the system.

\subsection{Novel analysis: Machine Learning}

Machine learning techniques have emerged as a valuable tool to analyze large data sets in a variety of scientific disciplines in recent years.  In the context of the Fermi-Hubbard model,  machine learning has first been employed as an unbiased way to compare experimental data to microscopic theories \cite{Bohrdt2019_ML}. 
As discussed in section \ref{higherOrderCorrs},  a large number of different correlation functions can be constructed.  
By using an artificial neural network for this task, one avoids choosing a specific correlator based on a given physical picture and instead lets the network make the decision which observable is best suited to distinguish between different theories. 
In order to analyze snapshots from quantum gas microscopes,  convolutional neural networks (CNNs) are particularly suited, since they constitute a well established network architecture for the analysis of two-dimensional images. Other applications of CNNs \cite{Khatami2020}, or extensions thereof \cite{Miles2020}, to the Fermi-Hubbard model focused on the question of interpretability of the results. Aside from the Hubbard model, machine learning techniques have also been applied to time-of-flight data in cold atom experiments \cite{Rem2018,Kaeming2021}.

The main approach of the following two subsections can be summarized as follows: a neural network is first trained to distinguish three classes of data: two competing theories,  and experimental (or exact numerical) data.  If the experimental data is well described by either one or both theories,  this training will yield a comparably low accuracy.  Next,  a neural network is trained to distinguish the two theories.  The performance of the neural network indicates how similar the snapshots generated from the two theories are.  Once the training process is finished,  experimental data is used as an input to the network, which is forced to assign one of the two theory labels to each experimental snapshot.  The resulting classification of experimental data thus shows which theory resembles the experiment closer. In a final step, the network architecture can be modified to \emph{interpret} the results. This allows to determine, in physical terms, how the data sets can best be distinguished.

\subsubsection{Case study: Doped 1D spin chains}\label{secML1D}

Here we consider the one-dimensional $t-J$ Hamiltonian, 

\begin{equation}\label{eq: t-J}
    \H = \PG \left[-t \sum_{i, \sigma} \l \cd_{i, \sigma} \c_{i+1, \sigma} + \hc \r + J \sum_{i} \l \S_{i} \cdot \S_{i+1} - \frac{\n_{i} \n_{i+1}}{4} \r \right] \PG,
\end{equation}

for which we generate groundstate snapshots for different dopings using three different approaches:
\begin{enumerate}
    \item[(i)] DMRG: Exact numerical data, where  we generate snapshots using DMRG with different values for $t/J = 1.0, 3.0$, and $5.0$. 
    \item[(ii)]  MF: Metropolis Monte Carlo sampling using a Gutzwiller projected mean-field approach \cite{Baskaran1987}:  $ \ket \psi = \PG \prod_{\sigma} \l \prod_{k} \cd_{k, \sigma} \ket 0 \r$. 
    We describe the system at a given filling as two independent Fermi seas of spin up and spin down spinons. During Metropolis Monte Carlo sampling of the Fock space snapshots, a Gutzwiller projection is directly applied to remove all doubly occupied sites in real space. This Gutzwiller projected mean-field approach is an accurate approximation of the ground state wave function of a Heisenberg chain at half-filling \cite{Gros1989}.
    \item [(iii)] SQ: A squeezed space approach \cite{Ogata1990,Kruis2004}: here   we factorize the wave function and describe the holes and the spins separately. Snapshots for the hole configuration are obtained by sampling the holes as non-interacting and spinless fermions on a chain. The spin configurations are sampled using the Gutzwiller projected approach for a Heisenberg chain of reduced length at half filling.
\end{enumerate}

As the snapshots are image-like data, convolutional neural networks are best suited for their classification.
Neural networks with a single convolutional layer and varying kernel sizes yield very good performances despite the simple architecture. The networks are either trained on the two classes of Monte Carlo data, i.e. class (ii) and (iii), or on all three classes with $t/J=1.0,~3.0~\text{or}~5.0$ for the DMRG data.

\begin{figure}[t]
  \centering
  \includegraphics[width=0.99\linewidth]{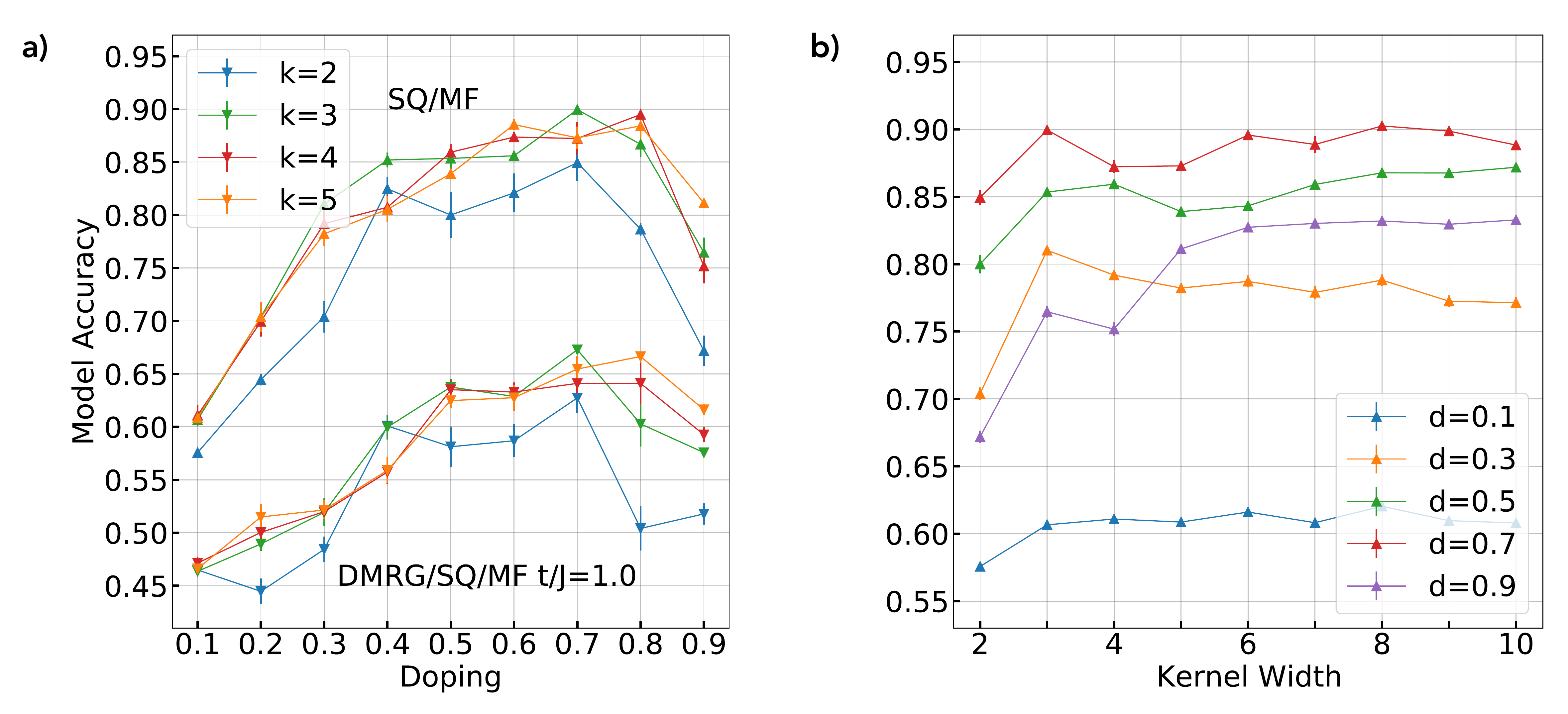}
  \caption{\textbf{Network performance in the classification of 1D $t-J$ model data.}
    a) The mean accuracy of networks trained on the two classes of Monte Carlo data (upper lines) and on all three classes (lower lines) is plotted as a function of the doping for different kernel sizes $k$. For the same kernel sizes there is a difference in accuracy of about $10 \%$ to $20 \%$ between the two cases. Therefore, the networks cannot distinguish the DMRG, SQ and MF classes as accurately as just the two Monte Carlo classes.
    b) The accuracy of the networks is plotted as a function of kernel size for different values of the doping $\delta$. Using kernels with two to three sites is sufficient to distinguish the squeezed space and Gutzwiller projected mean-field snapshots.
    }
    \label{fig:ACC}
\end{figure}

When training the network on all three sets of data the performance of the network is comparably low as seen in Fig.~\ref{fig:ACC} a). Even when considering more convolutional filters, the classification does not improve significantly. This indicates that at least two classes have a significant overlap in the features relevant for the networks. The networks trained on the two classes of Monte Carlo snapshots perform better which indicates that one of the theories could show similarities to the DMRG data. The performance in Fig.~\ref{fig:ACC} a) shows a maximum for a doping of around $0.7$. With decreasing doping, as well as for a doping larger than $0.8$, the performance of the network is worse. The simple networks yield comparably high accuracy and the increase of the kernel size does not result in a major improvement above three sites, see Fig.~\ref{fig:ACC} b). 

\begin{figure}
  \centering
    \includegraphics[width=0.9\linewidth]{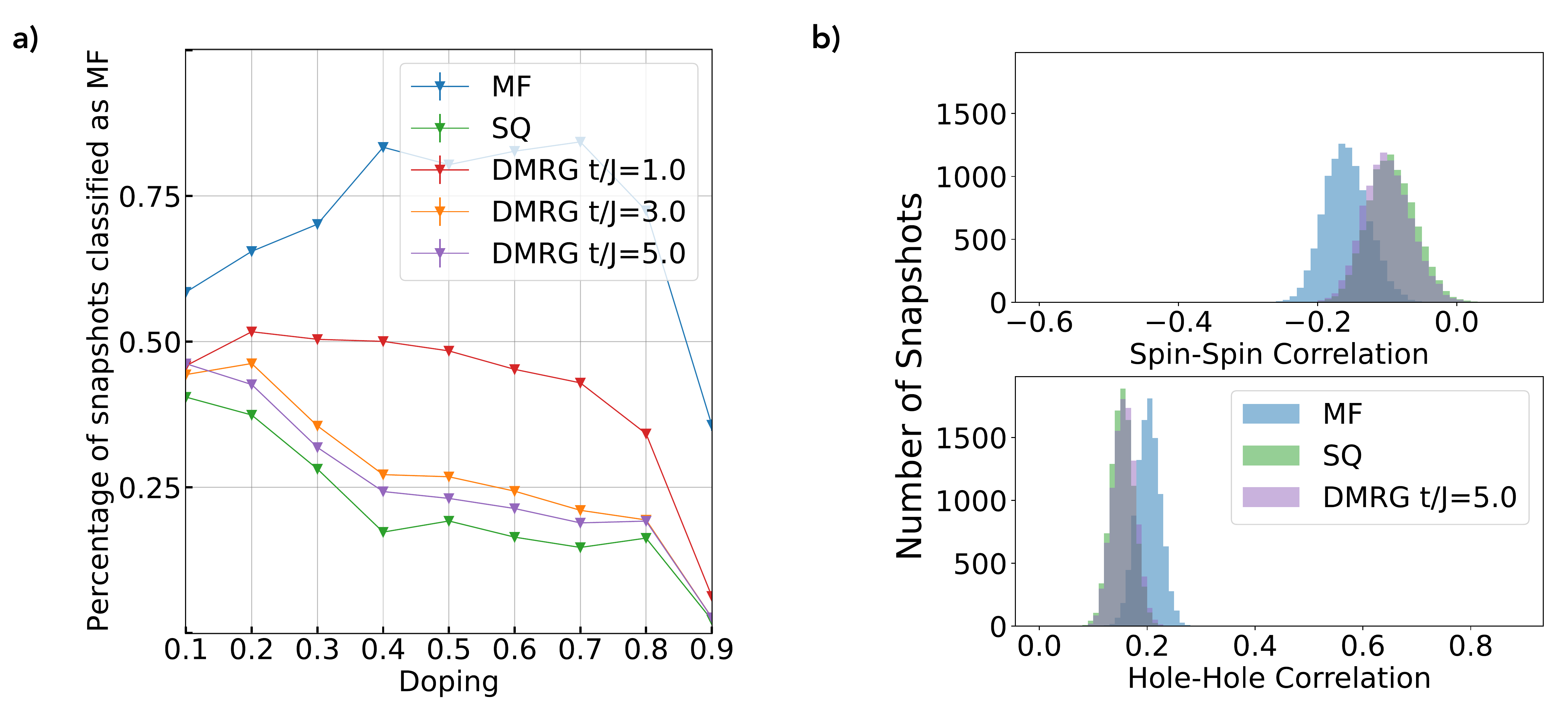}
  \caption{\textbf{Classification of DMRG data and comparison to theory.}
  a) The average percentage of snapshots of each class classified as MF is plotted. With increasing doping the DMRG data for larger values for $t/J$ are more similar to the SQ snapshots according to the networks classification. For $t/J=1.0$ neither of the two Monte Carlo approaches is favoured.  b) For each snapshot, the average values of the nearest neighbor spin-spin ($4 \langle \hat{S}^z_j \hat{S}^z_{j+1} \rangle$) and hole-hole ($\langle \hat{n}^{\rm h}_{j} \hat{n}^{\rm h}_{j+1} \rangle$) correlations are calculated, at doping $\delta=0.5$. The plot shows a histogram over all snapshots in each sample for the three different classes. A clear separation in the distributions between MF and the other two classes is visible, while SQ and DMRG overlap strongly.  All three figures indicate that the squeezed space approach is a better approximation for the DMRG data than the Gutzwiller projected mean-field approach.}
  \label{fig:CNN_1D_2}
\end{figure}

Due to the network architecture with a dense layer as output layer, we cannot simply read off a correlation function. However, we are able to get an idea of what correlation functions can be of interest. When analyzing the networks one finds similar patterns regardless of the kernel size and doping and, although the concrete weights and features may differ from one network to the other especially with varying kernel size, their interpretation remains similar: For kernel size equal to $2$ the convolutional layer has a finite output for a spin and a hole on neighbouring sites and two spins on neighbouring sites. For larger kernels more features may arise, as there are more possible combinations of spins and holes beneficial for classification. However, larger kernels do not yield a significant increase in performance and we arrive at the conclusion that possible correlation functions for distinguishing squeezed space and projected mean-field snapshots are the spin-spin and the hole-hole correlation.

The performance of the neural network significantly increases when distinguishing only the two Monte Carlo approaches as compared to the case where it is trained on all three classes of snapshots. Therefore, we conclude that the DMRG snapshots are rather similar to one of the two Monte Carlo approaches. In order to determine which of the two approaches captures the exact numerical results best, we classify the DMRG snapshots using the network trained on the two Monte Carlo classes. 
While for $t/J=1.0$ the two approaches capture the features of DMRG equally well, for the two larger values of $t/J$ and with increasing doping more snapshots are classified as squeezed space, see Fig.~\ref{fig:CNN_1D_2} a). 
When looking at the distribution of the mean value of the correlation functions per snapshot for each class of data, we see that the overlap for squeezed space and mean-field snapshots is small compared to the overlap for squeezed space and DMRG snapshots, see Fig.~\ref{fig:CNN_1D_2} b). Thus, using the spin-spin and hole-hole correlation functions we can reproduce a similar classification as with the neural network.

\subsubsection{Doped 2D Hubbard model: Classification}

\begin{figure}
\centering
  \includegraphics[width=0.99\linewidth]{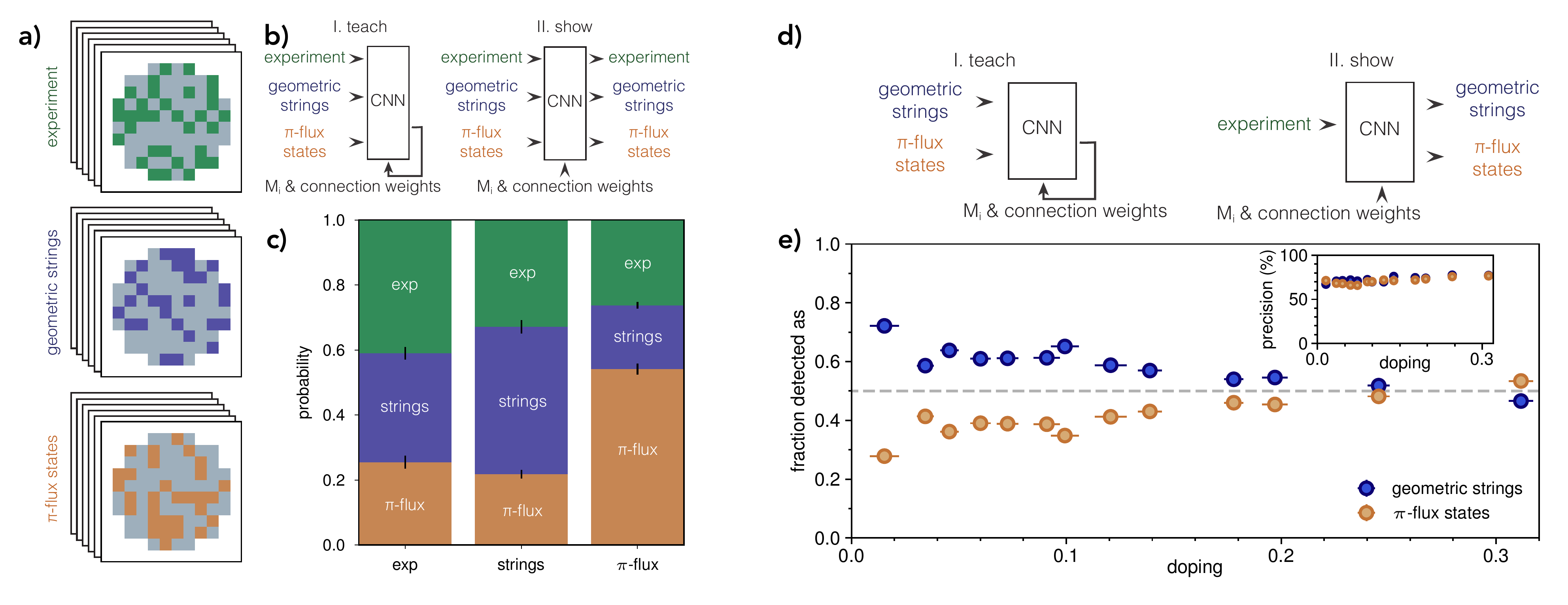}
\caption{\textbf{Machine learning analysis} of snapshots in the doped two-dimensional Fermi-Hubbard model. a) Snapshots from experiment and two different theories are classified using a neural network architecture b). c) The resulting classification of a test set at $9\%$ hole doping shows which data can be efficiently distinguished by the network. d) In a different approach, the neural network is trained to distinguish only between the two theories. After training, experimental images are then shown to the neural network. e) The resulting classification of experimental data as either of the two theories as a function of doping. Figure taken from \cite{Bohrdt2019_ML}.}
\label{fig:ML2D}
\end{figure}

Now we return to the 2D doped Hubbard model and experimental data. In \cite{Bohrdt2019_ML},  the $\pi$-flux and geometric string theory introduced in section \ref{higherOrderCorrs} were compared to experimental data of the two-dimensional Fermi-Hubbard model in a doping range of up to $35\%$.  As shown in \cite{Chiu2019Science}, these theories both lead to good agreement in terms of a range of observables, as discussed in sections \ref{ConventionalProbes} and \ref{higherOrderCorrs}, and thus provide a good starting point for further machine learning analysis. 

The experimental images used in \cite{Bohrdt2019_ML} contain information about only one spin species.  Before comparing theoretical images to experimental data,  the second spin species as well as doubly occupied sites were therefore converted to empty sites in the theory datasets as well.

Training a neural network to classify both theory as well as the experiment at $9\%$ doping leads to a poor accuracy of $47\%$. This indicates that the classification of experimental snapshots as one of the two theories is in principle meaningful, since otherwise the CNN would be able to clearly distinguish experimental data from both theories with a high accuracy.  As can be seen in Fig.~\ref{fig:ML2D} c), the main source of confusion for the CNN at $9\%$ doping is the similarity between  experiment and geometric string theory, while a differentiation of the $\pi$-flux theory is more successful. 

Next,  a neural network was trained to distinguish only between the two theories at a fixed doping value, yielding a significantly higher precision.  Subsequently,  the experimental data was used as an input to the neural network, see Fig.~\ref{fig:ML2D} d). The resulting classification reveals which theory matches the experiment more closely.  For small dopings, up to $\delta = 15\%$, the network classifies more experimental snapshots as geometric string theory, see Fig.~\ref{fig:ML2D} e), even though more conventional spin observables coincide equally well between the two different theories. For larger dopings, the experimental data cannot be unambiguously classified.

\subsubsection{Doped 2D Hubbard model: Interpretability}

An important step in gaining insights from a machine learning analysis lies in its interpretability: understanding what patterns or observables the network uses to make its decision. For the applications considered here, a first step is to simply look at the filters used by the convolutional neural network. This was done in Ref.~\cite{Khatami2020}, where a CNN with a single filter was trained to distinguish snapshots from a quantum gas microscope at high and low temperatures, $T \gg t$ and $T = 0.35 t$. At half filling, analyzing the filter showed directly that the network searches for signatures of antiferromagnetic correlations. Upon increasing the doping to $18\%$, the network still identified the lower temperature snapshots through antiferromagnetic correlations, but in this case it only considered short-ranged correlations.
However, in most cases the differences between the data sets under consideration will be more complex and thus harder to extract directly from the CNN. For example, in our new results presented in section \ref{secML1D}, the performance of the neural network as a function of the filter size provided additional insights. 

In Ref.~\cite{Miles2020}, a new network architecture, the correlator convolutional neural network (CCNN), was introduced. In this specific type of neural network, the standard non-linearity after the convolution with a filter is replaced by a order-by-order expansion of the convolution, which can be expressed in terms of the corresponding order of correlation functions. The order up to which the correlations are used by the CCNN is a hyperparameter specific to this network architecture.
In Ref.~\cite{Miles2020}, the network was trained to distinguish geometric string theory from an RVB state at $9\%$ doping. In this analysis the input contains three channels: spin up, spin down, and holes; the CCNN then allows for arbitrary correlations between spin and charge up to the order of choice. 

By tuning the allowed order of correlations, one can therefore investigate up to which order the correlations are important for the classification task at hand. If one, for example, compares two theories, the CCNN allows to directly determine up to which order of correlations the theories are indistinguishable. Without a CCNN, this task would require extensive analysis of countless types of higher-order correlations, including for example those discussed in Sec.~\ref{higherOrderCorrs}. Moreover, when comparing theory to experiment, one can make use of this network architecture to investigate up to which order the theory captures all experimental features.
Using a regularization path analysis allows furthermore to extract which correlation functions are most important in making the decision. In Ref.~\cite{Miles2020}, the dominant correlation function characterizing geometric string theory was found to be a four-point correlator remarkably close to the \emph{magnetic polaron correlator} discussed in section \ref{secMagneticPolaron}.

\section{New directions: mixed-dimensional bilayer systems}\label{secPairing}
In this section we focus on discussing a promising new direction for future experiments with ultracold atoms. The main idea is to propose simpler variants of the two-dimensional Hubbard model which should satisfy at least the following properties: (i) they have interesting and theoretically accessible limits, where close competition between different phases is avoided; (ii) they are continuously connected, by tuning just a few parameters, to the standard 2D Hubbard model. The hope is that together these properties allow to obtain a simple description deep inside some prototypical phase of matter which is also realized in the 2D Hubbard model, where stronger quantum fluctuations and competing ordered phases make its description and identification challenging. 

As an example, we mention the long-standing question about the pairing mechanism in high-temperature superconductors. Experimentally it has been established that pairing in the under-doped regime ($\delta \lesssim 20\%$) is not of BCS-type: phase fluctuations of the order parameter are believed to destroy superconductivity and the emergence of a pseudogap signals the possible existence of preformed, uncondensed cooper pairs above the critical temperature \cite{Emery1995,Keimer2015}. This phenomenology is closely related to that of the BEC-to-BCS crossover, but it remains difficult to draw a direct analogy to this phenomenon in the repulsive Hubbard model: At high doping ($\delta \gtrsim 20\%$) the system can be described by a weakly coupled BCS theory of superconductivity and the observed $d_{x^2-y^2}$ pairing symmetry is correctly obtained by a BCS analysis of the $t-J$ model~\cite{Coleman2015}. However, at low doping a strong pairing mechanism of holes into molecular pairs, which would be required deep in a BEC regime, remains elusive. Theoretically it has become clear that pairing is possible at low doping in the $t-J$ model \cite{White1997,Leung2002,Vidmar2013,Blomquist2021}, but it is not very robust to parameter changes. Recently broad numerical analysis by the Simons Collaboration has produced mounting evidence that the clean 2D Hubbard model may not support a superconducting ground state in the relevant regime, although the addition of next-nearest neighbor hopping terms $t'$ may support it \cite{Qin2019}.

Our main motivation here is to identify an experimentally accessible model, connected to the 2D Hubbard model, which supports a strongly paired phase in a BEC regime \footnote{We want to achieve this without introducing microscopically attractive interactions. These would easily yield pairing, but they would also completely change the remaining phenomenology of the model.}. Specifically, we propose to study a mixed-dimensional (mixD) bilayer $t-J$ model described by the following Hamiltonian, 
\begin{align}
\begin{split}
    \H = -\tpar \sum_{\substack{\ij, \\ \sigma, \mu}} \PG \left( \cd_{\vec{i},\sigma,\mu}\c_{\vec{j}, \sigma, \mu} + \mathrm{H.c.} \right)\PG &+ \Jpar\sum_{\ij,\mu} \left( \S_{\vec{i},\mu} \cdot \S_{\vec{j},\mu} - \frac{1}{4} \sum_{\alpha,\beta}\n_{\vec{i},\alpha,\mu}\n_{\vec{j},\beta,\mu} \right) \\ 
    &+ \Jperp\sum_{\vec{j}} \left( \S_{\vec{j},1} \cdot \S_{\vec{j},2} - \frac{1}{4} \sum_{\alpha,\beta}\n_{\vec{j},\alpha,1}\n_{\vec{j},\beta,2} \right).
    \label{eq:H_bilayer}
\end{split}
\end{align}
The first line corresponds to the usual $t-J$ model, realized independently in two layers; we introduced the layer index $\mu=1,2$. The second line introduces antiferromagnetic spin-exchange couplings $\propto J_\perp$ between the layers, see Fig.~\ref{fig1Bilayer}. Most importantly, inter-layer tunneling is completely suppressed. This last property renders our model mixed-dimensional \cite{Grusdt2018mixD}: While the spin channel is fully coupled between the two layers, the motion of charge excitations is restricted to their respective planes. From a statistical physics perspective, our model is two-dimensional, e.g. the Hohenberg-Mermin-Wagner theorem applies, preventing true long-range order at any non-zero temperature. However, at low temperatures, we can have a correlation length that becomes as large as the size of the experimental system. We will discuss possible experimental realizations of Eq.~\eqref{eq:H_bilayer} in the next subsection.

As we mentioned in the introduction, see Sec.~\ref{SecIntro} and Fig.~\ref{fig1Bilayer}, the mixD $t-J$ model in Eq.~\eqref{eq:H_bilayer} features a strong pairing mechanism for holes at low doping when $J_\perp$ dominates over $J_\parallel$. This implies that ground states deep in the BEC phase can be realized in this regime. As we will discuss below, other interesting states of spinons and chargons can also be expected in this model. To derive the ground states of the model we will distinguish between two limits of the couplings, defined respectively as
\begin{flalign}
    J_\perp \gg t_\parallel \geq J_\parallel, \qquad 0 \leq \delta \leq 1 \qquad \text{tight-binding}, \label{eq:TightBndgmDbiLay}\\
    t_\parallel \gg J_\perp, J_\parallel, \qquad \delta \ll 1  \qquad \text{strong ~coupling}, \label{eq:StrongCplgmDbiLay} \\
    t_\parallel \gg J_\perp, J_\parallel, \qquad  \delta \lesssim 1 \quad \qquad \text{weak ~coupling}. \label{eq:WeakCplgmDbiLay}
\end{flalign}

Before we continue our analysis of the mixD bilayer model, we briefly mention that studies of conventional bilayer setups \emph{with} tunneling $t_\perp$ between the layers also constitute another promising direction. A particular motivation to study such systems would be to directly emulate copper-oxide multi-layers, which are found in many high-temperature superconductors \cite{Mazin2010,Mukuda2012}. In this context it should be noted that a bilayer system differs significantly from a full three-dimensional (3D) system, since the formation of singlets between the layers is possible, whereas in a 3D system, instead the antiferromagnetic correlations in all directions are enhanced and long-range order can be established. On the other hand, nearest-neighbor correlations decrease as the coordination number is increased, as observed experimentally for different geometries in \cite{Greif2015} and as a function of the coupling in the $z$-direction for a three-dimensional system in \cite{Imriska2014} at a fixed entropy per particle. Numerically, it has been shown that at a fixed entropy per particle, the temperature increases in the crossover from two to three dimensions as the coupling $t_\perp/t$ in the third direction is increased from $0$ to $1$ \cite{Padilla2020}. Finally we note that $t_\perp$-terms could also be continuously re-introduced in the mixD Hamiltonian Eq.~\eqref{eq:H_bilayer} to connect to these conventional bilayer systems.

\subsection{Possible experimental realization}
Experimentally, the mixD bilayer $t-J$ model can be realized as an effective description of a bilayer Hubbard model with a tilt $\Delta$, corresponding to the Hamiltonian
\begin{equation}
    \H = -\tpar \sum_{\substack{\ij, \\ \sigma, \mu}} \left( \cd_{\vec{i},\sigma,\mu}\c_{\vec{j}, \sigma, \mu} + \mathrm{H.c.} \right) - t_\perp \sum_{\vec{j}, \sigma} \left( \cd_{\vec{j},\sigma,2}\c_{\vec{j}, \sigma, 1} + \mathrm{H.c.} \right) + U 
    \sum_{\vec{j},\mu} \hat{n}_{\vec{j},\uparrow,\mu} \hat{n}_{\vec{j},\downarrow,\mu} + \frac{\Delta}{2} \sum_{\vec{j},\sigma,\mu} (-1)^\mu \hat{n}_{\vec{j},\sigma,\mu}.
    \label{eq:HmixDHubbard}
\end{equation}
A standard Schrieffer-Wolff perturbation theory leads to the effective Hamiltonian in Eq.~\eqref{eq:H_bilayer} if next-nearest neighbor hopping terms of the holes (three-site terms) in the plane $\propto J_\parallel$ are neglected, as in the usual single-layer $t-J$ model \cite{Auerbach1998}. Assuming $|U \pm \Delta| \gg t_\perp$ and $U \gg t_\parallel$, the expressions for the couplings are
\begin{equation}
    J_\perp = \frac{2 t_\perp^2}{U + \Delta} + \frac{2 t_\perp^2}{U - \Delta}, \qquad \qquad J_\parallel = \frac{4 t_\parallel^2}{U}. 
    \label{eq:ParsMixDtJ}
\end{equation}
We consider a regime where $U > \Delta$, which leads to AFM couplings $J_\perp > 0$. Note however, that one could also start from a related spin-$1/2$ Bose-Hubbard model, where $U < \Delta$ would also lead to AFM couplings $J_\perp > 0$ between the layers \cite{Trotzky2008}. Strong tilts as proposed above have already successfully been used to suppress tunneling, including in the context of quantum magnetism, see e.g. \cite{Simon2011}.

Recently, several experiments have realized bilayer systems, where a pair of two-dimensional Fermi-Hubbard planes is coupled \cite{Koepsell2020,Hartke2020,Gall2021}. Such setups provide a natural starting point to realize the Hamiltonian in Eq.~\eqref{eq:HmixDHubbard}, where the additional gradient $\Delta$ can be added optically or magnetically.

The mixD bilayer $t_\parallel-J_\perp-J_\parallel$ model from Eq.~\eqref{eq:H_bilayer} has a global $U(1) \times U(1)$ symmetry corresponding to the conservation of total particle number individually in each layer. This symmetry characterizes the meta-stable excited states of the tilted Hubbard model, where any current between the layers is strongly suppressed by the gradient $\Delta \gg t_\perp$. For $U>\Delta$ the overall ground state of the Hubbard model corresponds to a situation where all holes are doped into the energetically higher layer, but the system still forms a Mott insulator at half filling. Hence, realizing states at finite doping requires an adequate preparation sequence experimentally, which ensures the correct particle number per layer. This can be achieved by making use of the individual tunability of $\Delta_\perp$ and $t_\perp$. The prepared excited states will be meta-stable; a description by the effective Hamiltonian in Eq.~\eqref{eq:H_bilayer} breaks down at times which are exponentially large in the ratio of $\Delta / t_\perp$ \cite{Strohmaier2010}.

Note that all three coupling regimes of the mixD $t-J$ model defined in Eqs.~\eqref{eq:TightBndgmDbiLay}, \eqref{eq:StrongCplgmDbiLay}, \eqref{eq:WeakCplgmDbiLay} involve $U \gg t_\parallel, t_\perp$ such that the underlying Hubbard model is always strongly coupled. We also emphasize that $t_\parallel \geq J_\parallel$ should always be assumed because otherwise the mapping to an effective $t-J$ model is not valid. On the other hand, $J_\perp / t_\parallel$ is freely tunable through the gradient $\Delta$. Using realistic experimental numbers which have already been realized, we expect that a regime $0.3 < t_\parallel / J_\perp < 3$ can easily be accessed in existing setups.

Finally we mention that lattice modulation at frequency $\omega=\Delta$ could be used to restore tunneling $t_\perp$ between the layers, with a strength tunable by the amplitude of the modulation and independent of $J_\perp$.

\subsection{Phase transition at zero doping}
\label{sec:BilayerTransition}
At zero doping, the mixD bilayer system is equivalent to a conventional bilayer system without the gradient $\Delta$, except for the renormalization of the superexchange coupling $J_\perp$ by $\Delta$. This bilayer spin system exhibits a zero-temperature phase transition from a disordered phase for large $J_\perp / J_\parallel$, corresponding to a fully symmetric valence bond solid (VBS) of rung-singlets, to a long-range ordered bilayer AFM when $J_\perp \ll J_\parallel$ \cite{Sandvik1994}. At non-zero temperatures $T>0$, the Hohenberg-Mermin-Wagner theorem precludes an ordered state in two dimensions. However the correlation length increases rapidly with inverse temperature, see Eq.~\eqref{eqxiTAFM}, which allows to see signatures of the transition for $T>0$ as well. 

\emph{Variational analysis.--} A simple theoretical understanding of this transition at $T=0$ can be gained from a variational description. To this end, we start by defining a local basis of singlets and three triplet states,
\begin{equation}
    \ket{s} = \frac{1}{\sqrt{2}} \l \ket{\! \uparrow_1 \downarrow_2} - \ket{\! \downarrow_1 \uparrow_2} \r, \quad \ket{t_+} = \ket{\! \uparrow_1 \uparrow_2}, \quad \ket{t_0} = \frac{1}{\sqrt{2}} \l \ket{\! \uparrow_1 \downarrow_2} + \ket{\! \downarrow_1 \uparrow_2} \r, \quad \ket{t_-} = \ket{\! \downarrow_1 \downarrow_2},
\end{equation}
where $\ket{\sigma^a_1 \sigma^b_2}$ denotes the state with spin $\sigma^a$ ($\sigma^b$) on layer $\mu=1$ ($\mu=2$). When $J_\perp \gg J_\parallel$, the triplet states can be discarded and the VBS state is fully captured by writing $\ket{\Psi_0^\perp} = \prod_{\vec{j}} \ket{s}_{\vec{j}}$. On the other hand, when $J_\perp \ll J_\parallel$ the classical N\'eel state pointing along $S^z$ can be used, $\ket{\Psi_0^\parallel} = \prod_{\vec{j}} \l \ket{s}_{\vec{j}} + (-1)^{j_x+j_y} \ket{t_0}_{\vec{j}} \r / \sqrt{2}$.

To capture the competition between the symmetric VBS phase and a symmetry-broken AFM phase, we will make an ansatz which allows for a N\'eel order parameter $\hat{\vec{\Omega}} = \sum_{\vec{j}} \sum_\mu (-1)^{j_x+j_y} (-1)^{\mu}  \hat{\vec{S}}_{\vec{j}}^\mu$ pointing in an arbitrary direction $\vec{n}$, with $\vec{n}^2 = 1$. For its construction it is useful to rotate the triplet states into another basis, which also forms a vector representation under spin rotations with the generator $\hat{\vec{S}}_{\vec{j}} = \sum_\mu \hat{\vec{S}}_{\vec{j}}^\mu$,
\begin{equation}
    \ket{t_z} = \ket{t_0}, \qquad \ket{t_x} = \frac{1}{\sqrt{2}} \l \ket{t_+} - \ket{t_-} \r, \qquad \ket{t_y} = \frac{i}{\sqrt{2}} \l \ket{t_+} + \ket{t_-} \r.
\end{equation}
As a trial state with variational parameters $\vec{n}$ and $\theta$, we choose:
\begin{equation}
    \ket{\Psi} = \prod_{\vec{j}} \ket{\psi_{\vec{j}}(\vec{n},\theta)}, \qquad \ket{\psi_{\vec{j}}(\vec{n},\theta)} = \cos \theta ~ \ket{s}_{\vec{j}} + (-1)^{j_x+j_y} \sin \theta  \sum_{a=x,y,z} n_a ~ \ket{t_a}_{\vec{j}}.
\end{equation}
Up to an overall constant, the evaluation of the variational energy of this state yields
\begin{equation}
    \langle \H \rangle = \l J_\perp - 4 J_\parallel \r \sin^2 \theta + 4 J_\parallel \sin^4 \theta,
    \label{eq:HvarMexHat}
\end{equation}
where $\H$ is the bilayer Heisenberg Hamiltonian, i.e. Eq.~\eqref{eq:H_bilayer} with $t_\parallel = 0$. Notably, the variational energy is independent of $\vec{n}$, reflecting the $SU(2)$ symmetry of the model. However, the variational state only depends on $\vec{n}$ when $\sin \theta \neq 0$. Minimizing Eq.~\eqref{eq:HvarMexHat} one recognizes the familiar Mexican hat potential, which only has a non-trivial solution $\sin \theta \neq 0$ if $J_\perp - 4 J_\parallel < 0$.

Summarizing, we find a ground state with broken $SU(2)$ symmetry for $J_\perp / J_\parallel < (J_\perp / J_\parallel)_c$. The variational calculation estimates the transition point to be at $(J_\perp / J_\parallel)_c \approx 4$. Quantum Monte Carlo calculations by Sandvik and Scalapino have pinpointed the  transition at $(J_\perp / J_\parallel)_c = 2.51(2)$ \cite{Sandvik1994}. 

\begin{figure}
\centering
  \includegraphics[width=0.95\linewidth]{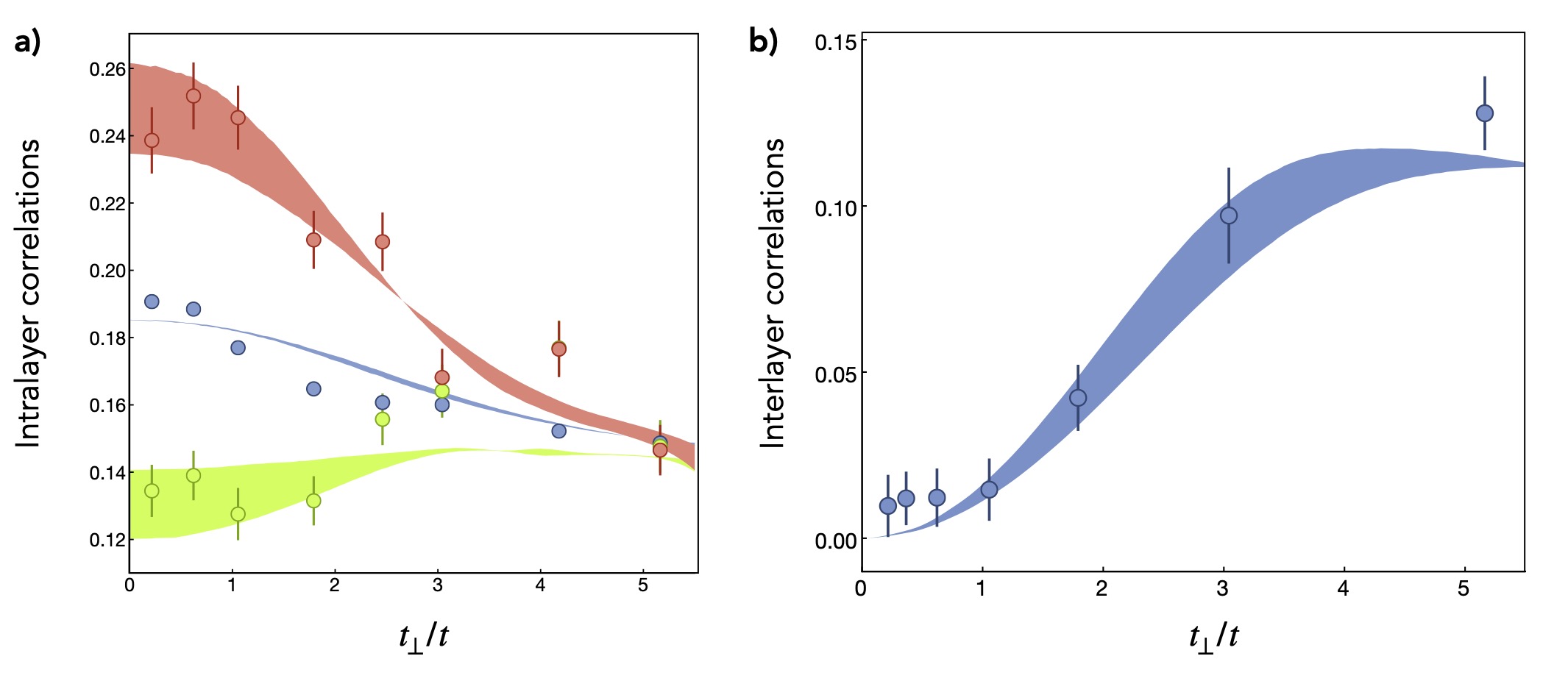}
\caption{\textbf{Cross-over from VBS to AFM} in a bilayer system upon tuning the interlayer hopping $t_\perp/t$. a) Intralayer correlations: The staggered structure factor (red) decays as the interlayer coupling is turned on. For comparison, the uniform structure factor at $\mathbf{q}=(0,0)$ (green) and the local magnetic moment $\langle (\hat{S}_i^z)^2\rangle - \langle \hat{S}_i^z\rangle^2$ (blue) are shown. b) Interlayer correlations: staggered spin correlations between the layers.
Shaded bands are DQMC calculations at $T/t=1.0-1.4$ and filling factor $n=0.4$. Figures extracted from \cite{Gall2021}.}
\label{fig:bilayer_Koehl}
\end{figure}

\emph{Experiments with ultracold atoms.--} The cross-over at finite temperatures from VBS to AFM has recently been observed using ultracold fermions in an optical lattice by the Bonn group \cite{Gall2021}. 
In this experiment, a homogeneous bilayer system, containing approximately $5600$ atoms per layer, was prepared by splitting a band-insulator into two layers using an optical superlattice in the vertical ($z$-) direction.  
The antiferromagnetic ordering within one layer was then measured using a Ramsey-type sequence as discussed in Sec.~\ref{secAFM}. The static structure factor at momentum $\mathbf{q} = (\pi,\pi)$ within the layers decays as the interlayer coupling $t_\perp/t$ increases, see Fig.~\ref{fig:bilayer_Koehl}a). 
The spin correlations along the $z$-direction were measured by merging the double-well into a single potential well. The corresponding ramp of the lattices was chosen such that the atoms end up in the vibrational ground state of the merged lattice only if they were in a spin-singlet configuration before. The occupation of the vibrational ground state, and thereby the probability for a spin-singlet along the vertical direction, was then detected using RF spectroscopy. As shown in Fig.~\ref{fig:bilayer_Koehl}b), the intralayer spin correlations increase with the intralayer hopping $t_\perp/t$.

\subsection{Tight-binding limit}
\label{secTightBndgLmt}
\begin{figure}
\centering
    \includegraphics[width=0.99\linewidth]{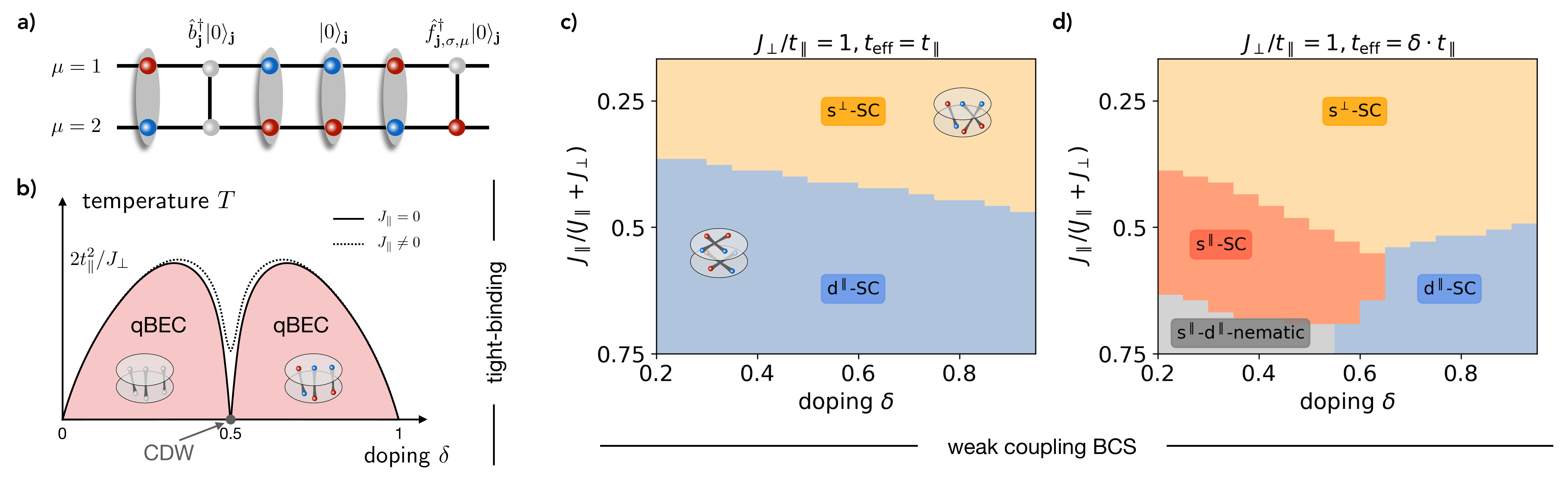}
  \caption{\textbf{Pairing in the mixed-dimensional bilayer system.}
      a) In the tight-binding limit, the mixD bilayer system can be described by a reduced local basis: Its vacuum $\ket{0}$ has rung singlets everywhere, and doped holes form local bosonic pairs $\bd_{\vec{j}} \ket{0}$. A charge imbalance between the two layers requires unpaired holes, which form spinon-chargon pairs $\fd_{\vec{j},\sigma,\mu}$ on a rung. b) The schematic phase diagram in the tight-binding limit, $J_\perp \gg t_\parallel, J_\parallel$, contains a quasi-condensate (qBEC) of pairs, separated by a BKT transition from the region with pre-formed pairs. Non-zero $J_\parallel \neq 0$ breaks the emergent $SU(2)$ symmetry at doping $\delta = 0.5$ down to $U(1)$. The plots in c) and d) show the mean-field BCS phase diagrams of the mixD $t-J$ bilayer model at weak coupling, $t_\parallel \gg J_\perp,J_\parallel$ and for high dopings, with intra-layer (inter-layer) interactions~$\Jpar$ ($\Jperp$) and intra-layer (inter-layer) hoppings~$t_{\parallel}$ ($t_\perp =0$)  at $T=0$. In c) we completely neglect any effect of Gutzwiller projections on the hopping term. In d) we approximate the effect of the Gutzwiller projector by working with an effective hopping~$\tpar^\mathrm{eff}=\delta\cdot\tpar$. }
  \label{fig:BilayerDiscussion}
\end{figure}

Now we consider the mixD bilayer model at non-zero doping, $\delta \neq 0$. We start in the conceptually simpler tight-binding limit where $J_\perp$ dominates, see Eq.~\eqref{eq:TightBndgmDbiLay}. In this case it is energetically favorable for the fermions to bind into inter-layer singlets along the rungs. Hence the low-energy physics can be described in a restricted local basis consisting of the following states,
\begin{equation}
    \ket{0}_{\vec{j}} = \ket{0}_{\vec{j},1} \otimes \ket{0}_{\vec{j},2} \qquad \ket{1}_{\vec{j}} \equiv \bd_{\vec{j}} \ket{0}_{\vec{j}} = \frac{1}{\sqrt{2}} \l \cd_{\vec{j},\uparrow,1} \cd_{\vec{j},\downarrow,2} - \cd_{\vec{j},\downarrow,1} \cd_{\vec{j},\uparrow,2} \r \ket{0}_{\vec{j}}.
\end{equation}
Here we introduced hard-core bosonic operators $\bd_{\vec{j}}$ creating the rung-singlet state. The remaining states in the local Hilbertspace of the mixD $t-J$ model, with only one fermion in either of the two layers, cost an excitation energy $J_\perp$ required to break up a singlet. They can be viewed as spinon-chargon bound states, and we write them as
\begin{equation}
    \fd_{\vec{j},\sigma,\mu} \ket{0} \equiv \cd_{\vec{j},\sigma,\mu} \ket{0}, \qquad \sum_{\sigma,\mu} \fd_{\vec{j},\sigma,\mu} \f_{\vec{j},\sigma,\mu} + \bd_{\vec{j}} \b_{\vec{j}} \leq 1,
\end{equation}
as summarized in Fig.~\ref{fig:BilayerDiscussion} a).

We start by discussing pairing in the low-doping regime, $\delta \ll 1$. To calculate the binding energy $E_{\rm bdg}$ of two holes into a pair occupying one rung, we calculate perturbatively in $t_\parallel$ the energy $E_0$ of the Mott insulating VBS state $\ket{1} = \prod_{\vec{j}} \bd_{\vec{j}} \ket{0}$, the energy $E_{\rm 1h}^\mu(\vec{k})$ of the one-hole doped spinon-chargon states with a mobile fermion of momentum $\vec{k}_1$ in layer $\mu$, $\fd_{\vec{k}_1,\sigma,\mu} \ket{0}$, and the energy $E_{\rm 2h}(\vec{k}_2)$ of a mobile pair $\b_{\vec{k}_2} \ket{1}$. Choosing the momenta $\vec{k}_{1,2}$ to minimize the kinetic energies resulting from effective nearest-neighbor tunnelings yields a binding energy
\begin{equation}
    E_{\rm bdg} = \l E_{\rm 2h} - E_0 \r - \l E_{\rm 1h}^1 + E_{\rm 1h}^2 - 2 E_0 \r = - J_\perp + z |t_\parallel| - \frac{5}{2} z \frac{t_\parallel^2}{J_\perp}
\end{equation}
where $z$ is the coordination number of the lattice in which sites $\vec{j}$ are defined. For example $z=4$ when two copies (labeled $\mu=1,2$) of a two-dimensional square lattice are stacked on top of each other, as illustrated in Fig.~\ref{fig1Bilayer}. For sufficiently small $z$, or sufficiently large $J_\perp$, we thus find strong binding, $E_{\rm bdg} < 0$, on an energy scale given by $J_\perp$. 

We emphasize the importance of working in mixed dimensions, where the inter-layer tunneling $t_\perp = 0$. Had we used the same states but included $t_\perp \neq 0$, as would be required in a \emph{standard} bilayer system, the one-hole spinon-chargon states would have each gained kinetic energy $- |t_\perp|$. To leading order the resulting binding energy $E_{\rm bdg} = 2 |t_\perp| - J_\perp + z |t_\parallel| > 0$ would always be positive since only $J_\perp \ll |t_\perp|$ can be realized starting from a strong-$U$ Hubbard model. Including quantum fluctuations stemming from higher-order corrections can still lead to $E_{\rm bdg}<0$ and a shallow bound state, for example in the two-leg ladder \cite{Noack1995,Balents1996,Dolfi2015,Zhu2014,Chen2018}, however with a binding energy significantly smaller in magnitude than $J_\perp$. This highlights the advantage of the mixD system, as well as the differences of the underlying pairing mechanisms.

The same perturbative analysis can be performed in the high-doping regime, $\delta \simeq 1$. Now the vacuum state is $\ket{0}$ and the mobile pair is described by $\bd_{\vec{k}} \ket{0}$. Again, the energetic cost $J_\perp$ to break up a singlet leads to a large binding energy  in the mixD bilayer setting when $J_\perp$ dominates over $t_\parallel$, namely $E_{\rm bdg} = - J_\perp + 2z |t_\parallel| (1 - |t_\parallel| / J_\perp) < 0$ where the second term reflects the difference in kinetic energy of the bound and unbound states. This regime realizes a BEC ground state of tightly-bound inter-layer pairs. Since the pairs carry zero angular momentum, we refer to the state as the $s^\perp$-BEC. As we show in Sec.~\ref{secWeakCplgBCS}, a corresponding BCS state with $s^\perp$-pairing symmetry exists when $t_\parallel$ dominates.

Finally we discuss the case of arbitrary doping, $0 \leq \delta \leq 1$. Neglecting the spinon-chargon states in the low-energy basis of pairs $\bd_{\vec{j}}$ allows us to derive an effective Hamiltonian which becomes
\begin{equation}
    \H_{\rm eff} = -t_{\rm eff} \sum_\ij \l \bd_{\vec{i}} \b_{\vec{j}} + \hc \r + V \sum_\ij \bd_{\vec{j}} \b_{\vec{j}} \bd_{\vec{i}} \b_{\vec{i}} + \mu_b \sum_{\vec{j}} \bd_{\vec{j}} \b_{\vec{j}}.
    \label{eqHhcbEff}
\end{equation}
It describes nearest-neighbor tunneling of the hard-core bosons $\b_{\vec{j}}$, with amplitude $t_{\rm eff} = 2 t_\parallel^2 / J_\perp$. Additionally, they experience a nearest-neighbor interaction $V = 4 t_\parallel^2 / J_\perp - J_\parallel / 2$, and the chemical potential is $\mu_b = - J_\perp - 2 z t_\parallel^2 / J_\perp$. By mapping the hard-core bosons to spins, with $\hat{J}^z_{\vec{j}} = \bd_{\vec{j}} \b_{\vec{j}} - 1/2$ -- not to be confused with the physical spins $\hat{\vec{S}}_{\vec{j}}$ in the bilayer Eq.~\eqref{eqHhcbEff} maps onto an XXZ model,
\begin{equation}
    \H_{\rm eff} = - K \sum_\ij \l \hat{J}^x_{\vec{i}} \hat{J}^x_{\vec{j}} + \hat{J}^y_{\vec{i}} \hat{J}^y_{\vec{j}} - \Delta \hat{J}^z_{\vec{i}} \hat{J}^z_{\vec{j}} \r, \qquad K = 4 \frac{t_\parallel^2}{J_\perp}, \qquad \Delta = 1 - \frac{J_\parallel}{2 K}.
    \label{eqXXZ}
\end{equation}
Here we neglected an overall energy offset $- J_\perp N_b - \frac{1}{2} (K + J_\parallel /2 ) L^d$, where $L^d$ is the volume and $N_b = L^d \delta$ the number of bosons. Assuming a bi-partite lattice, we can further introduce a unitary basis transformation on one sublattice which replaces $\hat{J}_{\vec{i}}^x \to (-1)^{\vec{i}} \hat{J}_{\vec{i}}^x$, $\hat{J}_{\vec{i}}^y \to (-1)^{\vec{i}} \hat{J}_{\vec{i}}^y$ and keeps $\hat{J}_{\vec{i}}^z \to  \hat{J}_{\vec{i}}^z$ unchanged everywhere. For small $J_\parallel$ this leads to an antiferromagnetic model.

The resulting phase diagram of the two-dimensional mixD square lattice bilayer system in the tight-binding limit is shown schematically in Fig.~\ref{fig:BilayerDiscussion} b). For $J_\parallel = 0$, the effective spin system has an emergent $SU(2)$ symmetry, which is broken down to a $U(1)$ symmetry by a non-zero magnetization $m = \delta - 1/2 \neq 0$. In direct analogy with the canted antiferromagnetism discussed in Sec.~\ref{secCantedAFM} we find a BKT transition to a quasi-condensate of pairs $\bd \ket{0}$ (qBEC) below a critical temperature $T_c$ on the order of $K$. This transition is absent at $\delta = 0.5$, unless the $SU(2)$ symmetry is broken down to $U(1)$ by the anisotropy $\Delta \neq 0$, which requires $J_\parallel \neq 0$. The $T=0$ ground state at $\delta=0.5$ corresponds to long-range charge-density wave (CDW) order within the planes. At low (high) doping the qBEC is best viewed as a quasi condensate with power-law correlations, of molecular hole- (fermion-) pairs.

\subsection{Weak coupling limit: BCS mean field analysis}
\label{secWeakCplgBCS}
In the previous section we have shown that a BEC phase with tightly bound fermion- or hole pairs forms at all dopings in the tight-binding limit, where $J_\perp$ dominates. Now we consider the situation where $t_\parallel$ dominates, and both $J_\perp$ and $J_\parallel$ can be treated as weak perturbations in the two-dimensional mixD bilayer. In this section we perform a BCS mean-field analysis of this situation, which, in the limit of $J_\perp \rightarrow 0$, includes the BCS analysis of the decoupled single-plane $t-J$ model where $d_{x^2-y^2}$ pairing is found \cite{Coleman2015}. Before starting, we discuss under which conditions the weak-coupling BCS ansatz is valid. 

Firstly, for $t_\parallel$ to dominate we require sizable doping values $\delta > 0$, since $t_\parallel$ is suppressed when the system forms a Mott insulator around $\delta = 0$. Secondly, in two dimensions an arbitrarily weak attractive interaction leads to the formation of a two-body bound state. Since $J_\perp$ still favors the formation of inter-layer singlets, it mediates an effective attraction between fermions from the top and bottom layers. Hence, for $1-\delta \ll 1$ the system is always in a BEC regime, namely when the typical inter-particle distance $(1-\delta)^{-1/2}$ is larger than the extend of the two-body bound state. 

The following fully self-consistent BCS mean-field analysis is thus constrained to the regime $\delta \lesssim 1$, as defined in Eq.~\eqref{eq:WeakCplgmDbiLay}. In practise we only consider the high-doping regime beyond $\delta>20\%$. We take into account all interaction terms from Hamiltonian~(\ref{eq:H_bilayer}) and also include the 3-site term arising to lowest order in $t_\parallel^2/U \propto J_\parallel$ in the effective Hamiltonian obtained from the Hubbard model \cite{Auerbach1998}. The latter is shown in the middle line of Tab.~\ref{tab:BCS_pairing}, and is often neglected in the $t-J$ model.

By applying an unrestricted Hartree-Fock approximation, we determine the different magnetic interaction channels with competing pairing symmetries, which are summarized in Tab.~\ref{tab:BCS_pairing} (see also Appendix~\ref{appBCS}).
The antiferromagnetic~$\Jperp$ interaction prefers singlet bonds between the layers manifested in isotropic point-like attractive interactions with $\mathrm{s}^\perp$-wave order parameter.
On the other hand, the intra-layer~$\Jpar$ interaction can lead to pairing in the anisotropic $\mathrm{s}^{\parallel}_{x^2+y^2}$, $\mathrm{p}^{\parallel}_{x\pm iy}$ and $\mathrm{d}^{\parallel}_{x^2-y^2}$ channels. The corresponding order parameters are defined in the Appendix in Eqs.~(\ref{eq:dperp_def},~\ref{eq:dpar_def}). 
By allowing the order parameters to be non-zero simultaneously, the mean-field approach can capture nematic or time-reversal symmetry broken phases within the layer, or co-existing phases between $\mathrm{s}^\perp$ and $(\mathrm{s}^{\parallel}, \mathrm{p}^{\parallel},\mathrm{d}^{\parallel})$-wave.

\begin{table}[t]
    \centering
    {\tabulinesep=1.2mm
    \begin{tabu}{|c|c|c|c|}
        \hline
        $\H_\mathrm{int}$ & pairing symmetry & spin & layer \\
        \hline $\Jpar\sum\limits_{\ij,\mu} \bigg( \S_{\vec{i},\mu} \cdot \S_{\vec{j},\mu} - \frac{1}{4} \sum\limits_{\alpha,\beta}\n_{\vec{i},\alpha,\mu}\n_{\vec{j},\beta,\mu} \bigg)$ & $s^\parallel$-wave /$d^\parallel$-wave & singlet & symmetric \\
        \hline $-\dfrac{\Jpar}{4}\sum\limits_{\substack{\ij, \langle i,j' \rangle \\ j\neq j', \alpha,\mu}} \bigg( \cd_{j', \alpha,\mu} \n_{i, \bar{\alpha},\mu} \c_{j, \alpha,\mu}+\cd_{j', \alpha,\mu} \cd_{i, \bar{\alpha}, \mu} \c_{i, \alpha,\mu} \c_{j, \bar{\alpha},\mu} \bigg)$ & $p^\parallel$-wave & triplet & symmetric \\
       \hline $\Jperp\sum\limits_{\vec{j}} \bigg( \S_{\vec{j},1} \cdot \S_{\vec{j},2} - \frac{1}{4} \sum\limits_{\alpha,\beta}\n_{\vec{j},\alpha,1}\n_{\vec{j},\beta,2} \bigg)$ & $s^\perp$-wave & singlet & symmetric \\ \hline
    \end{tabu}}
    \caption{\textbf{Pairing channels in the BCS analysis.} Magnetic interactions within and between the layers have competing pairing symmetries []order parameters defined in the Appendix, see Eqs.~\eqref{eq:dperp_def},~\eqref{eq:dpar_def}], which can be derived by taking all possible pairwise contractions of the fermionic operators. The combination of angular momentum, spin and layer exchange symmetry have to obey the Pauli principle, which is confirmed for all four interaction channels listed in the Table. }
    \label{tab:BCS_pairing}
\end{table}

At low dopings, our mean-field theory formulated in terms of the bare fermions can be replaced by a slave-boson mean-field theory with bosonic chargons and fermionic spinons \cite{Coleman1984,Lee2006}. In this case, a well-defined trial wavefunction can be obtained by applying a Gutzwiller projection to ensure that only one fermion can occupy each site. At high dopings, the Gutzwiller projection should not play a dominant role and we completely neglect it in the first version of our mean-field theory. In addition, we perform a second calculation with an effective, free fermion hopping term~$\tpar^\mathrm{eff}$ for comparison. We take into account the suppressed hopping of the fermionic spinons due to occupied sites and approximate~$\tpar^\mathrm{eff}=\delta\cdot\tpar$~\cite{Baskaran1987}, which interpolates between the exact limits for~$\delta=0$ and~$\delta=1$. The goal of this calculation is to understand on a qualitative, not a quantitative, level how the Gutzwiller projection may affect the results.

In our mean-field analysis, we consider $T=0$ and find a phase transition from $\mathrm{s}^\perp$ to $\mathrm{d}^{\parallel}_{x^2-y^2}$ ordering for increasing interactions~$\Jpar/\Jperp$, see Fig.~\ref{fig:BilayerDiscussion}c) and d). This reflects the different dominant magnetic ordering -- into local singlets and a long-range ordered state, respectively -- at half-filling (see Sec.~\ref{sec:BilayerTransition}).
Interestingly, for the renormalized fermion hopping, $\tpar^\mathrm{eff}=\delta\cdot\tpar$, the system favours $\mathrm{s}^{\parallel}$-wave superconductivity in an intermediate regime at moderate dopings and comparable interactions~$\Jpar \approx \Jperp$, see Fig.~\ref{fig:BilayerDiscussion}d). Additionally, for large enough intra-plane interactions~$\Jpar$, the systems shows nematic $(\mathrm{s}^\parallel+\mathrm{d}^{\parallel})$-wave order for low dopings, whereas pure $\mathrm{d}^{\parallel}$-wave order is obtained again for high dopings, again if the reduced hopping $\tpar^\mathrm{eff}$ is used, see Fig.~\ref{fig:BilayerDiscussion}d).
Moreover, by solving the self-consistency equations while minimizing the ground-state energy, we find throughout that $\mathrm{p}^{\parallel}$-wave order is not favored.

As in the case of half-filling, where tuning the ratio of couplings $\Jpar/\Jperp$ can drive a phase transition, we expect interesting physics to arise in the high-doping limit.
Since the system becomes $\mathrm{d}^{\parallel}$-wave superconducting at values~$\Jpar/\Jperp \gtrsim 1$, the mixed-dimensional bilayer system is appealing to study cuprate physics from a new perspective and it could give insights and impulses into the pairing mechanisms of high-Tc superconductivity when approaching the superconducting dome by decreasing $J_\perp$.

\subsection{Strong coupling limit}
Arguably the most interesting, but also the most challenging to describe, regime corresponds to strong coupling, see Eq.~\eqref{eq:StrongCplgmDbiLay}, where $t_\parallel$ dominates over $J_\perp$ and $J_\parallel$ while being at small doping, $\delta \ll 1$. As we show in a separate publication \cite{Bohrdt2021pairingPrep}, pairing of holes does not break down in this limit. Remarkably, even larger binding energies exceeding $|E_{\rm bdg}| > J_\perp$ can be realized for sufficiently large $t_\parallel / J_\perp$. More specifically, we show in Ref.~\cite{Bohrdt2021pairingPrep} that the binding energy in this limit scales as 
\begin{equation}
    E_{\rm bdg} \simeq - t_\parallel^{1/3} J_\perp^{2/3}.
\end{equation}

\begin{figure}
\centering
    \includegraphics[width=0.8\linewidth]{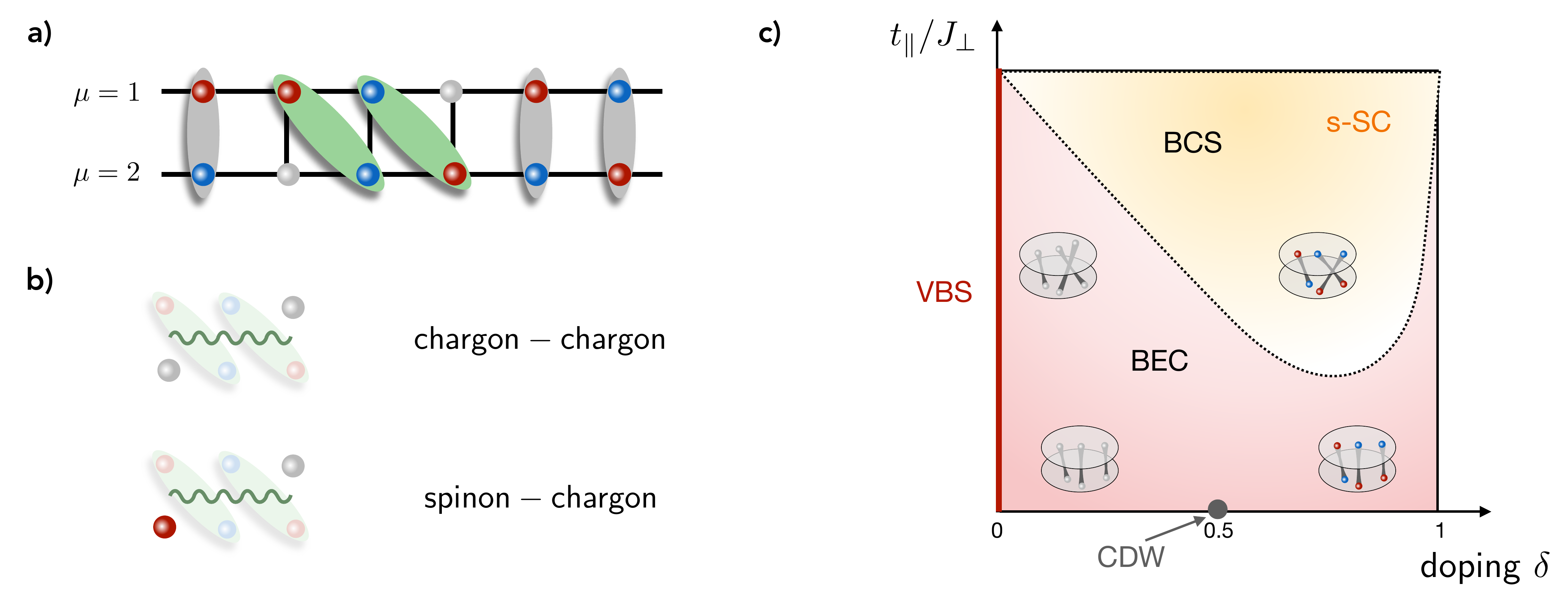}
  \caption{\textbf{Strong coupling regime and BEC-to-BCS cross-over.} a) In the strong coupling limit, holes are strongly bound, but their bound state extends over several lattice sites. Here a cut through the bilayer system is shown. b) The allowed states at strong coupling are best represented by their parton content. They can either form a spin-less meson consisting of two chargons, or a spin-$1/2$ spinon-chargon meson carriying the same quantum numbers as the underlying fermions. c) We show schematically the phase diagram of the mixD bilayer $t-J$ model for $J_\parallel =0$, which includes a BEC-to-BCS cross-over for sufficiently large values of $t_\parallel / J_\perp$ (dotted line). At $\delta = 0.5$ in the tight-binding regime, i.e. $t_\parallel / J_\perp \ll 1$, the system as an emergent $SU(2)$ symmetry and a charge-density wave (CDW) can form.
  }
  \label{fig:SCdiscussion}
\end{figure}

The underlying pairing mechanism relies on the strong frustration obtained by the hole motion in the regime where $J_\parallel = 0$. To move through the rung-singlet VBS background, an individual hole must tilt all singlets along its path, which costs an energy $J_\perp$ per singlet and leads to a strongly confining force, see Fig.~\ref{fig:SCdiscussion} a). To estimate the binding energy, one has to compare two scenarios: (i) when each hole binds to a spinon, i.e. an unpaired spin which is not part of a singlet; or (ii) when two holes are connected by a string, defined in the two-dimensional plane, of tilted singlets. These states are best characterized in terms of their parton content, (i) corresponds to a spinon-chargon meson with a spin-$1/2$ quantum number, while (ii) describes a spin-less chargon-chargon bound state; see Fig.~\ref{fig:SCdiscussion} b). The detailed analysis of these bound states can be found in Ref.~\cite{Bohrdt2021pairingPrep}.

At a non-zero but small density $\delta > 0$ of holes, doped with equal concentrations into both layers, we expect the formation of a quasi-condensate of molecular chargon-chargon (cc) mesons. In contrast to the tightly bound chargon pairs obtained in the tight-binding case, see Sec.~\ref{secTightBndgLmt}, the extended cc mesons in the strong coupling regime are highly mobile \cite{Bohrdt2021pairingPrep}. Hence we expect the formation of a quasicondensate below a BKT transition temperature $T_c$ which is significantly larger than in the tight-binding case. This is of significant importance for achieving a (quasi-) condensate of hole pairs experimentally with cold atoms. 

When the doping increases, the system crosses over from a BEC to a BCS regime. The location of the crossover is expected where the typical spacing between two cc mesons, $(\delta / 2)^{-1/2}$, becomes comparable to the average size of the chargon-chargon bound state. The latter increases with $t_\parallel$, which means the BEC-to-BCS cross-over shifts to smaller doping values as $t_\parallel / J_\perp$ increases. The expected phase diagram at finite doping and for $J_\parallel = 0$ is shown schematically in Fig.~\ref{fig:SCdiscussion} c).

\subsection{Speculations}
We close this section by speculating which additional features may show up in the finite-doping phase diagram when $J_\perp/J_\parallel$ is tuned. For $J_\perp=0$, two independent $t-J$ models are obtained, and their ground state phase diagram remains not fully understood. So naturally, the situation is even more dire in the central region, where $\delta \simeq 20\%$ and $J_\perp \simeq J_\parallel$ are comparable. In the following we will consider only the case when $t_\parallel/(J_\parallel+J_\perp) \simeq 3$ is sizable and around the value for $t_\parallel / J_\parallel$ expected in cuprate compounds. 

The results of our analysis so far, as expected in this parameter regime, are summarized schematically in Fig.~\ref{fig1Bilayer} b). At high doping, our BCS analysis predicts a transition where the pairing symmetry changes from inter-layer $s^\perp$ to intra-layer $d$-wave pairing. At intermediate doping, but still for $J_\perp / J_\parallel \approx 1$, we speculate that strong Hubbard interactions may lead to intra-layer $s$-wave pairing and possibly a nematic superconducting phase at lower lower $J_\perp / J_\parallel$. In this regime where $J_\parallel$ begins to dominate, we also expect a competition with the stripe, or charge-density wave, phases of the single-layer $t-J$ model. Their robustness to weak inter-layer coupling $J_\perp$ needs to be clarified in future studies. 

When $J_\perp \gg J_\parallel$, we predict a BEC-to-BCS cross-over around a doping $\delta_c$ which depends sensitively on $t_\parallel / J_\perp$, since the latter controls the size of the chargon-chargon bound states in the low-doping BEC regime. The physics of this cross-over is potentially very rich: not only do the chargon bound states start to overlap, but the strong and non-local spin-charge correlations underlying the binding mechanism at low doping are also expected to significantly modify the properties of the spin background. This, in turn, may drive the transition from the paired to the unpaired regime. 

One of the most interesting questions in this context is to which extend the BEC-to-BCS cross-over can be described by the parton picture. When $J_\perp \gg J_\parallel$ the system favors inter-layer singlet formation on both sides of the cross-over, which indeed suggests that the most relevant constituents may be chargon-chargon and spinon-chargon pairs in a background of rung-singlets. Hence an appealing scenario would be that the BEC-to-BCS cross-over may be related to a Feshbach resonance of spinon-chargon pairs. The large associated energy scales make an experimental exploration using ultracold atoms rather realistic in the immediate future.

The parton theory may also prove to be useful to study the properties of a few holes when $J_\perp / J_\parallel$ is tuned around $\delta \approx 0$. Both in the VBS and AFM spin-backgrounds, the few-body states can be described by the parton picture, either as spinon-chargon or chargon-chargon bound states. Therefore it is an interesting question to ask how they behave in the vicinity of the AFM-to-VBS transition. To gain some intuition, we note that for $J_\parallel =0$, spinon-chargon pairs connected by strings within one layer form, see Sec.~\ref{secMagneticPolaron}. Adding $J_\perp$ terms will favor parallel strings in the top- and bottom layers, which should effectively introduce an attraction between spinon-chargon pairs from the two different layers. As long as $J_\parallel$ dominates, the spinons and chargons are expected to carry well-defined layer indices. In this case, we speculate that a tetraquark bound state constituted by one spinon-chargon pair per layer may form, as indicated in Fig.~\ref{fig1Bilayer} b). In contrast, when $J_\perp \gg J_\parallel$ dominates, only a spinon and a chargon from opposite layers can form a bound state. More importantly, a chargon from the top layer can re-trace a string of tilted singlets created by a chargon in the bottom layer. This would transform a potential tetraquark state into the chargon-chargon meson when $J_\perp$ is increased, by continuously merging a pair of opposite spinons from the different layers into an inter-layer singlet.

\section{Summary and Outlook}\label{secSummary}
Quantum gas microscopy has started a new chapter in the experimental investigation of strongly correlated quantum matter, and the doped Hubbard model in particular. Already now, its key advantages are manifest: the ability to realize a broad range of microscopic models, with known and tunable parameters and without coupling to additional degrees of freedom such as phonons; and the availability of new probes which allow to characterize strongly correlated many-body states in ways that are not available in solid state systems.
In combination we believe that these two main factors will become driving forces to advance our theoretical understanding of the doped Hubbard model, and its cousin, the $t-J$ model. Other experimental approaches, e.g. emulations of the fermionic Hubbard model on qubit-based bosonic quantum simulators \cite{Arute2020,Suchsland2021}, can also be envisioned to play a role in the future, although  quantum gas microscopes currently appear to have an advantage.

Where is the field headed? We see countless promising directions as well as a few challenges ahead. Let us start by discussing the challenges, and possible ways to overcome them. So far the relatively high temperature of the systems compared to the relevant energy scales has been a major obstacle. To explore the ordered phases of the Hubbard model, temperatures well below the spin-exchange energy will be required, $T \lesssim J/10$. In the long term, we do not believe that reaching lower temperatures will become a fundamental issue. Already now, several encouraging results have been obtained which point towards the ability to implement powerful adiabatic preparation schemes that will allow to realize low entropy states. These include in particular schemes starting from an ultra-low entropy band insulator and transforming it into a local singlet covering of the system \cite{Chiu2018PRL}, or coupling to a second layer which may provide a gapless entropy sink and can be used for efficient cooling \cite{Kantian2018}. In the low-doping regime in particular, we view the classical N\'eel product state as a valuable starting point for adiabatic protocols to reach symmetry-broken ground states of the $SU(2)$-invariant Hubbard model. On the experimental side, issues related to atom loss \cite{Pichler2010} lead to some limitations. However, sympathetic cooling with a second ultracold atomic species also appears to be a promising approach to move forward. Aside from temperature, a short-term challenge will be to obtain larger system sizes and at the same time be able to eliminate and engineer the optical trapping potential without introducing disorder. However, these appear to be technical challenges, and already today several hundreds of sites can be studied in quantum gas microscopes, see e.g. \cite{Ji2020}. Another very recent development has been the use of optical tweezer arrays to improve and speed up the loading of optical lattices. This promises lower achievable entropies per particle and significantly increased experimental cycle times, and hence higher-quality data. 

As discussed in detail in the main part of our paper, we see many promising future directions that Fermi-Hubbard quantum gas microscopes can take. One general approach is to simplify the considered Hamiltonian in order to connect to theoretically more tractable regimes. Concrete examples include the use of Rydberg dressing in order to obtain anisotropic, and if desired long-ranged, spin-spin interactions \cite{GuardadoSanchez2020Ry}. This offers a route to implement the $t-J_z$ model where $SU(2)$ invariant Heisenberg couplings are completely replaced by Ising couplings between the spins. Despite being significantly simpler than the full-blown $t-J$ model, the physics of the $t-J_z$ Hamiltonian at finite doping remain poorly explored. Another example is provided by the use of gradients to suppress charge motion along one direction without affecting the super-exchange mechanism: this leads to mixed-dimensional systems. As we discussed extensively in this paper, this approach allows to reach binding energies between holes above the currently achievable temperatures in experiments. Therefore it opens the door to systematic explorations of possible pairing mechanism in doped antiferromagnets. In the context of a mixed-dimensional $t-J_z$ model, it was further shown that the critical temperature for stripe formation is of the order of $J_z$, bringing the ordered phases of the doped Hubbard model within reach of current experiments \cite{Grusdt2020}. Instead of simplifying terms in the Hamiltonian, one can also add additional couplings or modify the lattice geometry. For example, using dipolar atoms allows for nearest neighbour interactions, which may stabilize some ordered states. By working on a triangular \cite{Yang2021} or even Kagom\'e lattice \cite{Jo2012}, strongly frustrated spin systems can be realized. Doping these models is expected to lead to rich physics, and could potentially help theorists find a unified description of a larger class of doped antiferromagnets. 

How do quantum gas microscopes add to our theoretical understanding of the doped Hubbard model? In our view, the main strength of current experiments is to reveal, for the first time, in microscopic details the properties of the individual constituents. In this regard, the observation of a magnetic polaron and its dressing cloud in real space \cite{Koepsell2019} provides a milestone: From these measurements it has become clear that the charge carriers in the under-doped Hubbard model have a significant spatial extent, and potentially a rich internal structure. On the one hand, by performing spectroscopic experiments, such as ARPES \cite{Bohrdt2018,Brown2019}, we expect that quantum gas microscopes will soon be able to reveal this structure in more detail, with full energy resolution. The ability to perform lattice modulations and apply gradients will allow to go beyond the spectroscopic capabilities in solids, e.g. by studying rotational \cite{Bohrdt2021arXiv} or new time-dependent \cite{Schuckert2021} variants of ARPES. On the other hand, the recent experiments pose new theoretical challenges. For example, the properties of magnetic polarons at finite temperatures comparable to the super-exchange coupling $J$ remain poorly understood. A description of how the microscopic structure of a single charge carrier and its dressing cloud influences pairing and localization of holes at low doping also remains largely lacking. Here, too, we expect future quantum gas microscopy experiments to be able to shed more light on the microscopic mechanisms leading to the self-organization of holes into charge-density waves as well paired and condensed phases of matter. In a larger framework, theorists are still struggling to describe the regime around optimal doping, and the cross-over from a magnetic polaron regime to the Fermi liquid at high doping. This also includes the strange metal regime. First signatures of this cross-over and its peculiar transport properties have recently been observed with cold atoms \cite{Brown2019a,Koepsell2020_FL}. In the near future, we expect that quantum gas microscopy can study in even more detail how and where the structure of charge carriers changes as a function of doping. This will help us understand which key aspects to include in a microscopic description of the cross-over region.

\section{Acknowledgements}\label{Acknow}
We are grateful for countless inspiring discussions and many fruitful collaborations with our friends and colleagues. While listing everyone who made a contribution to this fascinating and rapidly evolving field is impossible, we would like to highlight in particular I. Bloch, M. Greiner, C. Gross and D. Greif, and not miss on expressing particular thanks to: W. Bakr, P. Bojovic, D. Bourgund, T. Chalopin, C. Chiu, I. Cirac, J. v. Delft, T. Esslinger, T. Hilker, S. Hirthe, C. Hubig, G. Ji, M. Kanasz-Nagy, L. Kendrick, W. Ketterle, E.-A. Kim, M. Knap, J. Koepsell,  M. Lebrat, I. Lovas, A. Mazurenko, C. Miles, A. Omran, M. Parsons, L. Pollet, F. Pollmann, M. Punk, S. Sachdev, G. Salomon, R. Schmidt, U. Schollw\"ock, T. Shi, P. Sompet, L. Tarruell, J. Vijayan, Y. Wang, K. Wohlfeld, M. Xu, Z. Zhu, and M. Zwierlein.

We acknowledge funding by the Deutsche Forschungsgemeinschaft (DFG, German Research Foundation) under Germany's Excellence Strategy -- EXC-2111 -- 390814868, the NSF through a grant for the Institute for Theoretical Atomic, Molecular, and Optical Physics at Harvard University and the Smithsonian Astrophysical Observatory. ED acknowledges support from the ARO grant number W911NF-20-1-0163, the National Science Foundation through grant No. OAC-1934714, the NSF Grant EAGER-QAC-QCH award number 2037687, Harvard-MIT CUA.

\section{Appendix}\label{appendix}

\subsection{BCS mean-field analysis: mixed-dimensional bilayer model at high doping}\label{appBCS}

Here we derive a BCS mean-field Hamiltonian from the mixed-dimensional bilayer $t-J$ model in Eq.~(\ref{eq:H_bilayer}) and solve it self-consistently by using a Bogoliubov transformation.
There are three magnetic interaction terms to consider: inter-layer~$\Jperp$ AFM interactions, intra-layer~$\Jpar$ AFM interactions, and the intra-layer~$\Jpar$ 3-site term, see Table~\ref{tab:BCS_pairing}.
In the following, we will use a Schwinger fermion representation of the spins, $\S_{j,\mu} = \frac{1}{2}\cd_{j,\alpha,\mu} \vec{\sigma}_{\alpha\beta} \c_{j,\beta,\mu}$, where~$\vec{\sigma}=(\sigma^x,\sigma^y,\sigma^z)$ are the Pauli matrices, $\alpha,\beta=\uparrow,\downarrow$ is the spin and $\mu=1,2$ is the layer index.

The inter-layer coupling can be exactly written in terms of singlet pair operators on lattice site~$\vec{j}$:
\begin{align}
    \H_{\Jperp} &= \Jperp \sum_{\vec{j}} \bigg( \S_{\vec{j},1} \cdot \S_{\vec{j},2} - \frac{1}{4} \sum_{\alpha,\beta}\n_{\vec{j},\alpha,1}\n_{\vec{j},\beta,2} \bigg) = -\Jperp \sum_{\vec{j}} \sd_{\vec{j}} \s_{\vec{j}} = -\Jperp N \sum_{\vec{k}} \sd_{\vec{k}} \s_{\vec{k}}
\end{align}
with
\begin{align}
    \sd_{\vec{j}} &= \frac{1}{\sqrt{2}} \left( \cd_{\vec{j},\uparrow,1}\cd_{\vec{j},\downarrow,2} - \cd_{\vec{j},\downarrow,1}\cd_{\vec{j},\uparrow,2} \right) \\
    \sd_{\vec{k}} &= \frac{1}{N\sqrt{2}} \sum_{\vec{q}}\left( \cd_{\vec{q}+\vec{k},\uparrow,1}\cd_{-\vec{q},\downarrow,2} - \cd_{\vec{q}+\vec{k},\downarrow,1}\cd_{-\vec{q},\uparrow,2} \right),
\end{align}
where~$N=L_x \times L_y$ denotes the system size in $2$D with lattice size~$L_x$ ($L_y$) in the $x$- ($y$-)direction, respectively.
Further, the fermion creation operators written in momentum space are $\c_{\vec{k},\alpha,\mu} = N^{-1/2} \sum_{\vec{j}} e^{-i \vec{j} \cdot \vec{k}}\c_{\vec{j},\alpha,\mu}$ with~$\vec{k} = (k_x,k_y)$ and~$k_j = -\pi /L_j,...,\pi /L_j$. 
By defining the $s^\perp$-wave mean-field order parameter
\begin{align}
    \Delta^\perp &= \frac{J_\perp}{\sqrt{2}} \langle \hat{s}_{\vec{k}=\vec{0}}\rangle = \frac{\Jperp}{2N}\sum_{\vec{q}} \langle \c_{\vec{q},\downarrow,2}\c_{-\vec{q},\uparrow,1} - \c_{\vec{q},\uparrow,2}\c_{-\vec{q},\downarrow,1} \rangle, \label{eq:dperp_def}
\end{align}
we can derive the mean-field Hamiltonian
\begin{align}
    \H^{\mathrm{MF}}_{\Jperp} = - \sum_{\vec{k}} \left[ \Delta^\perp\left( \cd_{\vec{k},\uparrow,1}\cd_{-\vec{k},\downarrow,2} - \cd_{\vec{k},\downarrow,1}\cd_{-\vec{k},\uparrow,2} \right)+ \mathrm{H.c.} \right] + \dfrac{2}{\Jperp}|\Delta^\perp|^2.
\end{align}
The order parameter has $s^\perp$-wave symmetry, is anti-symmetric under the exchange of spins (spin singlets) and must be -- due to Pauli's principle -- symmetric under layer exchange.

Next, we consider the intra-layer~$\Jpar$ AFM couplings, which can be written in momentum space as
\begin{align}
    \H_{\Jpar} = \frac{\Jpar}{2} \sum_{\vec{q},\mu} V(\vec{q}) \bigg( \S_{-\vec{q},\mu} \cdot \S_{\vec{q},\mu} - \frac{1}{4}\sum_{\alpha,\beta}\n_{-\vec{q},\alpha,\mu}\n_{\vec{q},\beta,\mu} \bigg)
\end{align}
with $V(\vec{q})=2\left[ \cos{(q_x)} + \cos{(q_y)} \right]$.
Normal ordering and collecting all terms, taking into account the fermionic exchange statistics, yields
\begin{align}
   \H_{\Jpar} &= -\frac{\Jpar}{4N}\sum_{\vec{q},\vec{k_1},\vec{k_2}}\sum_{\alpha,\beta,\mu} V(\vec{q})\left( \cd_{\vec{k_1}+\vec{q},\alpha,\mu} \cd_{\vec{k_2}-\vec{q},\beta,\mu} \c_{\vec{k_1},\beta,\mu} \c_{\vec{k_2},\alpha,\mu} + \cd_{\vec{k_1}+\vec{q},\alpha,\mu} \cd_{\vec{k_2}-\vec{q},\beta,\mu} \c_{\vec{k_2},\beta,\mu} \c_{\vec{k_1},\alpha,\mu} \right).
\end{align}
This exact Hamiltonian can now be used to apply mean-field theory and further we only consider Cooper pairs with net momentum zero.
With these approximations, the Hamiltonian becomes
\begin{align}
\begin{split}
   \H_{\Jpar}^{\mathrm{MF}} = -\frac{\Jpar}{4N}\sum_{\vec{k},\vec{k}'}\sum_{\mu}&\bigg\{ 2\big[ V(\vec{k}'-\vec{k}) + V(\vec{k}'+\vec{k}) \big] \langle \cd_{\vec{k}',\uparrow, \mu} \cd_{-\vec{k}',\downarrow, \mu} \rangle \c_{-\vec{k},\downarrow,\mu} \c_{\vec{k},\uparrow,\mu} \\ 
    & +\big[ V(\vec{k}'-\vec{k}) + V(\vec{k}'+\vec{k}) \big]\langle\cd_{\vec{k}',\uparrow, \mu} \cd_{-\vec{k}',\uparrow, \mu} \rangle \c_{-\vec{k},\uparrow,\mu} \c_{\vec{k},\uparrow,\mu} \\
    & +\big[ V(\vec{k}'-\vec{k}) + V(\vec{k}'+\vec{k}) \big] \langle\cd_{\vec{k}',\downarrow, \mu} \cd_{-\vec{k}',\downarrow, \mu} \rangle \c_{-\vec{k},\downarrow,\mu} \c_{\vec{k},\downarrow,\mu} 
    + \mathrm{H.c.} \bigg\} + \mathrm{const.} \label{eq:supp_Jpar_beforeMF},
\end{split}
\end{align}
where the constant term will be specified later.
We can see that the coupling amplitude is symmetric under~$\vec{k}' \leftrightarrow -\vec{k}'$ and thus the triplet terms must vanish under summation since they have odd angular momentum. 
In particular, the interaction can be decomposed into an~$s^\parallel$-wave and $d^\parallel$-wave contribution
\begin{align}
    V_{\vec{k},\vec{k}'} &= V^s_{\vec{k},\vec{k}'} + V^d_{\vec{k},\vec{k}'} = V(\vec{k}'-\vec{k}) + V(\vec{k}'+\vec{k}) \label{eq:Vkk} \\
    V^s_{\vec{k},\vec{k}'} &= 2 \left[ \cos{(k_x)} + \cos{(k_y)} \right] \left[ \cos{(k'_x)} + \cos{(k'_y)} \right] \\
    V^d_{\vec{k},\vec{k}'} &= 2 \left[ \cos{(k_x)} - \cos{(k_y)} \right] \left[ \cos{(k'_x)} - \cos{(k'_y)} \right].
\end{align}
Taking the symmetry in Eq.~(\ref{eq:supp_Jpar_beforeMF}) into account, the mean-field Hamiltonian can be written as
\begin{align}
   \H_{\Jpar}^{\mathrm{MF}} = - \sum_{\vec{k},\mu} \left[ \Delta^\parallel_{\vec{k},\mu} \cd_{\vec{k},\uparrow,\mu}\cd_{-\vec{k},\downarrow,\mu} + \mathrm{H.c.}  \right] + \frac{2}{\Jpar} \left( |\Delta^{s,\parallel}|^2 + |\Delta^{d,\parallel}|^2 \right), \label{eq:supp_HJpar_MF}
\end{align}
where we have defined the momentum dependent order parameter~$\Delta^{\parallel}_{\vec{k}} = \Delta^{s,\parallel}_{\vec{k}} + e^{i\beta_d}\Delta^{d,\parallel}_{\vec{k}}$ with phase~$\beta_d$ as
\begin{align}
    \Delta^\parallel_{\vec{k}} = \Delta^\parallel_{\vec{k},(\mu=1,2)} = \frac{\Jpar}{2N} \sum_{\vec{k}'} V_{\vec{k},\vec{k}'} \langle \c_{\vec{k}',\downarrow,\mu} \c_{-\vec{k}',\uparrow,\mu} \rangle, \label{eq:dpar_def}
\end{align}
where we drop the layer index~$\mu$ because~$\Delta^\parallel_{\vec{k},\mu}$ does not depend on~$\mu$ due to the symmetry between~$\mu=1 \leftrightarrow \mu=2$ in the system, which we will justify more rigorously below.
Moreover, the order parameter written without momentum subscript~$\vec{k}$ [Eq.~(\ref{eq:supp_HJpar_MF})] describes only the amplitude, e.g.\ $\Delta^{d,\parallel}_{\vec{k}} = \Delta^{d,\parallel} (\cos{k_x} - \cos{k_y})$.
Note that we have introduced the superscripts,~$\perp$ and $\parallel$, in the order parameters as well as in the spatial symmetry ($s^\perp$-, $s^\parallel$-, $d^\parallel$-wave) to point out whether the Cooper pair is a bound state between the layers ($\perp$) or within the layer ($\parallel$).

The last term to consider is the often neglected 3-site term, which arises in the derivation of the $t-J$ model in second-order perturbation theory from the Hubbard model.
A posteriori, we find that the 3-site term indeed does not lead to $p^\parallel$-wave order in the mean-field calculation.
The term is given by
\begin{align}
    \H_{3s} = -\frac{\Jpar}{4}\sum_{\substack{\ij, \langle \vec{i},\vec{j}' \rangle \\ \vec{j}\neq \vec{j}', \alpha}}\sum_\mu \left( \cd_{\vec{j}', \alpha,\mu} \n_{\vec{i}, \bar{\alpha},\mu} \c_{\vec{j}, \alpha,\mu}+\cd_{\vec{j}', \alpha,\mu} \cd_{\vec{i}, \bar{\alpha}, \mu} \c_{\vec{i}, \alpha,\mu} \c_{\vec{j}, \bar{\alpha},\mu} \right).
\end{align}
In momentum space, it can be written
\begin{align}
\begin{split}
    \H_{3s} = -\frac{\Jpar}{4N} \sum_{\alpha,\mu}\sum_{\vec{k}_1,\vec{k}_2,\vec{k}_3} &W(\vec{k}_1,\vec{k}_3)\times\\
    &\times \left[\cd_{\vec{k}_1,\alpha,\mu} \cd_{\vec{k}_2,\bar{\alpha},\mu}\c_{\vec{k}_1+\vec{k}_2-\vec{k}_3,\bar{\alpha},\mu}\c_{\vec{k}_3,\alpha,\mu} + \cd_{\vec{k}_1,\alpha,\mu} \cd_{\vec{k}_2,\bar{\alpha},\mu}\c_{\vec{k}_1+\vec{k}_2-\vec{k}_3,\alpha,\mu}\c_{\vec{k}_3,\bar{\alpha},\mu} \right]
\end{split}
\end{align}
with
\begin{align}
\begin{split}
    W(\vec{k}_1,\vec{k}_3) &= 4\left[ \cos{(k_3^x)} \cos{(k_1^y)} + \cos{(k_3^y)}\cos{(k_1^x)} \right] \\
    &+ 2\left[ \cos{(k_1^x+k_3^x)} + \cos{(k_1^y+k_3^y)} \right].
\end{split}
\end{align}
Again, when we only consider Cooper pairs with net momentum zero, we find
\begin{align}
    \H_{3s} = -\frac{\Jpar}{4N} \sum_{\alpha,\mu}\sum_{\vec{k},\vec{k}'} \left[ W(\vec{k},\vec{k}') - W(\vec{k},-\vec{k}') \right] \langle \cd_{\vec{k},\alpha,\mu} \cd_{-\vec{k},\bar{\alpha},\mu} \rangle \c_{-\vec{k}',\bar{\alpha},\mu}\c_{\vec{k}',\alpha,\mu} + \mathrm{const.}
\end{align}
Since the interaction strength~$W(\vec{k},\vec{k}') - W(\vec{k},-\vec{k}')$ is anti-symmetric under momentum exchange and the equation is symmetric under layer exchange, we can conclude that the order parameter describes symmetric coupling in the triplet channel to fulfill Pauli's principle.
Furthermore, the interaction potential can be re-written in terms of a $p^\parallel$-wave interaction
\begin{align}
    V^p_{\vec{k},\vec{k}'} &= W(\vec{k},\vec{k}') - W(\vec{k},-\vec{k}') = 2\left[ \sin{(k_x)} \sin{(k'_x)} + \sin{(k_y)}\sin{(k'_y)} \right] \\
    &= 2(1-i)\left[ Y_1^{1}(\vec{k})  Y_1^{1}(\vec{k}')  + Y_1^{-1}(\vec{k})  Y_1^{1}(\vec{k}') \right] + 2(1+i)\left[ Y_1^{1}(\vec{k})  Y_1^{-1}(\vec{k}') + Y_1^{-1}(\vec{k})  Y_1^{-1}(\vec{k}')  \right]
\end{align}
with $Y_1^{\pm1}(\vec{\vec{k}}) = \frac{1}{2}\left[\sin{(k_x)} \pm i\sin{(k_y)}\right]$.
The mean-field Hamiltonian then reads
\begin{align}
   \H_{3s}^{\mathrm{MF}} = - \sum_{\vec{k},\mu} \left[ \Delta^{p,\parallel}_{\vec{k},\mu} \cd_{\vec{k},\uparrow,\mu}\cd_{-\vec{k},\downarrow,\mu} + \mathrm{H.c.}  \right] + \frac{2}{\Jpar} |\Delta^{p,\parallel}|^2.
\end{align}

All together, the mean-field Hamiltonian becomes
\begin{align}
\begin{split}
    \H^{\mathrm{BCS}} = \sum_{\vec{k},\alpha,\mu} \epsilon_{\vec{k}} \cd_{\vec{k},\alpha,\mu}\c_{\vec{k},\alpha,\mu} &- \sum_{\vec{k}} \left[ \Delta^{\perp}\left( \cd_{\vec{k},\uparrow,1}\cd_{-\vec{k},\downarrow,2} - \cd_{\vec{k},\downarrow,1}\cd_{-\vec{k},\uparrow,2} + \mathrm{H.c.} \right) \right]+ \dfrac{2}{\Jperp}|\Delta^\perp|^2\\
    &- \sum_{\vec{k},\mu} \left[ \Delta^{\parallel}_{\vec{k},\mu} \cd_{\vec{k},\uparrow,\mu}\cd_{-\vec{k},\downarrow,\mu} + \mathrm{H.c.}  \right]+ \dfrac{2}{\Jpar}\sum_{\ell=s,p,d}|\Delta^{\ell,\parallel}|^2, \label{eq:H_BCS}
\end{split}
\end{align}
with the dispersion~$\epsilon_{\vec{k}}= -2t_{\mathrm{eff}}\big[ \cos{(k_x)} + \cos{(k_y)} \big] - \mu$ including the chemical potential~$\mu$.
The effective hopping $t_{\mathrm{eff}}=t$ in the first case (see Fig.~\ref{fig:BilayerDiscussion}c), or $t_{\mathrm{eff}}=\delta \cdot t$ depends on the hole doping $\delta$ to approximate the effect of the Gutzwiller projectors  (see Fig.~\ref{fig:BilayerDiscussion}d).
Note also that we have summarized the intra-layer couplings into a single term with~$\Delta^{\parallel}_{\vec{k}} = \Delta^{s,\parallel}_{\vec{k}} + e^{i\beta_d}\Delta^{d,\parallel}_{\vec{k}} + e^{i\beta_p}\Delta^{p,\parallel}_{\vec{k}}$ with complex phases between the different angular momentum channels.

In the following, we argue that we can express any self-consistent solution of the intra-layer and inter-layer order parameters in terms of ($\Delta^\perp$,$e^{i\chi}\Delta^\parallel_{\vec{k}}$) with $\Delta^\perp, \Delta^\parallel_{\vec{k}} \in \mathbb{R}$.
Let us consider the most general case and assume that the set of order parameters is $(e^{i\alpha_3}\Delta^\perp,e^{i\alpha_1}\Delta^\parallel_{\vec{k},1},e^{i\alpha_2}\Delta^\parallel_{\vec{k},2})$, where we distinguish the order parameters~$\Delta^\parallel_{\vec{k},\mu=1,2}$ of the two layers.
Now, we want to exploit the model's symmetries and consider the transformation of Hamiltonian~($\ref{eq:H_BCS}$) under unitary transformations. 
The system inherits a~$U(1)$ symmetry for each layer individually, which we can decompose into a global~$U(1)$ and relative layer~$U(1)$ symmetry.
\begin{enumerate}
    \item Global~$U(1)$ symmetry:
        \begin{align}
            \Ud_{\mathrm{gl}} \cd_{\vec{k},\alpha,\mu} \U_{\mathrm{gl}} = e^{i\beta_\mathrm{gl}}\cd_{\vec{k},\alpha,\mu}
        \end{align}
    \item Relative~$U(1)$ symmetry:
        \begin{align}
            \Ud_{\mathrm{rel}} \cd_{\vec{k},\alpha,\mu} \U_{\mathrm{rel}} = e^{i\beta_\mathrm{rel}(-1)^\mu}\cd_{\vec{k},\alpha,\mu}
        \end{align}
\end{enumerate}
Applying the two unitary transformations on Hamiltonian~($\ref{eq:H_BCS}$) yields
\begin{align}
\begin{split}
    &\Ud_{\mathrm{rel}}\Ud_{\mathrm{gl}}\H^{\mathrm{BCS}}\big[e^{i\alpha_3}\Delta^\perp, e^{i\alpha_1}\Delta^\parallel_{\vec{k},1},e^{i\alpha_2}\Delta^\parallel_{\vec{k},2} \big]\U_{\mathrm{gl}}\U_{\mathrm{rel}} \\
    &= \H^{\mathrm{BCS}}\big[e^{i(\alpha_3-\beta_{\mathrm{gl}})}\Delta^\perp, e^{i(\alpha_1-\beta_{\mathrm{gl}}-\beta_{\mathrm{rel}})}\Delta^\parallel_{\vec{k},1},e^{i(\alpha_2-\beta_{\mathrm{gl}}+\beta_{\mathrm{rel}})}\Delta^\parallel_{\vec{k},2} \big]\\
    &= \H^{\mathrm{BCS}}\big[\Delta^\perp,e^{i\chi} \Delta^\parallel_{\vec{k},1},e^{i\chi}\Delta^\parallel_{\vec{k},2} \big], \label{eq:H_BCS_unitary}
\end{split}
\end{align}
where we have chosen $\beta_{\mathrm{gl}}=\alpha_3$ and $\beta_{\mathrm{rel}}=(\alpha_1-\alpha_2)/2$ in the second equality.
Hence, we conclude that it is justified to take $e^{i\chi}\Delta^\parallel_{\vec{k},1}=e^{i\chi}\Delta^\parallel_{\vec{k},2}=e^{i\chi}\Delta^\parallel_{\vec{k}}$.

The BCS mean-field Hamiltonian~(\ref{eq:H_BCS}), which we want to solve self-consistently, is fully quadratic and can thus be solved exactly by first performing a rotation to decouple the two layers and then applying a Bogoliubov transformation.
The Hamiltonian in terms of the free fermionic Bogoliubov modes~$\hat{\gamma}_{\vec{k},\alpha,\mu}$ and dispersions~$E_{\vec{k}}^{\delta,\Delta}$ is given by
\begin{align}
    \H^{\mathrm{BCS}} &= \sum_{\vec{k},\alpha} \left( E_{\vec{k}}^\Delta\,\gamma^\dagger_{\vec{k},\alpha,1}\gamma_{\vec{k},\alpha,1} + E_{\vec{k}}^\delta\, \gamma^\dagger_{\vec{k},\alpha,2}\gamma_{\vec{k},\alpha,2} \right) + \dfrac{2N}{\Jperp}|\Delta^\perp|^2 + \dfrac{2N}{\Jpar}\sum_{\ell=s,p,d}|\Delta^{\ell,\parallel}|^2 - \sum_{\vec{k}} \left[ E_{\vec{k}}^\Delta + E_{\vec{k}}^\delta - \epsilon(\vec{k})\right] \label{eq:H_BCS_Bogol} \\
    E_{\vec{k}}^\Delta &= \sqrt{\epsilon^2(\vec{k}) + |\Delta_{\vec{k}}|^2}\\
    E_{\vec{k}}^\delta &= \sqrt{\epsilon^2(\vec{k}) + |\delta_{\vec{k}}|^2},
\end{align}
with~$\Delta_{\vec{k}} = |\Delta^{\perp}+\Delta^{\parallel}_{\vec{k}}|$ and~$\delta_{\vec{k}} = |\Delta^{\perp}-\Delta^{\parallel}_{\vec{k}}|$.
The mean-field energy at $T=0$ in the considered parameter regime is given by
\begin{align}
\begin{split}
    E^{\mathrm{BCS}} &= \dfrac{2N}{\Jperp}|\Delta^\perp|^2 + \dfrac{2N}{\Jpar}\sum_{\ell=s,p,d}|\Delta^{\ell,\parallel}|^2 \\
    &-\dfrac{N}{4\pi^2} \int_{\mathrm{BZ}} d^2\vec{k} \left( \sqrt{\epsilon^2(\vec{k}) + |\Delta^\perp + \Delta^\parallel_{\vec{k}}|^2} + \sqrt{\epsilon^2(\vec{k}) + |\Delta^\perp - \Delta^\parallel_{\vec{k}}|^2} - \epsilon(\vec{k}) \right) \label{eq:supp_MFenergy}
\end{split}
\end{align}
in the thermodynamic limit and integrating over the Brillouin zone (BZ).

Hamiltonian~(\ref{eq:H_BCS_Bogol}) is solved by the Fermi-Dirac distribution and for $T=0$, we can write down a set of self-consistency equations given a fixed particle number (hole density~$\delta$), which can be tuned via the chemical potential~$\mu$.
W.l.o.g. we fix the gauge such that~$\Delta^\perp, \Delta^{\parallel}_{\vec{k}} \in\mathbb{R}$ with relative phase $(\Delta^\perp, e^{i\chi}\Delta^{\parallel}_{\vec{k}})$ as discussed above:
\begin{align}
    \delta &= \frac{1}{8\pi^2}\int_{BZ} d^2\vec{k} \left[ \frac{\epsilon_{\vec{k}}}{2E_{\vec{k}}^\delta} + \frac{\epsilon_{\vec{k}}}{2E_{\vec{k}}^\Delta} \right]\label{eq:mu_selfconsist} \\
    \Delta^\perp &= \frac{\Jperp}{16\pi^2} \int_{BZ} d^2 \vec{k} \left[ \frac{\Delta^\perp + \Delta^\parallel_{\vec{k}}}{E_{\vec{k}}^\Delta} + \frac{\Delta^\perp - \Delta^\parallel_{\vec{k}}}{E_{\vec{k}}^\delta} \right] \label{eq:Dperp_selfconsist}\\
    \Delta^{\ell,\parallel}_{\vec{k}} &= \frac{\Jpar}{32\pi^2} \int_{BZ} d^2 \vec{k}'\, V^\ell_{\vec{k},\vec{k}'} \left[ \frac{\Delta^\perp + \Delta^\parallel_{\vec{k}'}}{E_{\vec{k}'}^\Delta} - \frac{\Delta^\perp - \Delta^\parallel_{\vec{k}'}}{E_{\vec{k}'}^\delta} \right]~~\mathrm{with}~\ell=s,p,d~,\label{eq:Dpar_selfconsist}
\end{align}
Let us now consider the possibility of solutions with simultaneously broken symmetries. 

\begin{itemize}
    \item $s^\perp+e^{i\chi}\ell^\parallel$ ($\ell=s,p,d$) co-existence: The l.h.s. of Eq.~(\ref{eq:Dperp_selfconsist}) is always purely real. Thus from requiring the imaginary part of the r.h.s. to vanish, it follows that $\chi = 0, \pm \pi/2, \pi$. However, we do not find evidence for a co-existing intra- and inter-layer superconducting phase in the numerical calculations.
    \item $\ell^\parallel \pm i \ell^{'\parallel}$ ($\ell,\ell'=s,p,d$) time-reversal symmetry broken phase: The symmetries in the model allow us to choose $\ell^\parallel\in\mathbb{R}$. Inserting $\ell^\parallel \pm i \ell^{'\parallel}$ into Eq.~(\ref{eq:Dpar_selfconsist}) gives separate equations for each coupling sector~$\ell,\ell'$. However, the l.h.s. of the equation for $\Delta^{\parallel,\ell}_{\vec{k}}$ is purely real whereas the r.h.s. is complex by construction. Thus, the order parameters must be strictly zero prohibiting a time-reversal symmetry broken phase.
    \item $\ell^\parallel\pm \ell^{'\parallel}$ ($\ell,\ell'=s,p,d$) nematic phase: The self-consistency equations allow for the co-existence of two intra-plane order parameters. To have a time-reversal broken/nematic phase, we additionally require the ground states $\ell^\parallel\pm \ell^{'\parallel}$ to be degenerate. We can see the degeneracy from  Eq.~(\ref{eq:supp_MFenergy}) by considering the transformation under~$C4$ rotational symmetry, $ E_{\vec{k}}^{\delta,\Delta}(\ell^\parallel+\ell^{'\parallel}) \rightarrow E_{\vec{k}}^{\delta,\Delta}(\ell^\parallel-\ell^{'\parallel})$. Since the integration measure is invariant under $C4$ rotation, the two configurations $\ell^\parallel\pm\ell^{'\parallel}$ yield equal mean-field energies and are thus degenerate. By numerically solving the self-consistency equations, we indeed find a nematic regime in the case $t_{\mathrm{eff}}=\delta\cdot t$ as shown in Fig.~\ref{fig:BilayerDiscussion}d) for low dopings.
\end{itemize}

The self-consistency Eqs.~(\ref{eq:mu_selfconsist}-\ref{eq:Dpar_selfconsist}) define recursive equations, which can be solved numerically, to find the mean-field solution of the ground state.
Thus, the superconducting order for a fixed set of parameters $(t,\Jpar,\Jperp,\delta)$ can be determined.
The solutions are plotted in Fig.~\ref{fig:BilayerDiscussion}c), d) in the main text.

%

\end{document}